\pdfoutput=1
\documentclass[12pt,english]{article}
\usepackage[latin1]{inputenc}
\usepackage[a4paper]{geometry}
\usepackage{babel}
\usepackage{amsmath}
\usepackage{amssymb}
\usepackage{graphicx}
\usepackage{setspace}
\usepackage{comment}
\usepackage[unicode=true, bookmarks=true,bookmarksnumbered=true,bookmarksopen=true,bookmarksopenlevel=1, breaklinks=true,pdfborder={0 0 1},colorlinks=false,hyperfootnotes=false]{hyperref}
\hypersetup{
	bookmarksnumbered=true,
	colorlinks=true,
	linkcolor=blue,
	citecolor=blue,
	urlcolor=blue,
	menucolor=blue
}
\usepackage[hang,flushmargin]{footmisc}

\makeatletter

\providecommand{\tabularnewline}{\\}

\usepackage{amscd}
\usepackage{amsthm}
\usepackage{mathrsfs}
\usepackage{amsfonts}
\usepackage[natbibapa]{apacite}
\usepackage{color}
\usepackage{parskip}
\usepackage{caption}
\usepackage{subcaption}
\usepackage{array}
\usepackage{babel}
\usepackage[colorinlistoftodos,prependcaption,textsize=tiny]{todonotes}
\usepackage{enumitem}
\usepackage{url}
\usepackage{booktabs}
\usepackage{adjustbox}
\usepackage{siunitx}
\usepackage[document]{ragged2e}
\usepackage{graphicx}

\usepackage{flafter}

\usepackage{lineno}

\setlist{nosep}

\usepackage{times}

\DeclareMathOperator*{\argmin}{arg\,min}

\newcommand{\footremember}[2]{%
	\footnote{#2}
	\newcounter{#1}
	\setcounter{#1}{\value{footnote}}%
}
\newcommand{\footrecall}[1]{%
	\footnotemark[\value{#1}]%
} 

\newcommand*{\doi}[1]{\href{https://doi.org/#1}{#1}}


\makeatother

\begin{document}
\title{\vspace{-10mm}
	Hidden Group Time Profiles: \\Heterogeneous Drawdown Behaviours in Retirement 
\vspace{-4mm}} 
\author{Igor Balnozan\footremember{UNSW}{University of New South Wales, UNSW Sydney, NSW 2052, Australia.}\footnote{Correspondence: i.balnozan@unsw.edu.au / West Lobby Level 4,
		UNSW Business School Building, UNSW	Sydney, NSW 2052, Australia.}, Denzil G. Fiebig\footrecall{UNSW},
Anthony Asher\footrecall{UNSW}, Robert Kohn\footrecall{UNSW}, Scott A. Sisson\footrecall{UNSW}
}

\date{Version: 19 March 2025\vspace{-4mm}}
\maketitle

\begin{abstract}
	\begin{onehalfspace}
	    	This article investigates retirement decumulation behaviours using the Grouped Fixed-Effects (GFE) estimator applied to Australian panel data on drawdowns from phased withdrawal retirement income products. Behaviours exhibited by the distinct latent groups identified suggest that retirees may adopt simple heuristics determining how they draw down their accumulated wealth. Two extensions to the original GFE methodology are proposed: a latent group label-matching procedure which broadens bootstrap inference to include the time profile estimates, and a modified estimation procedure for models with time-invariant additive fixed effects estimated using unbalanced data.
	\end{onehalfspace}

\end{abstract}

\vfill

\begin{singlespace}
	{\footnotesize 
		
		\textbf{Keywords}
		\vspace{-4mm}
		
		Panel data, discrete heterogeneity, microeconomics, retirement
		
		\textbf{JEL codes}
		\vspace{-4mm}
		
		C51, D14, G40

		\textbf{Acknowledgements}
		\vspace{-4mm}
		
Balnozan is grateful for the support provided by the Commonwealth Government of Australia through the provision of an Australian Government Research Training Program Scholarship, and to State Super for provision of the State Super Academic Scholarship.
Balnozan, Kohn and Sisson are partially supported by the Australian Research Council through the Australian Centre of Excellence in Mathematical and Statistical Frontiers (ACEMS; CE140100049).
We
thank Plan For Life, Actuaries \& Researchers, who collected, cleaned and allowed
us to analyse the data used in this research; this data capture forms
one part of a broader survey into retirement incomes, commissioned
by the Institute of Actuaries of Australia.
This research includes computations using the computational cluster
Katana supported by Research Technology Services at UNSW Sydney. Katana DOI: 10.26190/669x-a286

		\textbf{Conflict of interest statement}
				\vspace{-4mm}
				
		The authors have no conflict of interest to declare.

	}
\end{singlespace}

\pagebreak
\doublespacing

\section{\label{sec:Introduction}Introduction}

The importance and prevalence of preference heterogeneity is a pervasive
feature of microeconomic modelling. Such heterogeneity
is usually unobserved, remaining unexplained after controlling
for the observable characteristics of individuals. Where the effects of unobservables have clear economic interpretations, methods of identifying latent groups of individuals that share common behaviours can provide insights into economic phenomena. 

There is a need to improve understanding of the behaviours of retirees who fund their consumption by drawing down savings accumulated in personal Defined
Contribution accounts. Such understanding can inform appropriate financial advice and product development. 
Existing theoretical work by \citet{bateman2008choices} 
gives reason to expect distinct behavioural groups, where group behaviours correspond to strategies that individuals employ when accessing their retirement savings using phased withdrawal products. 

This article investigates drawdown behaviours in phased withdrawal
retirement income products by studying latent time profiles estimated using the Grouped Fixed-Effects (GFE) estimator of \citet{bonhomme2015grouped}. Of primary interest is testing for the presence of multiple behavioural groups in the data, and characterising any groups found. Furthermore, in this application, using the GFE estimator also allows testing whether retirees were heterogeneous
in their responses to the Global Financial
Crisis and retirement incomes policy changes, demonstrating the value of the GFE estimator to such event studies. 

In the presence of group-level time-varying
unobservable heterogeneity, using the standard two-way fixed-effects (2WFE) model defined in Section \ref{sec:Methodology} generally gives biased covariate effect estimates 
and unrepresentative time effect estimates. This general result is consistent with the findings of the present study: estimates for the time effects in the standard 2WFE model, which allows for only one set of time effects, obscure the different latent group-level
effects which suggest retirees may adopt simple, distinct heuristics when 
determining how they draw down their accumulated wealth during
retirement. The importance of this comparison with the 2WFE model results motivates our modification to the GFE estimator described in Section \ref{sec:Methodology}.

A key novelty in this application of the GFE estimator is 
the 
focus 
on performing statistical inference on the estimated effects of the latent
heterogeneity. To robustify the GFE estimator in this scenario,
this article proposes 
an extension to the bootstrap method
outlined in the Online Supplement to \citet{bonhomme2015grouped}. While
these authors describe a procedure for using the bootstrap to obtain standard
errors for covariate effects, a group label switching problem prevents finding
standard errors for the effects of group-level time-varying unobservable
heterogeneity. A method to match 
group labels across independent GFE estimations is required so that the bootstrap can
also explore uncertainty in the latent heterogeneity effect estimates.

Related to this, the fixed-$T$ variance estimate formula given in the
Online Supplement to \citet{bonhomme2015grouped} provides an alternative approach
to performing inference on the effects of the latent heterogeneity, where $T$ is the maximum number of observations on each unit in the sample.
The proposed label-matching method also allows studying the
properties of the standard errors for the latent group effects derived
from this analytical formula, by observing their distribution across
a large number of simulated datasets. This is useful in determining the
potential applicability of the analytical formula for obtaining standard
errors of the latent group effects in finite samples, explored in the Online Supplement to this article.

The rest of the paper is organised as follows. Section \ref{sec:Related-Literature} summarises the literature to which this article contributes. Section \ref{sec:Superannuation-Drawdowns-Dataset} describes the available data for our application to retirement decumulation behaviours. Section \ref{sec:Methodology} outlines the GFE estimator from \citet{bonhomme2015grouped}, and explains our proposed extensions. Section \ref{sec:Main-Results} presents the main results from applying the GFE estimator to the available data. Section \ref{sec:Discussion} concludes by examining the implications for retirement incomes research and policy. 
This article has an Online Supplement which provides robustness checks, a simulation exercise, descriptive analysis of the data, and further details on the second proposed methodological extension.

\section{\label{sec:Related-Literature}Related literature}

\subsection{\label{subsec:Drawdown-Behaviours-in}Drawdown behaviours in phased withdrawal retirement income products}

Recent 
work
studying drawdown behaviours builds on \citet{horneff2008following}. These authors compare three rate-based drawdown strategies against a level guaranteed
lifetime annuity, using a model of retiree utility that incorporates
stochastic interest returns and retiree lifetimes. The first strategy
draws a fixed proportion of the account balance each year; the
second determines a terminal time horizon $T$ and draws a proportion $1/T$ of the account balance
in the first year, $1/(T-1)$ in the second year, and so on, continuing until
the $T^{th}$ year of the plan when all the remaining funds are drawn
down; the third draws a proportion of the balance that updates each year based on the surviving retiree's expected remaining lifetime.

\citet{bateman2008choices} extend 
\citet{horneff2008following}, motivated by newly legislated minimum annual drawdown rates
for a phased withdrawal retirement income product in Australia known
as the account-based pension. Table \ref{tab:Age-based-Minimum-Drawdown-Rates}
details these age-based rates, which start at 4\% of the account balance
for retirees under age 65, and increase as a step function of age
to a maximum of 14\% for retirees aged 95 or above; these are specified in
Schedule 7 of the Superannuation Industry (Supervision) Regulations
1994. 
The rates, when applied to a retiree's account balance at the
start of the relevant financial year, give the dollar amount the retiree
must draw down from their account during that financial year.

\citet{bateman2008choices} use a stochastic lifecycle
model to compare five drawdown strategies. Three are based on \citet{horneff2008following}; the remaining two are based on legislated
minimum drawdown rates in Australia: always drawing at the newly legislated
minimum drawdown rates effective from 1 July 2007;\footnote{Superannuation Industry (Supervision) Amendment Regulations 2007 (No.~1) Schedule 3.}
and drawing at the minima previously in effect. \citet{bateman2008choices} also use their
model to derive the implied optimal drawdown plan,
and examine the sensitivity of these results to changes in the model
assumptions. The fixed-rate strategy
they consider draws a fixed proportion of
the remaining account balance each year, where this fixed proportion is
the annuity payout rate the retiree would have received in
the market at the time of writing by purchasing a guaranteed lifetime
annuity using their account balance at retirement. The authors find that while the newly legislated minima outperform the second and third rate-based strategies of \citet{horneff2008following} described earlier, in most cases, simulated utility is increased by initially drawing at a fixed rate higher than the minimum, then switching to the new minimum drawdown schedule once the minimum rate rises higher than the fixed rate. 

Based on this theoretical literature alone, researchers can expect, \emph{a priori}, that analysing drawdowns data will reveal groups
of retirees exhibiting distinct drawdown behaviours based
on simple rules, or heuristics. The present article uses recent advances in panel
data methods to empirically investigate whether such groups exist, and if so, what behaviours they display.
Section \ref{subsec:Capturing-Time-varying-Unobserve} summarises recent methodological progress in panel
data econometrics.

\subsection{\label{subsec:Capturing-Time-varying-Unobserve}Capturing time-varying unobserved heterogeneity}

\citet{bonhomme2015grouped} present the Grouped Fixed-Effects (GFE)
estimator and apply it to a linear panel data model that allows for group-level time-varying unobserved heterogeneity and unknown group membership. Importantly, their method allows for correlation between the latent group effects and the observed regressors. 
Being able to control for, and estimate, the varying latent group time profiles is the key advantage of using this model over the traditional fixed-effects model, which captures
only time-constant unit-level unobservable heterogeneity. 

Factor-model approaches to capturing unobserved heterogeneity \citep[e.g.,][]{bai2009panel, su2018identifying} provide more flexible specifications for the individual-specific heterogeneity. These models assume the presence of common, but unobserved, time-varying factors to which individuals respond heterogeneously. 
\citet{bonhomme2015grouped} show that their linear panel model, to which they apply the GFE estimator, is a special type of latent-factor model, where the latent factors are group-specific time effects and the factor loadings are group membership indicator variables. In applications where group time profiles are of economic interest, having this interpretation of the time-varying unobservable heterogeneity is an advantage of using the GFE estimator over a factor-model approach. Furthermore, compared to competing latent-factor model methods, the GFE estimator converges faster to the true parameter values with $T$, and so has better finite-sample performance \citep{bonhomme2015grouped}. This makes GFE the preferred estimator when analysing a typical microeconomic panel, which may have a large number of units $N$, but rarely more than a moderate length $T$. 

A related research area concerns finite
mixture models, which can incorporate unobservable heterogeneity when the heterogeneity is constant over time. Finite mixture models are parsimonious solutions to modelling unobserved
heterogeneity in populations, wherein data from latent subpopulations
can be drawn from different distributions, and the true allocation
of specific individuals to subpopulations is unobserved. \citet{deb2013finite} develop finite mixture models with
time-constant fixed effects. Critically, these models cannot specify the likelihood of being in
a particular group as a function of the covariates. By contrast, one
of the primary advantages of using the GFE estimator over a standard 2WFE model estimator is the ability to control for correlation between latent group effects and the
included covariates. Mixture-of-experts
models \citep[e.g.,][]{jacobs1991adaptive, jordan1994hierarchical} generalise mixture models by
parametrically specifying a link between group identity and covariate
values; however, these models do not control for unit-level fixed effects
in panel data.

The discussion above suggests that 
the GFE 
estimator 
is the most appropriate for the present application, because: the available dataset only has 
a moderate panel length $T$; a review of the theoretical drawdowns literature
suggests that there may be a finite number
of distinct behavioural groups in the population, consistent with
the GFE assumption that there are a finite number of latent group time profiles; the time profiles recovered
by the GFE estimator are likely to have economically meaningful interpretations. 

\section{\label{sec:Superannuation-Drawdowns-Dataset}Data on superannuation drawdowns}

The available superannuation dataset contains information on drawdowns from account-based
pensions (ABPs), a phased withdrawal retirement income product
in Australia. Using ABPs, retirees generate personal income streams or receive
lump-sum payments by drawing down their accumulated savings during
retirement. Throughout, the balance remains invested in financial
markets based on each retiree's chosen combination of safe and risky
exposures.

Following the Global Financial Crisis (GFC), the Australian government introduced
temporary minimum drawdown rate reductions as per Schedule
7 of the Superannuation Industry (Supervision) Regulations 1994. These resulted in `concessional' minimum drawdown
rates, which came into effect for the financial year ended 30 June
2009. For financial years 2009--2011 inclusive, the reduction
was 50\%, so that any retiree's minimum drawdown rate over this period
was halved relative to the standard schedule in Table \ref{tab:Age-based-Minimum-Drawdown-Rates}.
For financial years 2012 and 2013, the reduction decreased to 25\%,
meaning the minimum drawdown rates were three-quarters of those given
in the table. For subsequent financial years, the rates returned
to their nominal values.

To generate an income stream from an ABP, retirees can elect a payment
amount and a frequency, for example \$1000 monthly, which the superannuation
fund follows in paying the retiree from their account balance.
This type of drawdown, which is specified by the beginning
of a given financial year, is referred to here as a `regular' drawdown, and is the primary
object of interest in this study. These regular drawdowns are distinct
from `ad-hoc' drawdowns, referring to any additional
lump-sum payments requested by the retiree during the financial year.
A rigorous analysis of ad-hoc drawdowns is crucial
in understanding retirees' needs for flexibility and insurance, but
is not the focus of this paper. 

The dependent variable studied here is $y_{it}:=\ln {\rm DR}_{it}$; ${\rm DR}_{it}$ is the rate of drawdown for retiree $i$ over financial year $t$, which is a ratio of the regular drawdown amount over the account balance: ${\rm DR}_{it}:={\rm DA}_{it}/{\rm AB}_{it}$. The dependent variable is the logarithm of this rate because the raw rate variable is strictly positive and right-skewed. Appendix \ref{sec:p1-appendix} gives further details on the relationship between the drawdown rate and the account balance, as well as histograms of the dependent variable and the raw rate variable. 

Qualitatively,
one interpretation of the drawdown rate is the speed at which retirees deplete
their accumulated savings throughout retirement. In ABPs, retirees
are exposed to the risk of outliving their savings and their retirement income becoming reliant
on non-superannuation assets or taxpayer-funded transfer payments known as the age pension.
Policy and product design should support retiree needs
for regular income, and so understanding the drawdown behaviours that
manifest in this flexible withdrawal product has the potential to
inform policymakers, financial advisors and superannuation fund trustees.

The superannuation dataset contains $N=9516$ individuals,
combining source data from multiple large industry and retail superannuation funds. Due to the small number of funds in the sample whose
data permitted determining the regular drawdown rate, the economic results
in this paper may not be representative of the Australian superannuation
system as a whole.

The data capture window spans the financial years 2004 to 2015 inclusive, for $T=12$ annual observations. The main results use a balanced subset of each superannuation fund's data, removing individuals with missing observations.
However, as the data capture window is not the same for each fund, when combining these balanced subsets, the resulting dataset is unbalanced. 
The Online Supplement shows that the results are robust to using a fully balanced subsample of all the available data. 

For 2.4\% of the records in the sample, missing distinctions between regular and ad-hoc drawdown amounts can be imputed based on observed movements in the corresponding total drawdown amounts. The imputation method involves calculating, for each individual--time combination, the individual's average monthly total drawdown amount in that financial year. Comparing these averages with the exact monthly total drawdown amounts, if an individual's exact drawdown amount is less than or equal to 105\% of the average drawdown amount, the exact drawdown amount is treated as a regular drawdown, and the corresponding ad-hoc drawdown amount for that month is zero. Otherwise, if an individual's exact drawdown amount is more than 105\% of the average drawdown amount, the regular drawdown amount for that period is set to the computed average, and the excess is treated as an ad-hoc drawdown in that month. In periods where the individual makes an ad-hoc drawdown, this method may underestimate the ad-hoc drawdown amount. 

Individuals who die during the sample observation period are dropped from the data. 

The superannuation dataset is derived from administrative data, and so has limited demographic
information on members. Table \ref{tab:agg-time-invariant-variables}
provides summary statistics for characteristics that vary across individuals but not time; Table \ref{tab:agg-time-varying-variables} summarises the time-varying variables in the dataset. The two covariates considered are the minimum
drawdown rate for each individual at each time point, and their account
balance at the beginning of the respective financial year, both on the log scale.
Thus the model, with regular drawdown rate on the log scale as the
dependent variable, estimates the elasticity of the regular drawdown
rate with respect to the minima and account balances. The remainder
of the variation in log regular drawdown rates is attributed to the
latent group effects estimated by the GFE procedure, and residual noise.

The lack of demographic information, particularly health and marital
status, is also a data-based motivation for using GFE as an estimation
procedure. These unobserved characteristics are a potential source of omitted variable bias; they are possibly correlated with the dependent variable as well as the included covariates. Because relevant characteristics such as health and marital status may vary over time, controlling for time-constant
unobserved heterogeneity by using a standard fixed-effects model may not remove the biasing effect of all relevant unobservables. For this reason, a model allowing for sources of time-varying unobserved heterogeneity,
such as the linear panel model to which the GFE estimator is applied, may be more appropriate; the results in
Section \ref{subsec:Common-Parameter-Estimates} provide evidence
for this proposition.

\section{\label{sec:Methodology}Methodology}

\subsection{\label{subsec:Grouped-Fixed-Effects}The Grouped Fixed-Effects (GFE) Estimator}

The GFE estimator introduced by \citet{bonhomme2015grouped} considers a linear model of the
form
\begin{equation}
	y_{it}=x_{it}'\theta+\alpha_{g_{i}t}+v_{it};\label{eq:GFE-model-vanilla}
\end{equation}
in our application $i=1,2,...,N$ indexes individual retirees; $t=1,2,...,T$ indexes financial
years; $y_{it}$ is the dependent variable; $x_{it}$ is a vector
of covariates; $\theta$ are the partial effects of covariates
on the dependent variable after controlling for group-level time profiles;
$g_{i}=1,2,...,G$ identifies the group membership for unit
$i$, where $G$ is the chosen number of latent groups in the sample; $\alpha_{g_{i}t}$ is a term representing time-varying,
group-specific unobserved heterogeneity---these are time-varying model intercepts that can differ across individuals, depending on the value of the individual's group identifier, $g_i$; $v_{it}$ represents
the residual effect of all other unobserved determinants of the dependent
variable. 

For the estimator consistency and asymptotic Normality results of \cite{bonhomme2015grouped} to apply, the distribution of the errors, $v_{it}$, must have finite fourth moment and be uncorrelated with the covariates, $x_{it}$; errors can be correlated across cross-sectional units if there is a finite limit on the magnitude of the correlation; the unit-level time-sequences of errors can be autocorrelated if the sequences are strongly mixing processes; the tails of the error distribution must exhibit a faster-than-polynomial decay rate. The time profile values, $\alpha_{gt}$, may be correlated with $x_{it}$ but not $v_{it}$. For the complete list of assumptions required for the GFE estimator, see \cite{bonhomme2015grouped}, pp.~1156--1159.

The linear model in (\ref{eq:GFE-model-vanilla}) can be extended to include time-constant, individual-specific
fixed effects, $c_{i}$, so that
\begin{equation}
	y_{it}=x_{it}'\theta+c_{i}+\alpha_{g_{i}t}+v_{it},\label{eq:GFE-model-with-c_i}
\end{equation}
because applying the within transformation (centering the variables around individual-specific means) reduces (\ref{eq:GFE-model-with-c_i}) to the form of (\ref{eq:GFE-model-vanilla}). To see
this, for any variable $z_{it}$, define $\dot{z}_{it}:=z_{it}-\bar{z}_{i}$, where $\bar{z}_{i}:=T^{-1}\sum_{t=1}^{T}z_{it}$; then, 
\begin{align}
	\dot{y}_{it} =\dot{x}_{it}'\theta+\dot{\alpha}_{g_{i}t}+\dot{v}_{it},\label{eq:GFE-model-with-c_i-within-transformed}
\end{align}
which has the same form as (\ref{eq:GFE-model-vanilla}). In (\ref{eq:GFE-model-with-c_i-within-transformed}), the time profiles, $\dot{\alpha}_{g_{i}t}$, have a different economic interpretation to the $\alpha_{g_{i}t}$ from (\ref{eq:GFE-model-vanilla}); Section \ref{subsec:time-profs-in-trans-model} explains the treatment and interpretation of the group time profile estimators in the transformed model (\ref{eq:GFE-model-with-c_i-within-transformed}). 

Throughout this paper, `GFE model' refers to (\ref{eq:GFE-model-with-c_i}).  A 
useful way to consider the GFE model is as a generalisation of the 2WFE model,
\begin{equation}
	y_{it}=x_{it}'\theta+c_{i}+\alpha_{t}+v_{it},\label{eq:standard-two-way-fixed-effects-model}
\end{equation}
where the GFE model allows distinct time profiles for $G$ groups, with group membership unobserved. 

\citet{bonhomme2015grouped} state the assumptions needed for large-$N$, large-$T$ consistency and asymptotic normality of the GFE estimator, in 
models both with, and without, time-invariant fixed effects. Moreover, the asymptotics show that the estimator converges in distribution even when $T$ grows substantially more slowly than $N$. For any fixed value of $T$, however, the $\theta$ and $\alpha$ estimators are root-$N$ consistent not for the true population parameters, but for `pseudo-true' parameter values, and although these pseudo-true values may differ from the true population values for any fixed $T$, any difference vanishes rapidly as $T$ increases. Specifically, the pseudo-true and true values may differ when group classification is imperfect; because of this possibility, the authors derive a fixed-$T$ consistent variance estimate formula which takes into account potential group misclassification, providing a practical alternative to the large-$T$ variance estimate formula \citep[][p.~1161]{bonhomme2015grouped}. To examine the finite-$T$ properties of the GFE estimator, \citet{bonhomme2015grouped}
use a Monte Carlo exercise with simulated datasets that match the $N=90$,
$T=7$ panel used in their empirical application. 
In the present study, the superannuation dataset has large $N=9516$ but moderate $T=12$; the Online Supplement finds that the GFE estimator and the fixed-$T$ consistent variance estimator perform well on simulated datasets of this size. 

There are two problems in using the standard 2WFE model when there is group-level time-varying unobservable heterogeneity,
$\alpha_{g_{i}t}$, correlated with observable characteristics,
$x_{it}$: first, estimates of the covariate effects, $\theta$, may suffer
from omitted variable bias; 
second, the group time
profiles, which in many applications have interesting economic
interpretations, remain hidden. 
The 2WFE 
model is a special case of the GFE model with $G=1$; hence, the GFE model is an appealing alternative to the 2WFE 
model in applications where there is the possibility
of unobservable time-varying heterogeneity in addition to time-invariant unobservable heterogeneity. While
determining the precise number of groups, $G$, lacks a general
solution, estimating the GFE model can provide evidence for
whether only controlling for one set of time effects, as in the 2WFE 
model, is adequate; Section \ref{subsec:Common-Parameter-Estimates} discusses this issue.

The GFE estimator for (\ref{eq:GFE-model-vanilla}) minimises the sum of squared residuals, giving
\begin{equation}
	(\widehat{\theta},\widehat{\alpha},\widehat{\gamma})=\argmin_{(\theta,\alpha,\gamma)\in\mathit{\Theta}\times\mathcal{A}^{GT}\times\mathit{\Gamma}_{G}}\sum_{i=1}^{N}\sum_{t=1}^{T}(y_{it}-x_{it}'\theta-\alpha_{g_{i}t})^{2},\label{eq:GFE-estimator-definition}
\end{equation}
where the vector $\gamma=(g_{1},g_{2},...,g_{N})$ defines the
grouping of each of the $N$ units into one of the $G$ groups; $\mathit{\Gamma}_{G}$
is the set of all possible groupings of $N$ units into $G$ or fewer
groups; $\mathit{\Theta}$ is a subset of $\mathbb{R}^{k}$, the $k$-dimensional real space, where $k$ is
the number of covariates or equivalently the dimension of $x_{it}$;
$\mathcal{A}$ is a subset of $\mathbb{R}$; the $\alpha_{gt}$ parameters belong to $\mathcal{A}^{GT}$.

\citet{bonhomme2015grouped} present an iterative algorithm for estimating (\ref{eq:GFE-estimator-definition}). The
algorithm initialises values for the grouping vector $\gamma$ and
the model parameters $(\theta,\alpha)$; it then alternates between the
following two steps until convergence: 1) a grouping update step, which
allocates each unit to the group minimising the sum of squared
residuals given the most recent estimate of the model parameters $\theta$
and $\alpha$; 2) a parameter update step, which estimates the parameters $(\theta,\alpha)$
conditioning on the most recent estimate of the grouping vector $\gamma$.

Estimating the GFE model does not require a balanced panel; the algorithm can be adjusted to run even when the sample contains unit--period combinations with missing data. 
However, when estimating the GFE model with unbalanced data, preserving the equivalence between the 2WFE model (\ref{eq:standard-two-way-fixed-effects-model}) and the GFE model (\ref{eq:GFE-model-with-c_i}) with $G = 1$ requires an adjustment to the original GFE estimator. 

The original GFE estimator for (\ref{eq:GFE-model-with-c_i}) with $G=1$ gives the same estimates as applying the within transformation to the data and running a least-squares regression on the time-demeaned data, including $T$ time dummy variables and no constant term. 
In balanced samples, this returns the same numerical results as standard implementations of the 2WFE 
model: for example \texttt{Stata}'s 
\texttt{xtreg} command with the \texttt{fe} option and time dummy variables as covariates.
However, in unbalanced panels, the results differ.

With unbalanced data, obtaining the results from standard implementations of the 2WFE model requires time-demeaning the time dummy variables at the unit level. Hence, this article proposes and utilises a modified GFE estimator that for $G=1$ recovers precisely the same estimates as a standard 2WFE estimator even when the data is unbalanced and the model includes time-invariant unit-level unobservable heterogeneity; see Section \ref{subsec:Extension:-Alternative-Estimation-Method}. When the panel is balanced, or when there are no time-invariant unit-level unobservables, the results from this method are identical to the unmodified GFE estimation results. However, panel data models in microeconomic applications generally control for time-invariant unit-level unobservable heterogeneity.
Hence, this is a useful extension when applying the GFE estimator to microeconomic applications more broadly. 
To examine the sensitivity of the present results to using the modified algorithm, the Online Supplement provides a robustness check using the unmodified GFE estimation procedure in place of the proposed modified algorithm.

\citet{bonhomme2015grouped} explain that in the absence of covariates, their algorithm outlined above reduces to $k$-means clustering. Similarly to standard implementations of $k$-means clustering
algorithms, the results depend on the starting values. Running the algorithm multiple times with randomly generated starting values increases the likelihood of finding
solutions corresponding to smaller values of the objective function in 
(\ref{eq:GFE-estimator-definition}).

In the present study, the algorithm is initialised by drawing starting values for each covariate effect from independent
Gaussian distributions, centred at the coefficient estimates from a 2WFE 
regression fit to the data, 
and
with standard deviations equal to the magnitude of these coefficient
estimates. Results for the application
take the most optimal solution across 1000 independent runs of the
algorithm using randomly drawn starting values to initialise
the parameters. The Online Supplement also tests the sensitivity of the results to the number
of starting values used.

To facilitate inference on the group time profiles, we implement the fixed-$T$
variance estimate formula in the Online Supplement to \citet{bonhomme2015grouped}, as well as an extension to their non-parametric bootstrap method. 
Notably, approximating the variability in time profile estimates
using the non-parametric bootstrap method described by \citet{bonhomme2015grouped}
requires solving a label-switching problem; see Section \ref{subsec:Extension:-Bootstrapping-Time}.

\subsection{\label{subsec:time-profs-in-trans-model}Time profiles in the transformed model}

Within-transforming the data to
remove the effect of any time-constant unobservable heterogeneity
results in time-demeaned group time profiles, $\dot{\alpha}_{gt}$. As the $\dot{\alpha}_{gt}$ contain only
information about changes in the group effects over time relative to their mean, and no information
about the absolute level of the effect, the estimates of the $\dot{\alpha}_{gt}$  are modified in
order to obtain the desired economic interpretations. All
estimated time profiles are shifted to begin at a value of 0 at $t=1$,
so that the interpretations of the values are as changes relative
to the first time period. This is analogous to estimating a set of
$T-1$ coefficients for the time effects in the standard 2WFE 
model, where each estimate is interpreted as a change
relative to the omitted reference period.

This study uses two methods to estimate the uncertainty in the shifted time-demeaned group time profiles,
defined as $\tilde{\dot{\alpha}}_{gt}:=\dot{\alpha}_{gt}-\dot{\alpha}_{g,1}$,
for all $g$ and $t$. 
First, Normal-approximation 95\% confidence intervals are constructed for the $\tilde{\dot{\alpha}}_{gt}$, using variance estimates which are computed from the variance--covariance matrix components for the time-demeaned time profiles, $Var(\dot{\alpha}_{gt})$, $Var(\dot{\alpha}_{g,1})$ and $Cov(\dot{\alpha}_{gt},\dot{\alpha}_{g,1})$; these objects are obtained by applying the fixed-$T$ variance estimate
formula after estimating the GFE model on the within-transformed
data. Second, these confidence intervals are compared to the corresponding intervals obtained from the bootstrap, using the proposed extension for matching group labels across bootstrap replications.

A comparison of the time profile standard errors
derived from the fixed-$T$ variance estimate formula versus simulated standard
errors from a Monte Carlo exercise shows that the fixed-$T$ variance estimate formula performs well on simulated data calibrated
to the superannuation dataset; see the Online Supplement for details. Moreover, in the present study, confidence intervals
constructed from the bootstrap procedure results are broadly similar to those derived from the fixed-$T$ variance estimate formula, implying practically equivalent
inference on the estimated parameters. For smaller datasets, where
asymptotic results are less likely to provide accurate standard
errors, and where the computational cost of running the bootstrap
is lower, it may be preferable to use the bootstrap. However, the results
suggest that in larger datasets, using the fixed-$T$ variance estimate formula may be sufficiently accurate while keeping the
computational burden low compared to using the bootstrap.

\subsection{\label{subsec:selecting-G}Selecting $G$}

\citet{bonhomme2015grouped}
propose an information criterion that correctly selects the number
of groups 
when $N$ and $T$ are both large and similar in magnitude; however, for large-$N$ moderate-$T$ panels,
like the superannuation dataset, this criterion may overestimate the true value of $G$ \citep{bonhomme2015grouped}.
An alternative method to choose the number of groups, which \citet{bonhomme2015grouped}
employ in their empirical application, uses the fact that underestimating
$G$ leads to a type of omitted-variable bias in the $\theta$ parameters, to the extent that the unobserved effects are correlated with the included covariates. 
Conversely, overestimating $G$ does not
bias the $\theta$ estimates, although it does increase the number of parameters
to estimate and the parameter space of possible groupings, resulting
in noisier estimates for all model parameters. 

Taken together, these properties suggest that observing a plot of
how the coefficient estimates change as $G$ increases 
can reveal the point at which the coefficients have
been de-biased relative to the $G=1$ case. Visually, the coefficients
might vary significantly between $G=1$ and some higher value, $G^\star$, after
which they may appear to stabilise, albeit becoming noisier as $G$
continues to increase. One can then select $G^\star$ as the number of groups suggested by the data.\footnote{This method would not apply to a model with group-specific covariate effects.} This is how \citet{bonhomme2015grouped}
decide on $G=4$ in their application, rather than $G=10$ as suggested by the information
criterion---which, given the dimensions of their panel, may
overestimate the true number of groups. 

For applications that focus only on estimating unbiased effects of the covariates, the above suggests choosing the number of groups by selecting the smallest value of $G$ that places the covariate effect estimates close to their stable values.
However, the present application is interested in the time profiles for all latent groups in the sample.
It is possible that stopping
at the first value of $G$ that seems to successfully de-bias the
regression coefficients will result in at least one estimated group whose
time profile is an average over distinct time profiles of constituent subgroups.
For the purposes of obtaining unbiased regression coefficients alone,
separating out these mixed groups by further increasing $G$ may unnecessarily increase the complexity of the model. However, identifying all distinct behavioural groups with  
economically meaningful interpretations may require searching beyond the smallest $G$ that de-biases the regression coefficients. 

On the other hand, overestimating $G$ can result in the separation of a particular time profile into multiple biased representations of itself, where the difference between the representations is spuriously driven by noise \citep{bonhomme2015grouped}. This means that the more similar a pair of time profiles appear, the more sceptical the researcher should be in regarding them as distinct effects. 

Hence, 
this study posits a selection rule that picks the greater of two values
of $G$: the smallest $G$ that appears to result in unbiased estimates
of the regression coefficients; or the last value of $G$ for which
the marginal change from $G-1$ to $G$ groups finds an economically
important behaviour, sufficiently distinct from all others and exhibited by a non-trivial portion of the sample. 
Section \ref{sec:Main-Results} discusses this selection process alongside the main results.

\subsection{\label{subsec:Extension:-Bootstrapping-Time}Extension 1: Bootstrapping time profile estimates}

\citet{bonhomme2015grouped} address how to obtain bootstrap standard errors for the covariate effects, but not the time profiles. As the covariate effects are unambiguously labelled across multiple
runs of the GFE procedure, estimating their standard errors by comparing
estimates across a large number 
of bootstrap replications is standard.
By contrast, group-level results from the procedure are unique only up to a relabelling of the groups, and the group labels are determined at random during estimation. This means that the same economic group may receive different labels across multiple independent estimation runs, e.g., when performing the bootstrap. Hence, estimating the standard
errors of group-level time effects by comparing time profile estimates across bootstrap replications requires a method to match labels across runs.

The bootstrap procedure outlined in the Online Supplement to \citet{bonhomme2015grouped}
involves running the GFE procedure
$B$ times, each time on a different bootstrap replicate dataset. 
To create the bootstrap
replicate datasets, all $N$ units in the original
data are sampled with replacement $N$ times, taking all observations corresponding
to a given unit into the replicate dataset when that unit is sampled.
Each of the $B$ model runs produces a set of coefficient estimates
for the model covariates, as well as an estimated grouping and a set
of time profiles for each group. 

To match group labels across replications, we propose a
label-matching procedure, 
similar to the solution \citet{hofmans2015added}
use for the analogous problem in $k$-means clustering; see the Online Supplement for details. 

This method matches time profiles across different
bootstrap replications in the main results presented in Section \ref{sec:Main-Results},
as well as across different simulated datasets and bootstrap replications
in the simulation exercise reported in the Online Supplement. In general, it could be seen as an issue that time profile bootstrap results are contingent on some method to match labels, because, in principle, the bootstrap results could prove sensitive to the choice of label-matching method. The results here show that in this application, the bootstrap and analytical formula results are similar, and so little is gained by using the bootstrap over the analytical formula. Thus, we did not pursue this question further, but this general problem may be of interest to future research.

While deriving standard errors for the shifted time-demeaned group time
profiles, $\tilde{\dot{\alpha}}_{gt}$, using the fixed-$T$ variance estimate formula requires combining elements from the variance--covariance matrix of the time-demeaned time profile
estimators, $\dot{\alpha}_{gt}$, as described in Section \ref{subsec:time-profs-in-trans-model}, the bootstrap procedure can directly estimate the variability of the $\tilde{\dot{\alpha}}_{gt}$ terms.

\subsection{\label{subsec:Extension:-Alternative-Estimation-Method}Extension 2: Alternative estimation method for unbalanced data}
Consider the model  in (\ref{eq:standard-two-way-fixed-effects-model}), which, after time-demeaning, has the form
\begin{equation}
	\dot{y}_{it}=\dot{x}_{it}'\theta+\dot{\alpha}_{t}+\dot{v}_{it}.\label{eq:2WFE-using-GFE-notation}
\end{equation}
The primary estimands of interest are the $T-1$ relative effects $\tilde{{\dot{\alpha}}}_{t}=\dot{\alpha}_{t}-\dot{\alpha}_{1}$ for $t=2,3,...,T$. Since $\dot{\alpha}_{t}:=\alpha_{t}-\bar{\alpha}$, where $\bar{\alpha}=T^{-1}\sum_{t=1}^{T}\alpha_{t}$, the set of $T$ values of $\dot{\alpha}_{t}$ satisfy $\sum_{t=1}^{T}\dot{\alpha}_{t}=0$.

In balanced panels, the following three methods obtain identical estimates for the $\tilde{{\dot{\alpha}}}_{t}$:\footnote{This claim is justified in the Online Supplement.}
\begin{enumerate}
	\item 
	regress $\dot{y}_{it}$ on $\dot{x}_{it}$ and $T$ time dummy variables with no constant term to estimate all $T$ values for $\dot{\alpha}_{t}$ as the coefficients of the time dummies; then compute $\tilde{{\dot{\alpha}}}_{t}=\dot{\alpha}_{t}-\dot{\alpha}_{1}$ for $t=2,3,...,T$; 
	\item 
	regress $\dot{y}_{it}$ on $\dot{x}_{it}$ and $T-1$ time dummy variables excluding the dummy variable for the first time period, including a constant term, after which the $T-1$ desired estimates of $\tilde{{\dot{\alpha}}}_{t}$ are the estimated coefficients of the included $T-1$ time dummy variables; 
	\item 
	first, time-demean the time dummy variables corresponding to periods $2,3,...,T$; then, regress $\dot{y}_{it}$ on $\dot{x}_{it}$ and the $T-1$ time-demeaned time dummy variables, not including a constant term, after which the $T-1$ desired estimates of $\tilde{{\dot{\alpha}}}_{t}$ are the estimated coefficients of the included $T-1$ time dummy variables. 
\end{enumerate}
When the panel is unbalanced, the choice of estimation strategy requires further consideration. With unbalanced data, methods 1 and 2 described above for obtaining estimates of $\tilde{{\dot{\alpha}}}_{t}$ no longer guarantee that the corresponding
values of $\dot{\alpha}_{t}$ sum to zero; a nonzero sum contradicts a property of the model given by (\ref{eq:2WFE-using-GFE-notation}). Conversely, method 3 obtains the desired estimates while ensuring that the $T$ estimates of $\dot{\alpha}_{t}$, derived from the $T-1$ estimates for $\tilde{{\dot{\alpha}}}_{t}$, 
sum to 0, consistent with the model. 

Using the GFE estimator implementation provided by \citet{bonhomme2015grouped} with $G=1$ on unbalanced data gives estimates for the time effects in line with those produced by methods 1 and 2, whereas standard implementations of the 2WFE 
model give the same estimates as method 3. This motivates our modification to the GFE methodology, which aligns the $G=1$ results with the standard 2WFE 
model estimates, regardless of whether the panel is balanced or not: in the modified procedure's parameter update step, the algorithm interacts the group identity variables with $T-1$ time-demeaned time dummy variables, analogous to using method 3 for $G=1$.\footnote{The Online Supplement provides further details of this modification, as well as results using the unmodified GFE estimation procedure to compare with the results presented here.} 

\section{\label{sec:Main-Results}Main results}

\subsection{\label{subsec:Common-Parameter-Estimates}Covariate effects}

\begin{figure}
	\caption{\small \label{fig:covrt-ests-vs-G-for-G-1-to-large-sdsm2}Point estimates and 95\% confidence intervals for partial
		effects of log minimum
		drawdown rate and log account balance covariates on log regular drawdown
		rate, controlling for group-level time-varying unobservable heterogeneity
		for a range of values of $G$. Shaded regions denote 95\% confidence intervals constructed using standard
		errors derived from fixed-$T$ variance
		estimate formula in the Online Supplement to \citet{bonhomme2015grouped}.}
	\begin{center}
		\includegraphics[width=0.8\textwidth]{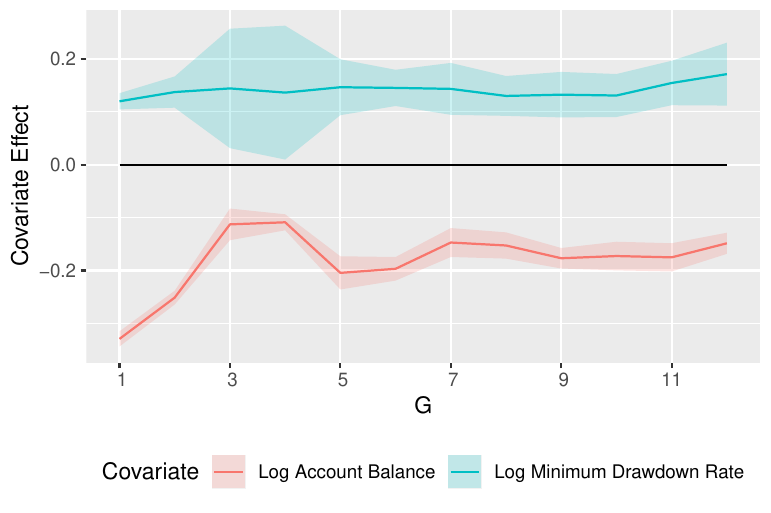}
	\end{center}
\end{figure}

Figure \ref{fig:covrt-ests-vs-G-for-G-1-to-large-sdsm2} plots the covariate effect estimates versus $G$ for the GFE model fitted to the superannuation drawdowns dataset. 
The values of $G$ on the horizontal axis correspond to the number of latent groups specified; the vertical axis denotes partial effect values for the two included covariates: log minimum drawdown rate and
log account balance. The lines connect covariate effect estimates, while the
shaded regions around these correspond to 95\% confidence intervals
constructed using standard errors derived from the fixed-$T$ variance
estimate formula in the Online Supplement to \citet{bonhomme2015grouped}.

As the dependent variable and covariates are on the log scale,
the covariate estimates represent the elasticity
of regular drawdown rates to changes in account balances and the minimum
drawdown rates. Section \ref{subsec:Choice-of-G} provides the economic interpretation for the
corresponding estimates after choosing
the value of $G$.

A critical feature of Figure \ref{fig:covrt-ests-vs-G-for-G-1-to-large-sdsm2}
is that as $G$ increases, changes in the log account balance covariate effect estimates are large relative to the confidence intervals for the first few values of $G$. After $G=7$, the point estimates stabilise and
successive 95\% confidence intervals show significant overlap, suggesting that from
$G=7$, the GFE procedure removes the bias arising from correlation
between group-level latent time effects and the included covariates.

Using the proposed modified GFE model estimation procedure, the point estimates corresponding to $G=1$ are identical to the results from a standard 2WFE 
estimation on the drawdowns dataset. 
Hence, Figure \ref{fig:covrt-ests-vs-G-for-G-1-to-large-sdsm2} implies that the estimators using a more traditional analysis including only one time profile are biased compared to the estimators in the GFE model for $G\geq7$. The Online Supplement shows that this result holds 
in the fully balanced case, where both the modified and unmodified GFE procedures give results identical to the 2WFE model when $G=1$. 

\subsection{\label{subsec:Choice-of-G}Choice of $G$}

\begin{figure}
	\caption{\small \label{fig:Geq4-to-9-time-profile-plots}Point estimates for effects
		of group-level time-varying unobservable heterogeneity on log regular
		drawdown rates 
		assuming $G=4,5,...,9$. Estimated time-demeaned group time profiles shifted
		to begin at 0 on the vertical axis. 
	}
	\begin{center}
		\includegraphics[width=1\textwidth]{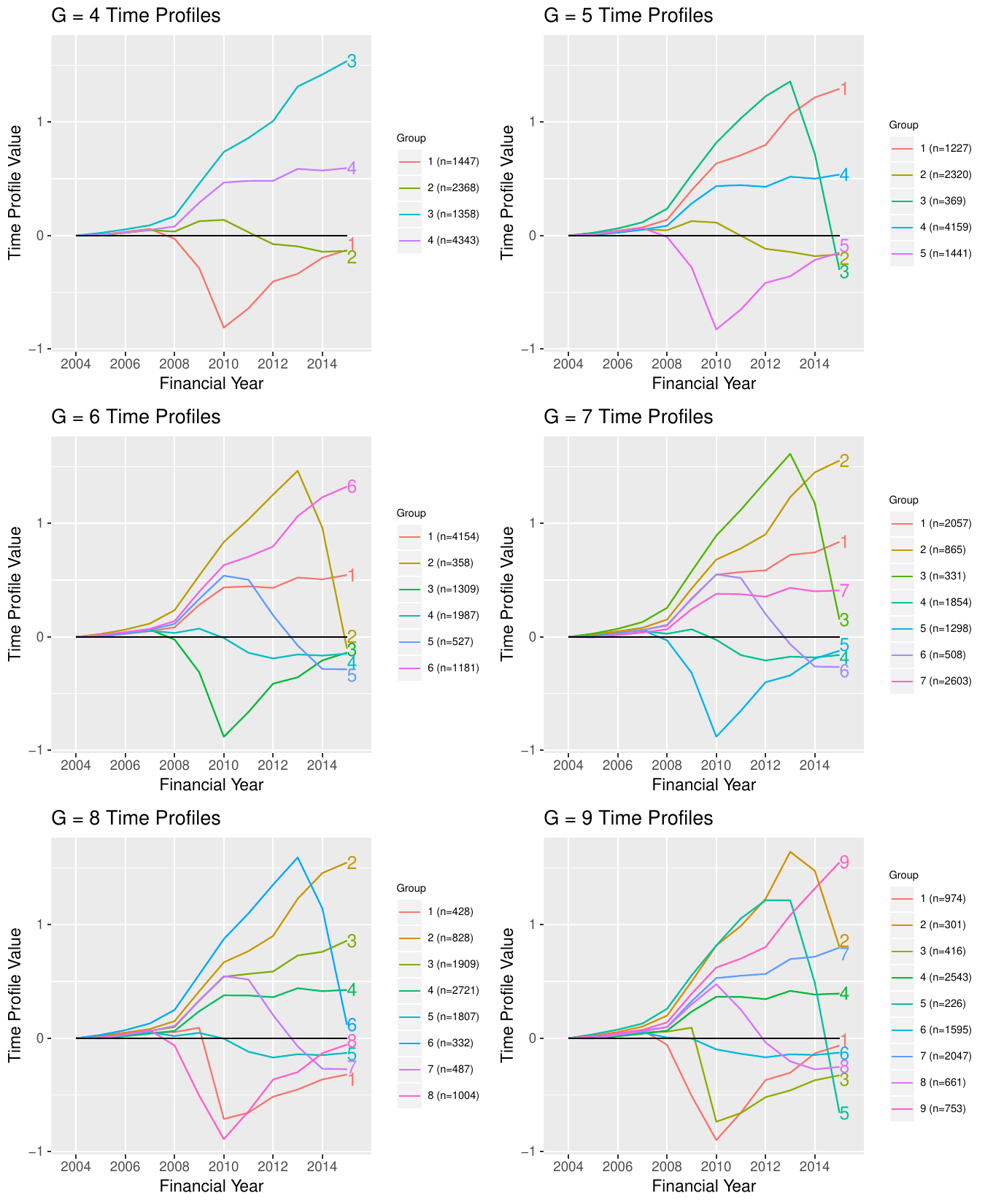}
	\end{center}
\end{figure}

Figure \ref{fig:Geq4-to-9-time-profile-plots} plots the
time profiles estimated for $G=4,5,...,9$, all shifted to begin
at a $y$-axis value of 0, and omitting confidence interval bounds
for clarity. The figure suggests that while increases in $G$ initially produce new and economically distinct time profiles, at $G=7$, incremental moves to a larger value of $G$ split existing time profiles into highly similar representations of the same trajectory. This is consistent with a theoretical result from \citet{bonhomme2015grouped} that implies overestimating $G$ can create spurious copies of the true time profiles, differing by random noise.
Hence, the number of latent groups to use
in the model is set to $G=7$. Section \ref{subsec:Time-Profile-Estimates}
explains the economic interpretations of the resulting time profile
estimates.

Selecting $G=7$ groups corresponds to estimated covariate effects of 0.144
and $-$0.147 for the log minimum drawdown rate and log account balance,
respectively. These are statistically significant at the usual levels, e.g. 5\%, with respective standard errors of 0.0251 and 0.0140.
As the dependent variable and covariates are on the log scale,
the covariate effects represent partial elasticities of the regular drawdown rate
with respect to the minimum drawdown rate and account balances. For
example, consider an increase of 0.1 in the log minimum drawdown
rate, which translates to a proportional increase in the level of
the minimum drawdown rate of approximately 10.5\%. This increase in the log minimum drawdown
rate is expected to raise regular
drawdown rates proportionally by an estimated $e^{0.0144}-1=1.45$\%
on average, holding account balance equal and controlling for group-level
time-varying heterogeneity. A similar
proportional increase in a retiree's account balance is expected to effect
a proportional decrease of $1.46$\% ($1-e^{-0.0147}$) in their regular drawdown
rate on average, holding the minimum drawdown rate constant and controlling
for group effects.

\subsection{\label{subsec:Time-Profile-Estimates}Time profiles}

\begin{figure}
	
	\centering

		\caption{\small \label{fig:fixed-T-variance-formula-results-for-time-profile-plot-for-Geq7-model}Point
			estimates and 95\% confidence intervals from analytical formula for
			effects of group-level time-varying unobservable heterogeneity on
			log regular drawdown rates 
			assuming $G=7$. Shaded regions denote 95\% element-wise confidence intervals constructed using standard
			errors derived from fixed-$T$ variance estimate formula in the Online Supplement to \citet{bonhomme2015grouped}. Time-demeaned
			group time profiles shifted to begin at 0 on the vertical axis.
		}
		\begin{center}
			\includegraphics[width=0.8\textwidth]{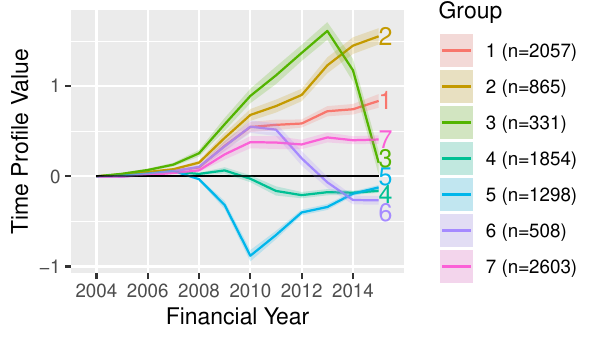}
		\end{center}
  \end{figure}

\begin{figure}
		\caption{\small \label{fig:bootstrap-B-eq-1000-results-for-the-G-eq-7-model-where-the-bootrep-datasets-are-drawn-from-the-actual-sample}
			Point
			estimates and 95\% bootstrap confidence intervals for effects of
			group-level time-varying unobservable heterogeneity on log regular
			drawdown rates assuming $G=7$. Shaded regions denote 95\% element-wise confidence intervals computed from empirical percentiles 
			across 1000 bootstrap replications. Time-demeaned group time profiles
			shifted to begin at 0 on the vertical axis. 
		}
		\begin{center}
			\includegraphics[width=0.8\textwidth]{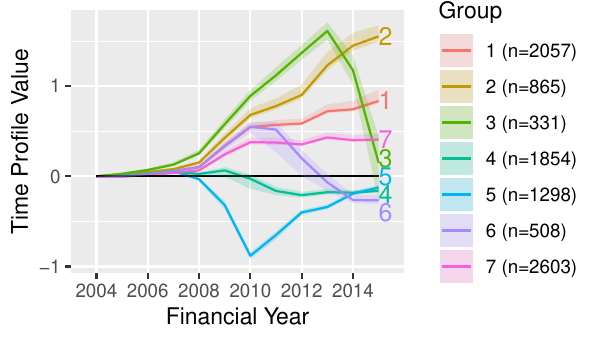}
		\end{center}
	
\end{figure}

Figure \ref{fig:fixed-T-variance-formula-results-for-time-profile-plot-for-Geq7-model} shows
the time profiles estimated using the selected value of $G=7$ with element-wise confidence interval bounds constructed using standard
errors derived from the fixed-$T$ variance estimate formula. 
Figure \ref{fig:bootstrap-B-eq-1000-results-for-the-G-eq-7-model-where-the-bootrep-datasets-are-drawn-from-the-actual-sample}
presents the equivalent plot with confidence interval bounds computed
from the empirical 2.5 and 97.5 percentiles of the time profile estimates
from 1000 bootstrap replications.\footnote{Each bootstrap replication uses distinct sets of randomly generated starting values for the estimation algorithm.} While qualitatively similar overall to the intervals constructed using standard errors derived from the fixed-$T$ variance estimate formula, a comparison highlights some differences. 
In particular, 
the intervals are wider around financial years 2011--2013 for group 6; 
the interval is wider for financial year 2010 for group 4; 
the intervals are tighter for group 5 throughout; 
the interval is wider in the terminal financial year for group 3; and 
the point estimates from the original sample tend towards the lower end of the 95\% bootstrap confidence intervals for groups 1, 2 and 7. 
However, due to the estimated effect magnitudes, these numerical differences in the sets of confidence intervals do not meaningfully change inferential conclusions regarding the estimates or their economic interpretation.

The values of the time profiles represent behavioural effects on the proportional
changes in a retiree's regular drawdown rate, relative to their own
rate of drawdown in 2004.  
As
the dependent variable is on the log scale, a value of, say, $0.5$
on the $y$-axis represents a proportional change in the regular drawdown
rate of $e^{0.5}\approx1.65$, or roughly a 65\% increase in the drawdown
rate, relative to 2004 levels.

\begin{figure}
	\caption{\small \label{fig:Time-Profiles-Geq1model-CIs-from-formula}Point
		estimates and 95\% confidence intervals for
		effects of group-level time-varying unobservable heterogeneity on
		log regular drawdown rates assuming 
		$G=1$. Shaded regions (indistinguishable from point estimates in plot) denote 95\% element-wise confidence intervals constructed using standard
		errors derived from fixed-$T$ variance estimate formula in the Online Supplement to \citet{bonhomme2015grouped}. Time-demeaned
		time profile shifted to begin at 0 on the vertical axis. Owing to the modification of the GFE model estimation algorithm, point estimates here are identical to those recovered from estimating a standard 2WFE 
		model on the data. The Online Supplement gives the corresponding results using the unmodified GFE model estimation procedure as a comparison.
	}
	\begin{center}
		\includegraphics[width=0.8\textwidth]{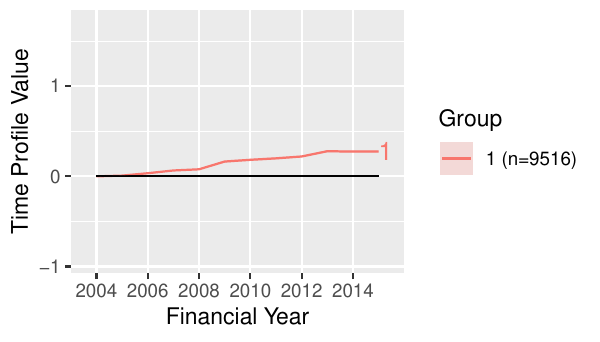}
	\end{center}
\end{figure}

Figure \ref{fig:Time-Profiles-Geq1model-CIs-from-formula} shows the time profile estimates for the $G=1$ model estimated using the modified GFE procedure, with the $y$-axis scale set to the same scale used
for the $G=7$ model group time profile plot (Figure \ref{fig:fixed-T-variance-formula-results-for-time-profile-plot-for-Geq7-model}).
Comparing these shows that a model controlling for only one set of time effects 
fails to capture the shape and magnitude of any of the time profiles
in the seven-group model. Visually, the resulting single time profile
appears to `average over' the seven markedly distinct time profiles. 
Hence, not only are
the covariate effect estimates biased when $G=1$, 
but the
single time profile is entirely unrepresentative of the latent group-level
time profiles.

\subsection{\label{subsec:Characterising-Groups}Behavioural interpretations}

The following refers to the group time profiles in Figure
\ref{fig:fixed-T-variance-formula-results-for-time-profile-plot-for-Geq7-model}.
All the time profiles exhibit similar, slowly rising
trajectories until the 2008 financial year, after which
they diverge in statistically and economically significant ways.

\begin{figure}
	\caption{\small \label{fig:g5-log-rdr-vs-fy}Panel plot showing all individual time
		series for group 5 of the time-demeaned (TD) log regular drawdown
		rate variable vs. financial year.}
	\begin{center}
		\includegraphics[width=0.8\textwidth]{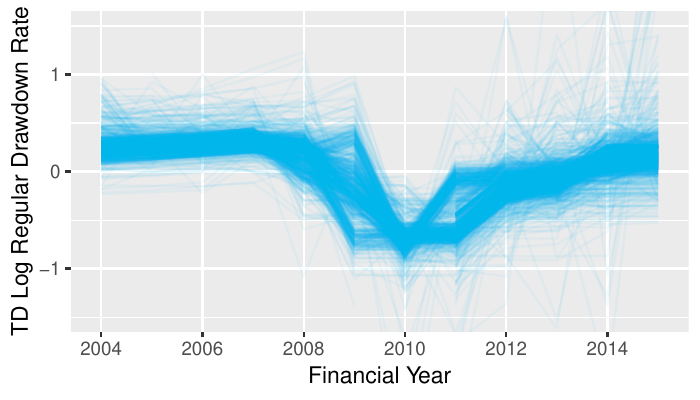}
	\end{center}
\end{figure}

Consider group 5, whose time profile describes a rapid reduction
in drawdown rates between financial years 2008 and 2010, and then
a gradual return towards pre-reduction levels. The timing of these movements follows closely the progression
of concessional minimum drawdown rates (see Section \ref{sec:Superannuation-Drawdowns-Dataset}).
This implies that members of group 5 were the most responsive to
the changing minimum drawdown rates over this period, compared to
the rest of the sample. Figure \ref{fig:g5-log-rdr-vs-fy} confirms that many individuals in group 5 show a step-like pattern
in their log regular drawdown rate series during the second half of
the observation period. This pattern is suggestive of the new step-function
schedule for minimum drawdown rates which came into effect 1 July
2007.

\begin{figure}
	\caption{\small \label{fig:some-g-log-rda-and-log-ab-vs-fy}Panel plots showing all
		individual time series for groups 1, 2, 3, 6 and 7 (top to bottom) of
		the time-demeaned (TD) log regular drawdown amount (left column) and
		TD log account balance (right column) variables on the vertical axes
		vs. financial year on the horizontal axes. Plots in each column maintain same $y$-axis scale for comparability.}
	\begin{center}
		\includegraphics[width=1\textwidth]{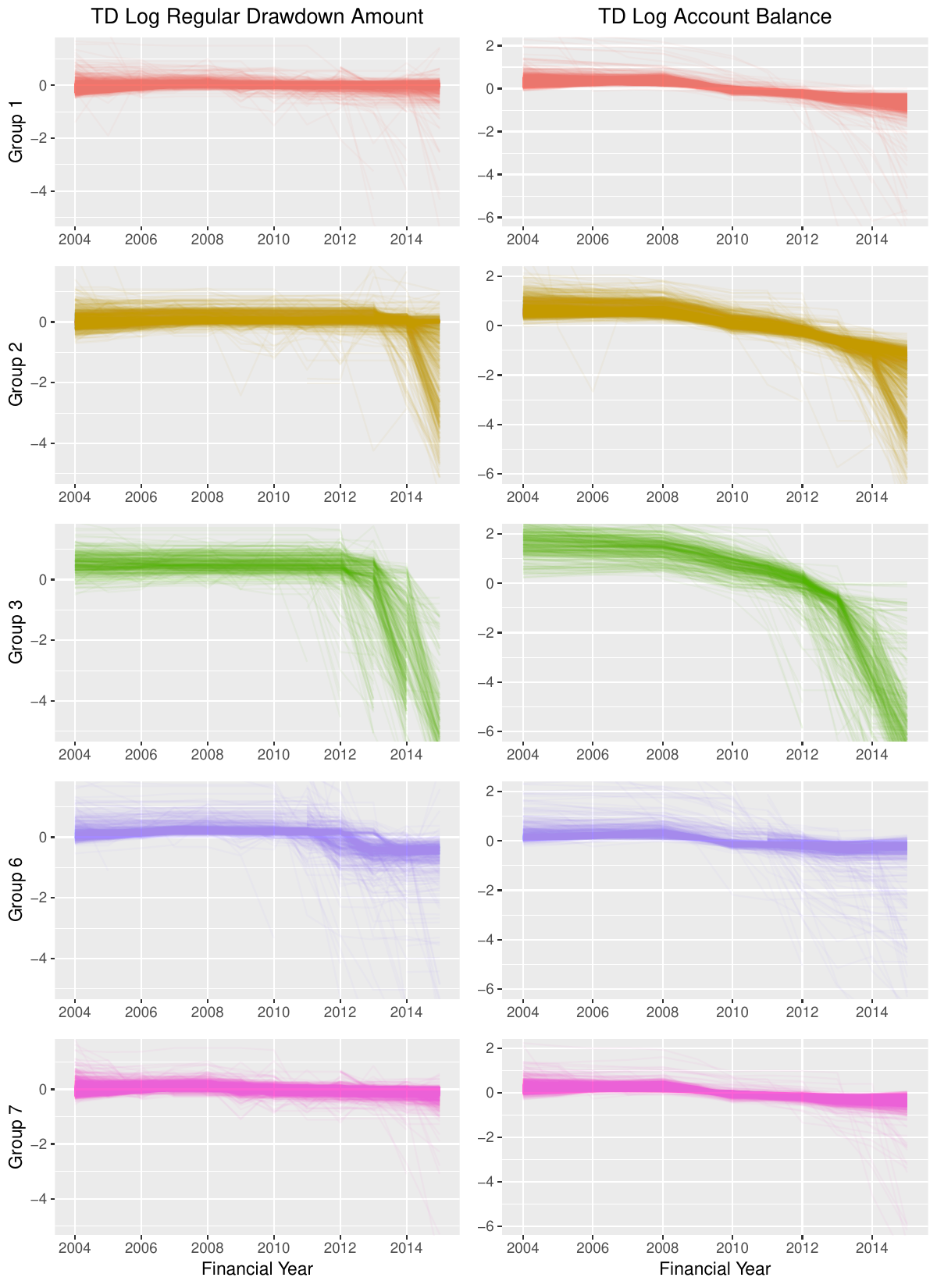}
	\end{center}
\end{figure}

Groups 1 and 2 display time profiles rising steadily following financial
year 2008 for the remainder of the sample. Group 7 follows a similar
rising trend with a reduced magnitude for the first few financial years
after 2008, before stabilising for the remainder of the observation
window. The panel plots in Figure \ref{fig:some-g-log-rda-and-log-ab-vs-fy} suggest a tendency
for individuals in these groups to draw constant dollar amounts, while
their account balances gradually decline over time. 
The corresponding plots for group 3 show a similar preference for constant dollar amounts, but after financial year 2013, individuals in this group suffer
a rapid decline in both the amounts drawn down and account balances.
Many members of group 6 undergo a similar evolution, characterised
by mostly constant drawdowns initially and a subsequent reduction in
the level in later years. The magnitudes of the downwards revisions appear smaller
for this group, and occur earlier for many retirees. Moreover, while
group 3 members continue to revise down over successive
financial years, it appears that many individuals in group 6 revise
down once and then continue to draw at the new, reduced level.
The group 7 plots show gradual downward revisions in the amounts year on year.

Thus, groups 1, 2, 3, 6 and 7 appear to be similar in that members draw mostly constant dollar amounts
over time; however, these groups seem to differ in how many members
make a downwards revision in the level of their drawdown amounts, and whether this
revision is once-off or the beginning of a downward trend. Crucially, the timing of downwards revisions often appears
to align with periods where account balances begin falling at an accelerated
rate. Hence, the heuristic of drawing constant amounts over time
may be adversely impacting retiree financial security by contributing
to a premature exhaustion of account balances. 

Figure \ref{fig:fixed-T-variance-formula-results-for-time-profile-plot-for-Geq7-model} shows that the time profile for group 4 closely follows 
the zero line before financial year 2011. 
A time profile with all values near zero suggests the model for the group's data is well approximated by $y_{it}=x_{it}'\theta+c_{i}+v_{it}$. 
Correspondingly, a possible interpretation for group 4's behaviour is that before financial year 2011, they were constantly tuning their drawdown rates as they progressively faced higher minimum
drawdown rates and as their account balances changed. This can potentially
represent a group of `engaged' retirees, who regularly occupy themselves
with determining their desired rate of drawdown; however, it is not possible to further investigate this hypothesis using this dataset.

Group 4's
time profile drops significantly below 0 after financial year 2010. 
These movements, while of a smaller magnitude compared to the time profile values of other groups, are still economically significant; they suggest a behavioural response, after netting out the effect of covariates, of reducing drawdown rates relative to earlier levels from 2011 onwards. 

The Online Supplement provides summary statistics and descriptive plots for all seven groups. Based on these, some of the ways in which the groups differ in terms of observable characteristics are:
the proportion of males in groups 1 to 3 are 62\%, 62\% and 69\% respectively, while the sample on aggregate is 56\% male;
the members of groups 1 and 3 tend to be older, with median retirees aged 81 at 31 Dec 2015;
members in groups 4, 5 and 6 have the youngest median retirees, aged 78 at 31 Dec 2015; 
group 5 members have the highest median risk appetite, a variable defined in the Online Supplement as a summary measure of the relative sizes of derived equity returns within individual accounts compared to the S\&P/ASX 200 market index; 
group 7 members have the lowest median risk appetite;
group 7 members make ad-hoc drawdowns least frequently, in roughly 3\% of person--years observed, followed by group 1, for 5\% of person--years observed---for all other groups this frequency was 9--11\%;
in years where they make ad-hoc drawdowns, group 3 members tend to draw down the largest proportions of their account balance using ad-hoc drawdowns over the course of a year---on average, these ad-hoc drawdowns over the year amount to 35\% of their account balance at the start of the respective financial year.

\subsection{Prior expectations}
\begin{figure}
	\caption{\small \label{fig:Time-Profiles-Geq2-model-CIs-from-formula}Point
		estimates and 95\% confidence intervals from analytical formula for
		effects of group-level time-varying unobservable heterogeneity on
		log regular drawdown rates assuming $G=2$. Shaded regions (indistinguishable from point estimates in plot) denote 95\% element-wise confidence intervals constructed using standard
		errors derived from fixed-$T$ variance estimate formula in the Online Supplement to \citet{bonhomme2015grouped}. Time-demeaned
		group time profiles shifted to begin at 0 on the vertical axis.}
	\begin{center}
		\includegraphics[width=0.8\textwidth]{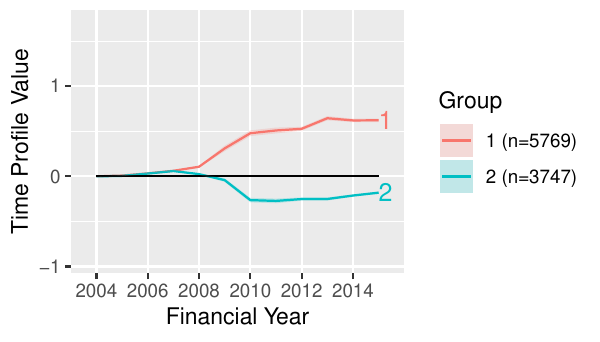}
	\end{center}
\end{figure}

Section \ref{sec:Introduction} presents this study's prior expectations
for finding at least two types of strategies in the data: drawing constant dollar amounts, and following closely
the minimum drawdown rates. Most of the group behaviours found under the
seven-group assumption can be interpreted as cases of these two hypothesised behaviours. Moreover,
a restricted model assuming only two groups already begins to show
evidence for the existence of both of these groups.
Figure \ref{fig:Time-Profiles-Geq2-model-CIs-from-formula}
shows the time profiles for the two-group model. As in Figure \ref{fig:Time-Profiles-Geq1model-CIs-from-formula},
the $y$-axis scale allows for direct comparison of the estimated
magnitudes with the $G=7$ model results. 
The time profile of the first group appears similar to those for the groups in the seven-group model that seem to target
constant dollar amounts for the regular drawdowns; the second profile shows a dip comparable to the group
that seems to follow the minimum drawdown rate.

The Online Supplement also provides descriptive panel
plots for these two groups. These plots are broadly consistent with
the interpretation that the first group often attempts to hold drawdown
amounts constant, although with downward revisions in the amount common,
as well as rapid declines in account balance towards the end of the
period; the second group mainly makes decisions regarding
their drawdown rate. This analysis, alongside the observation that moving to a
two-group model already removes much of the bias in the covariate
effect estimates, suggests that accounting for both rate-based and
amount-based strategies seems to be the most important step in controlling
for unobservable heterogeneity in the data.

\section{\label{sec:Discussion}Discussion}

\subsection{\label{subsec:Retirement-Incomes}Retirement incomes}

A key implication of our work for retirement incomes research and policy is that there is now a statistical basis for empirical results that indicate different behavioural stances explain much of the variation in the drawdowns from phased withdrawal retirement income products. The significant change in covariate effect estimates and the emergence of distinct time profiles, as the number of groups assumed by the GFE estimator increases, is evidence for the presence of multiple behavioural groups in the data. The magnitudes of the estimated group time profile values suggest a sizeable contribution of this behavioural component to the observed drawdowns. 

Retirees need to trade-off current against future consumption in the face of investment, longevity, and other, background risks, including inflation rates and health expenditure \citep{yang2009impact}. Retirement fund trustees and financial advisors can use the findings from this research when aiding retirees who are navigating the trade-offs in apportioning their accumulated retirement savings between consumption, income-generating assets and precautionary assets. The results do not find clear evidence for a group following the optimal heuristic strategy from \citet{bateman2008choices}, which is to initially draw at a fixed rate until the rising minima require drawing larger proportions. Two-thirds of the sample belong to the groups exhibiting a preference for drawing down a constant dollar amount over time. We do not have evidence as to whether the initial level is optimal, but two of the groups targeting level income streams had their account balances begin a rapid decline following the GFC; retirees with this preference who are unable to respond to periods of low or negative investment returns by curtailing their income streams may be at risk of prematurely depleting their account balances. 14\% of the sample belong to the group who most closely follow the next-most optimal heuristic strategy, which is to always draw at the legislated minimum rates. 

Those providing financial advice may wish to convey these findings to retirees in the form of cautionary tales, or as part of a standard set of recommendations presented to retirees. Practitioners and policymakers may also wish to consider these findings when designing retirement income products and policy to align with the Comprehensive Income Products for Retirement framework, which places an emphasis on generating income streams without compromising the ability to retain superannuation savings until late into retirement by including some life annuity products \citep{Treas2016}. Financial advisors may benefit retirees by predicting their individual needs for income and precautionary savings later in life, monitoring their actual experience relative to this forecast, and recommending a reduction in expenditure before poorer experience exhausts their accounts.

The present study clarifies how a robust behavioural analysis
of drawdowns data is vital for informing the various stakeholders in the
retirement incomes space, including retirees, the government, and
the financial services industry. However, Section \ref{sec:Superannuation-Drawdowns-Dataset}
notes that despite having data from multiple superannuation
funds, due to the small number of funds in the estimation
sample, the results may not describe the population of
Australian retirees in general. In particular, using the current dataset
may overlook behavioural patterns present
in funds not included in the sample. 

\subsection{\label{subsec:GFE-in-Behavioural}GFE in behavioural microeconomics and event studies}

We believe that the GFE model is valuable for research in behavioural microeconomics and event studies. While
the present paper focuses on the behavioural interpretations
of the latent group effects, the superannuation data 
covers a period in which multiple common macroeconomic and policy
shocks affect all members in the sample. Hence, the superannuation application
also invites an event-study interpretation, although it is impossible
to disentangle the effects of several common shocks occurring
in close proximity. These include the Global Financial Crisis, the introduction of a new
schedule of minimum drawdown rates for account-based pensions, and
temporary tweaks made to the minimum drawdown rates. Taken together,
however, the time profile plots presented in Figure \ref{fig:fixed-T-variance-formula-results-for-time-profile-plot-for-Geq7-model}
clearly show how otherwise homogeneous-looking trends diverge radically,
plausibly driven by one or more of these shocks. Hence, the results
broadly show how researchers can utilise the same methodology
in an event-study framework.

In the superannuation application, the GFE assumption of a group structure to the time-varying
unobservable heterogeneity is tenable because it was \emph{a priori} expected that an important source
of heterogeneity in the drawdowns arises from the latent motivations
that drive individuals to choose one of a finite number of drawdown
strategies. In general, other behavioural economics studies
with panel data may have analogous motivations to assume 
that there is some true, finite number of groups in the population, which the GFE estimator attempts to characterise. This prior assumption may
be important as it is not known precisely how the estimator behaves when a group structure only approximates, rather
than accurately captures, the nature of the unobservable heterogeneity.
For a discussion on approximating more general forms
of heterogeneity, see e.g. \citet{bonhomme2021discretizing}.


\section{Appendix: Account balance decomposition} \label{sec:p1-appendix}

\begin{figure}
	\caption{\small \label{fig:rdr-lrdr-histos}Histograms of regular drawdown rate variable and its logarithm in superannuation data.}
	\begin{center}
		\includegraphics[width=0.8\textwidth]{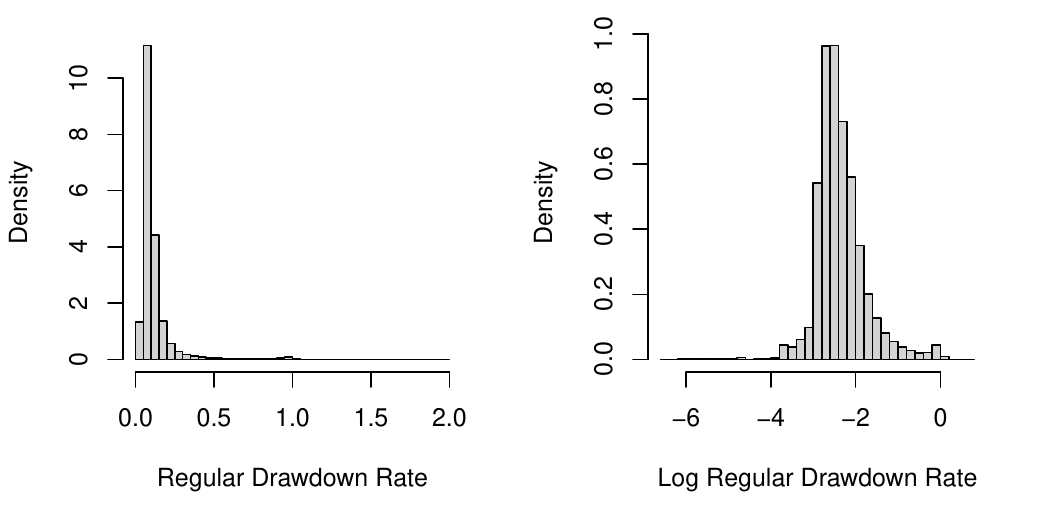}
	\end{center}
\end{figure}

The ABP account balance at 1 July of calendar year $t+1$ for an arbitrary retiree ($i$ subscripts omitted) is given by
$${\rm AB}_{t+1}={\rm AB}_{t}-{\rm DA}_{t}-{\rm ADA}_{t}+{\rm NIRA}_{t}-{\rm AF}_{t}+{\rm NT}_{t},$$
where:
\begin{itemize}
	\item $t$ indexes financial years;
	\item ${\rm AB}_{t}$ is account balance at start of financial year $t$;
	\item ${\rm DA}_{t}$ is the regular drawdown amount. This is the total dollar amount,
	determined in advance of the financial year, received over the course of the financial year
	as regular payments. E.g., the retiree nominates, before a financial year starts,
	to receive \$1000, once a month, each month, giving ${\rm DA}_{t}=$ 12,000;
	\item ${\rm ADA}_{t}$ is the ad-hoc drawdown amount. This is the total dollar value
	of any lump-sum commutations made throughout the year, not including
	the regular payment stream nominated;
	\item ${\rm NIRA}_{t}$ are net investment returns, in dollars, over financial year
	$t$;
	\item ${\rm AF}_{t}$ are account fees deducted;
	\item ${\rm NT}_{t}$ are net transfers, e.g., rollover amount from another super fund, or receipt
	of spouse's death benefit.
\end{itemize}
The dependent variable in the estimated model is $y_{it} := \ln {\rm DR}_{it}$, where ${\rm DR}_{it}={\rm DA}_{it}/{\rm AB}_{it}$; ${\rm DA}_{it}\neq {\rm AB}_{it}-{\rm AB}_{i,t+1}$. Figure \ref{fig:rdr-lrdr-histos} shows histograms of $ {\rm DR}_{it}$ and $ \ln {\rm DR}_{it}$ for all non-missing individual--time combinations in the superannuation data. 


\phantomsection

\addcontentsline{toc}{section}{References}

\bibliographystyle{apacite}
\bibliography{paper1-standalone}

\begin{table}
	\caption{\small \label{tab:Age-based-Minimum-Drawdown-Rates}Minimum drawdown rates by age for account-based pensions, effective 1 July 2007.}
	\begin{tabular}{|c|c|c|c|c|c|c|c|}
		\hline 
		Age at financial year start & \textless 65 & 65--74 & 75--79 & 80--84 & 85--89 & 90--94 & 95+\tabularnewline
		\hline 
		Minimum drawdown rate & 0.04 & 0.05 & 0.06 & 0.07 & 0.09 & 0.11 & 0.14\tabularnewline
		\hline 
	\end{tabular}
\end{table}

\begin{table}
	\caption{\label{tab:agg-time-invariant-variables}Summary statistics -- Time-invariant
		variables.}
	
	\resizebox{\textwidth}{!}{
		
		\begin{tabular}{|c|c|c|c|c|}
			\hline 
			& Age at 31 December 2015 & Age at Account Open & Sex: Male & Risk Appetite\tabularnewline
			\hline 
			Mean & 79.4 & 63.57 & 0.56 & 0.41\tabularnewline
			\hline 
			SD & 5.22 & 4.17 & 0.5 & 0.21\tabularnewline
			\hline 
			Median & 79.78 & 64.23 & 1 & 0.46\tabularnewline
			\hline 
			Q1 & 76.37 & 60.9 & 0 & 0.25\tabularnewline
			\hline 
			Q3 & 83.03 & 65.39 & 1 & 0.54\tabularnewline
			\hline 
			Min & 60.66 & 48.48 & 0 & 0\tabularnewline
			\hline 
			Max & 101.46 & 85.44 & 1 & 1.88\tabularnewline
			\hline 
			Count & 9516 & 9516 & 9516 & 9507\tabularnewline
			\hline 
		\end{tabular}
		
	}
	
	\vspace{4pt} \small The age at 31 December 2015 represents the individual's cohort, equivalent
	to measuring a year-of-birth variable. The age at account opening is the age when the retiree
	initiates a phased withdrawal product and begins drawing down from the
	account. The sex indicator variable equals 1 if the retiree
	is male, and 0 otherwise. The risk appetite variable is a proxy for the returns in the account relative
	to the reference S\&P/ASX 200 index; see Online Supplement for details. SD refers to the sample standard deviation of the variable. Q1 and Q3 refer to the empirical 25th and 75th percentiles, respectively.
\end{table}

\begin{table}
	\caption{\label{tab:agg-time-varying-variables}Summary statistics -- Time-varying
		variables.}
	
	\resizebox{\textwidth}{!}{
		
		\begin{tabular}{|c|c|c|c|c|c|c|}
			\hline 
			& Regular Drawdown & Regular Drawdown & Ad-hoc Drawdown & Ad-hoc Drawdown & Ad-hoc Drawdown & Account Balance\tabularnewline & Rate & Amount & Indicator & Rate & Amount &  \tabularnewline
			\hline 
			Mean & 0.12 & 6435.91 & 0.07 & 0.15 & 10,216.56 & 72,686.55\tabularnewline
			\hline 
			SD & 0.12 & 6121.76 & 0.25 & 0.21 & 24,672.68 & 78,721.39\tabularnewline
			\hline 
			Median & 0.09 & 4800 & 0 & 0.07 & 4655.9 & 52,063\tabularnewline
			\hline 
			Q1 & 0.07 & 2992 & 0 & 0.02 & 1132.33 & 30,532.5\tabularnewline
			\hline 
			Q3 & 0.12 & 7728 & 0 & 0.18 & 10,000 & 87,427\tabularnewline
			\hline 
			Min & 0 & 1 & 0 & 0 & 1 & 1\tabularnewline
			\hline 
			Max & 2 & 166,695 & 1 & 0.9 & 600,000 & 2,427,083\tabularnewline
			\hline 
			Count & 107,935 & 107,975 & 108,717 & 7450 & 7454 & 108,635\tabularnewline
			\hline 
		\end{tabular}

	}
	
	\vspace{4pt} \small The regular drawdown rate is the variable of interest; subsequent regressions use the natural logarithm of the regular drawdown rate as the dependent variable. The account balance is defined as the position at the start of the relevant financial year; this is the denominator for the rate computations. The ad-hoc drawdown indicator variable equals 1 if the retiree
	made an ad-hoc withdrawal from their account during a given
	financial year. SD refers to the sample standard deviation of the variable. Q1 and Q3 refer to the empirical 25th and 75th percentiles, respectively. 
\end{table}

\end{document}


\title{\vspace{-20mm}Online Supplement for \\`Hidden Group Time Profiles: Heterogeneous
Drawdown Behaviours in Retirement'\vspace{-4mm}}
\author{Igor Balnozan\footremember{UNSW}{University of New South Wales, UNSW Sydney, NSW 2052, Australia}\footnote{Correspondence: i.balnozan@unsw.edu.au / West Lobby Level 4, 		UNSW Business School Building, UNSW	Sydney, NSW 2052, Australia.}, Denzil G. Fiebig\footrecall{UNSW}, Anthony Asher\footrecall{UNSW}, Robert Kohn\footrecall{UNSW}, Scott A. Sisson\footrecall{UNSW}}
\date{Version: 19 March 2025\vspace{-4mm}}
\maketitle
\begin{abstract}
		\begin{onehalfspace}
This material supplements the paper
with: 
robustness checks to examine the sensitivity of the main results to changes in the dataset and estimation strategy; 
a simulation exercise to compare simulated standard errors with those estimated from the analytical formula and the bootstrap;
additional details on the superannuation drawdowns dataset; further details on the proposed modification to the estimation algorithm for unbalanced data;
additional descriptive results for the seven-group model;
and panel plots for the two-group model.
		\end{onehalfspace}
\end{abstract}
\vfill

\begin{singlespace} 	
{\footnotesize  	
		\textbf{Keywords} 	
	\vspace{-4mm} 
	
		Panel data, discrete heterogeneity, microeconomics, retirement 
	
		\textbf{JEL codes} 	
	\vspace{-4mm} 	

		C51, D14, G40 	
	
		\textbf{Acknowledgements} 	
	\vspace{-4mm} 	

Balnozan is grateful for the support provided by the Commonwealth Government of Australia through the provision of an Australian Government Research Training Program Scholarship, and to State Super for provision of the State Super Academic Scholarship.
Balnozan, Kohn and Sisson are partially supported by the Australian Research Council through the Australian Centre of Excellence in Mathematical and Statistical Frontiers (ACEMS; CE140100049).
We
thank Plan For Life, Actuaries \& Researchers, who collected, cleaned and allowed
us to analyse the data used in this research; this data capture forms
one part of a broader survey into retirement incomes, commissioned
by the Institute of Actuaries of Australia.
This research includes computations using the computational cluster
Katana supported by Research Technology Services at UNSW Sydney. Katana DOI: 10.26190/669x-a286

		\textbf{Conflict of interest statement} 		
		\vspace{-4mm} 			

		The authors have no conflict of interest to declare.

	} \end{singlespace}

\newpage{}
\doublespacing

\section{\label{sec:label-matching-alg-details}Label-matching procedure}

The following label-matching procedure is used in the paper to fix group labels across independent bootstrap replications, based on the method \citet{hofmans2015added} apply to a similar problem in $k$-means clustering.

\begin{enumerate}
	\item Fix the labelling generated by the GFE model estimation completed on the original sample. 
	These labels act as the reference labels to which all subsequent estimates will align their group labels. 
	Define the initial time profile estimates labelled using the reference labels as $\widehat{\alpha}_{g}^{r}:=\left( \widehat{\alpha}_{g,1}^{r}, \widehat{\alpha}_{g,2}^{r}, ..., \widehat{\alpha}_{g,T}^{r} \right)'$ for $g=1,2,...,G$.
	\item For $b=1,2,...,B$, find the permutation of group labels for the $b^{th}$ bootstrap run
	which minimises the sum of Euclidean distances, aggregated
	across all $G$ time profiles, between the time profiles estimated
	in the $b^{th}$ bootstrap run and the original sample estimates as
	identified by the reference labels; i.e., for $b=1,2,...,B$, do:
	\begin{enumerate}
		\item 
		Using the unmodified labels generated by the $b^{th}$ bootstrap run, define the estimated time profiles from the $b^{th}$ bootstrap run with these labels as the length-$T$ vectors $\widehat{\alpha}_{g}^{b,0}$ for $g=1,2,...,G$.
		\item 
		Index the $G!$ permutations of the label sequence $\left(1,2,...,G\right)$ by $p=1,2,...,G!$. 
		Each permutation $p$ for the $b^{th}$ bootstrap run defines a permuted set of time profiles, $\widehat{\alpha}_{g}^{b,p}$, for $g=1,2,...,G$, and a corresponding mapping function,\\ $m_{p}:\{1,2,...,G\}\rightarrow\{1,2,...,G\}$, such that $m_{p}\left(g\right)$ is the $g^{th}$ element of the $p^{th}$ permuted label sequence, and $\widehat{\alpha}_{g}^{b,p}=\widehat{\alpha}_{m_{p}\left(g\right)}^{b,0}$.
		\item For $p=1,2,...,G!$, do:\footnote{If $G!$ is too large to check each permutation in an acceptable amount of time, an alternative is to use the more computationally efficient approach given in \citet{munkres1957algorithms} and used by \citet{hofmans2015added}.}
		\begin{enumerate}
			\item For $g=1,2,...,G$, compute the Euclidean distance between the length-$T$ vectors $\widehat{\alpha}_{g}^{b,p}$
			and $\widehat{\alpha}_{g}^{r}$, given by $\lVert\widehat{\alpha}_{g}^{b,p}-\widehat{\alpha}_{g}^{r}\rVert$; here $ \lVert \cdot \rVert $ is the $L_2$ norm.
			\item Sum all $G$ Euclidean distances to compute an aggregate distance metric for
			that permutation, $\sum_{g=1}^{G}\lVert\widehat{\alpha}_{g}^{b,p}-\widehat{\alpha}_{g}^{r}\rVert$.
		\end{enumerate}
		\item Select the permutation, $p^\star$, with the smallest aggregate distance metric:
		\begin{align}
			p^\star=\argmin_{p\in\{1,2,...,G!\}}\sum_{g=1}^{G}\lVert\widehat{\alpha}_{g}^{b,p}-\widehat{\alpha}_{g}^{r}\rVert.
		\end{align}  
		Relabel the $G$ groups in the $b^{th}$ bootstrap run using the labelling given by the
		selected permutation.
	\end{enumerate}
\end{enumerate}

\section{\label{sec:p1-ext2-extra-details}More details on `Extension 2: Alternative estimation method for unbalanced data'}

We now justify the claims made in Section 4.5 of the paper about the relationship between the three methods for estimating time profiles with balanced and unbalanced data.

\paragraph{Lemma 1.}

Methods 1--3 estimate the same time profiles in the balanced
data case.

\paragraph{Proof.}

The regression equation being estimated by method 3 (no constant term;
$T-1$ time-demeaned time dummy variables, $\dot{\delta}_{s}$) is
\begin{equation}
	\dot{y}_{it}=\dot{x}_{it}'\theta+\sum_{s=2}^{T}\tilde{\dot{\alpha}}_{s}\dot{\delta}_{s}+\dot{v}_{it},\label{eq:method-3-reg-eqn} 
\end{equation}
where $\dot{\delta}_{s}:=\delta_{s}-T^{-1}\sum_{l=1}^{T}\delta_{l}$;
$\delta_{l}=\mathbb{I}\left\{ t=l\right\} $. $\dot{\delta}_{s}=\delta_{s}-T^{-1}$,
as $\sum_{l=1}^{T}\delta_{l}=1$ for all $t$. Thus, (\ref{eq:method-3-reg-eqn})
can be rewritten using time dummy variables, $\delta_{s}$, instead
of time-demeaned time dummy variables, $\dot{\delta}_{s}$, as
\begin{equation}
	\dot{y}_{it}=\dot{x}_{it}'\theta+\dot{\alpha}_{1}+\sum_{s=2}^{T}\tilde{\dot{\alpha}}_{s}\delta_{s}+\dot{v}_{it},\label{eq:method-2-reg-eqn} 
\end{equation}
where $\dot{\alpha}_{1}:=-T^{-1}\sum_{s=2}^{T}\tilde{\dot{\alpha}}_{s}$. 

(\ref{eq:method-2-reg-eqn}) is precisely the regression equation
being estimated by method 2 (include constant term; $T-1$ time dummy
variables, $\delta_{s}$). Defining $\dot{\alpha}_{s}:=\dot{\alpha}_{1}+\tilde{\dot{\alpha}}_{s}$
for $s=1,2,...,T$, (\ref{eq:method-2-reg-eqn}) can be rewritten
as
\begin{equation}
	\dot{y}_{it}=\dot{x}_{it}'\theta+\sum_{s=1}^{T}\dot{\alpha}_{s}\delta_{s}+\dot{v}_{it},\label{eq:method-1-reg-eqn} 
\end{equation}
because $\dot{\alpha}_{1}+\sum_{s=2}^{T}\tilde{\dot{\alpha}}_{s}\delta_{s}=\dot{\alpha}_{1}\delta_{1}+\sum_{s=2}^{T}\dot{\alpha}_{s}\delta_{s}$. 

(\ref{eq:method-1-reg-eqn}) is the regression equation estimated
by method 1 (no constant term; $T$ time dummy variables, $\delta_{s}$).

Combining $\dot{\alpha}_{1}=-T^{-1}\sum_{s=2}^{T}\tilde{\dot{\alpha}}_{s}$
with $\dot{\alpha}_{s}=\dot{\alpha}_{1}+\tilde{\dot{\alpha}}_{s}$
implies $\sum_{t=1}^{T}\dot{\alpha}_{t}=\dot{\alpha}_{1}+\sum_{t=2}^{T}\dot{\alpha}_{t}$
$=\dot{\alpha}_{1}+\sum_{t=2}^{T}\left(\dot{\alpha}_{1}+\tilde{\dot{\alpha}}_{t}\right)$
$=\dot{\alpha}_{1}+\left(T-1\right)\dot{\alpha}_{1}-T\dot{\alpha}_{1}=0$.

The conversion between $\dot{\alpha}_{s}$ and $\tilde{\dot{\alpha}}_{s}$ terms is similar to the transformation used in \citet[][p.~178]{suits1984dummy}, but applied to a model with no constant term.

\paragraph{Lemma 2.}

With unbalanced data, method 1 does not guarantee that $\sum_{t=1}^{T}\widehat{\dot{\alpha}}_{t}=0$.

\paragraph{Proof.}

Consider a special case with no covariates and $T=2$ as an illustration.
The corresponding regression equation is
\[
\dot{y}_{it}=\dot{\alpha}_{1}\delta_{1}+\dot{\alpha}_{2}\delta_{2}+\dot{v}_{it},
\]
or equivalently, $\dot{y}_{i1}=\dot{\alpha}_{1}\delta_{1}+\dot{v}_{i1}$
and $\dot{y}_{i2}=\dot{\alpha}_{2}\delta_{2}+\dot{v}_{i2}$. 

The least-squares estimators of $\dot{\alpha}_{1}$ and $\dot{\alpha}_{2}$
are $\widehat{\dot{\alpha}}_{s}=N_{s}^{-1}\sum_{i\in{\cal I}_{s}}\left(y_{i,s}-\bar{y}_{i}\right)$
for $s=1,2$, where ${\cal I}_{s}$ is the set of units observed in
period $s$; $N_{s}$ is the number of elements in ${\cal I}_{s}$;
$\bar{y}_{i}:=T_{i}^{-1}\sum_{t\in\tau_{i}}y_{it}$; $\tau_{i}$ is
the set of periods in which unit $i$ is observed; $T_{i}$ is the
number of elements in $\tau_{i}$. Then, $\widehat{\dot{\alpha}}_{1}+\widehat{\dot{\alpha}}_{2}=:A-B$,
where $A:=N_{1}^{-1}\sum_{i\in{\cal I}_{1}}y_{i,1}+N_{2}^{-1}\sum_{i\in{\cal I}_{2}}y_{i,2}$,
and $B:=N_{1}^{-1}\sum_{i\in{\cal I}_{1}}\bar{y}_{i}+N_{2}^{-1}\sum_{i\in{\cal I}_{2}}\bar{y}_{i}$.

When the panel is balanced, $N_{1}=N_{2}=N$, ${\cal I}_{1}={\cal I}_{2}=\left\{ 1,2,...,N\right\} $,
and $T_{i}=T=2$ for all $i$. Then, $A=B=T\bar{\bar{y}}$, where
$\bar{\bar{y}}:=\left(N_{1}+N_{2}\right)^{-1}\left(\sum_{i\in{\cal I}_{1}}y_{i,1}+\sum_{i\in{\cal I}_{2}}y_{i,2}\right)$
$=\left(NT\right)^{-1}\left(\sum_{i=1}^{N}\sum_{t=1}^{T}y_{it}\right)$,
so that $\widehat{\dot{\alpha}}_{1}+\widehat{\dot{\alpha}}_{2}=0$.
In general, however, with unbalanced data this zero-sum is not guaranteed: a counter-example
is the dataset $$\left\{ y_{it}\right\} =\left[\begin{array}{cc}
	1 & 2\\
	3 & .\\
	4 & 5
\end{array}\right],$$ which has $A=6+\frac{1}{6}$ $\neq B=6+1.5$ $\neq T\bar{\bar{y}}=6$.

\paragraph{Lemma 3.}

Method 3 can be used on unbalanced data while ensuring that $\sum_{t=1}^{T}\widehat{\dot{\alpha}}_{t}=0$.

\paragraph{Proof.}

By construction, method 3 estimates $T-1$ terms, $\widehat{\tilde{\dot{\alpha}}}_{t}$
for $t=2,3,...,T$, then computes $\widehat{\dot{\alpha}}_{1}=-T^{-1}\sum_{t=2}^{T}\widehat{\tilde{\dot{\alpha}}}_{t}$ and $\widehat{\dot{\alpha}}_{t}=\widehat{\dot{\alpha}}_{1}+\widehat{\tilde{\dot{\alpha}}}_{t}$ for $t=2,3,...,T$,
which ensures that $\sum_{t=1}^{T}\widehat{\dot{\alpha}}_{t}=0$.

\paragraph{Other comments.}

The unmodified GFE model estimate obtains the time profiles arising from interacting group identity with a set of $T$ time dummy variables, analogous to using method 1 in the $G=1$ case. In the modified procedure's parameter update step, the algorithm interacts the group identity with $T-1$ time-demeaned time dummy variables, generating $G(T-1)$ estimates of relative effects $\tilde{{\dot{\alpha}}}_{g t}:=\dot{\alpha}_{g t} - \dot{\alpha}_{g, 1}$; this is analogous to method 3 for $G=1$. As the GFE algorithm requires $GT$ estimates of the $\dot{\alpha}_{gt}$ to compute the sum of squared residuals across all $i$ and $t$, it is necessary to perform the conversion step from the $T-1$ estimates of the ${\tilde{\dot{\alpha}}}_{gt}$ to $T$ estimates of the ${\dot{\alpha}}_{gt}$, for each of the $G$ time profiles.

\section{\label{sec:Robustness-Checks}Robustness checks}

This section explores the sensitivity of the main results to changes
in data composition and estimation methodology.

\subsection{\label{subsec:Data-Composition}Data composition}

The first robustness check determines whether the results depend materially
on using a balanced subsample of each fund's data. It uses the same model as the main results,
but the data is filtered down to retain
a fully balanced sample, leaving $N=8274$ units in the sample.

\subsubsection*{Covariate effects}

\begin{figure}
	\caption{\label{fig:baln-subs-covrt-ests-vs-G-1-to-16}Fully balanced subsample
		-- Point estimates and 95\% confidence
		intervals for partial effects of log minimum drawdown rate and log account balance covariates on log
		regular drawdown rate, controlling for group-level time-varying unobservable
		heterogeneity assuming $G=1,2,...,16$. Shaded regions denote confidence intervals constructed
		using standard errors derived from fixed-$T$ variance estimate formula.}
	
	\includegraphics[width=1\textwidth]{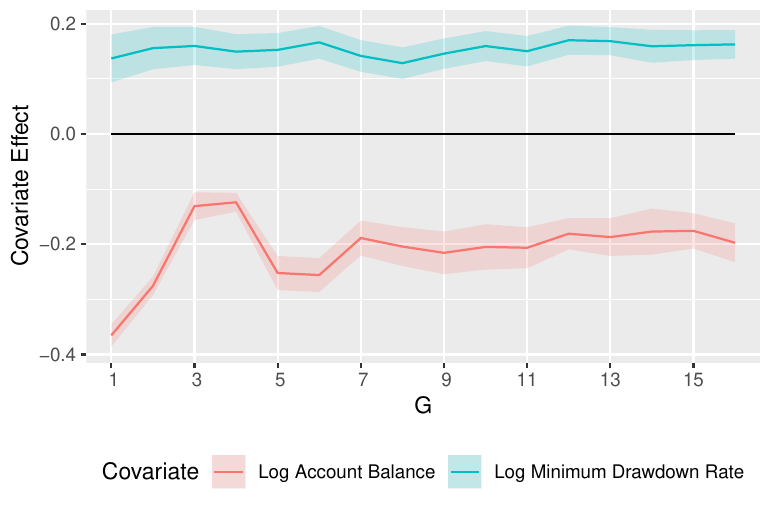}
\end{figure}

Figure \ref{fig:baln-subs-covrt-ests-vs-G-1-to-16} plots covariate
effect estimates for different values of $G$ using the fully balanced
subsample. Like the dataset used for the main results in the paper,
the fully balanced subsample supports a seven-group model using the
selection method in the paper.

\subsubsection*{Time profiles}

\begin{figure}
	\caption{\label{fig:baln-subs-Geq4-to-9-time-profile-plots}Fully balanced
		subsample -- Point estimates for effects of group-level time-varying
		unobservable heterogeneity on log regular drawdown rates assuming
		$G=4,5,...,9$. Estimated time-demeaned group time profiles shifted
		to begin at 0 on the vertical axis.}
	
	\includegraphics[width=1\textwidth]{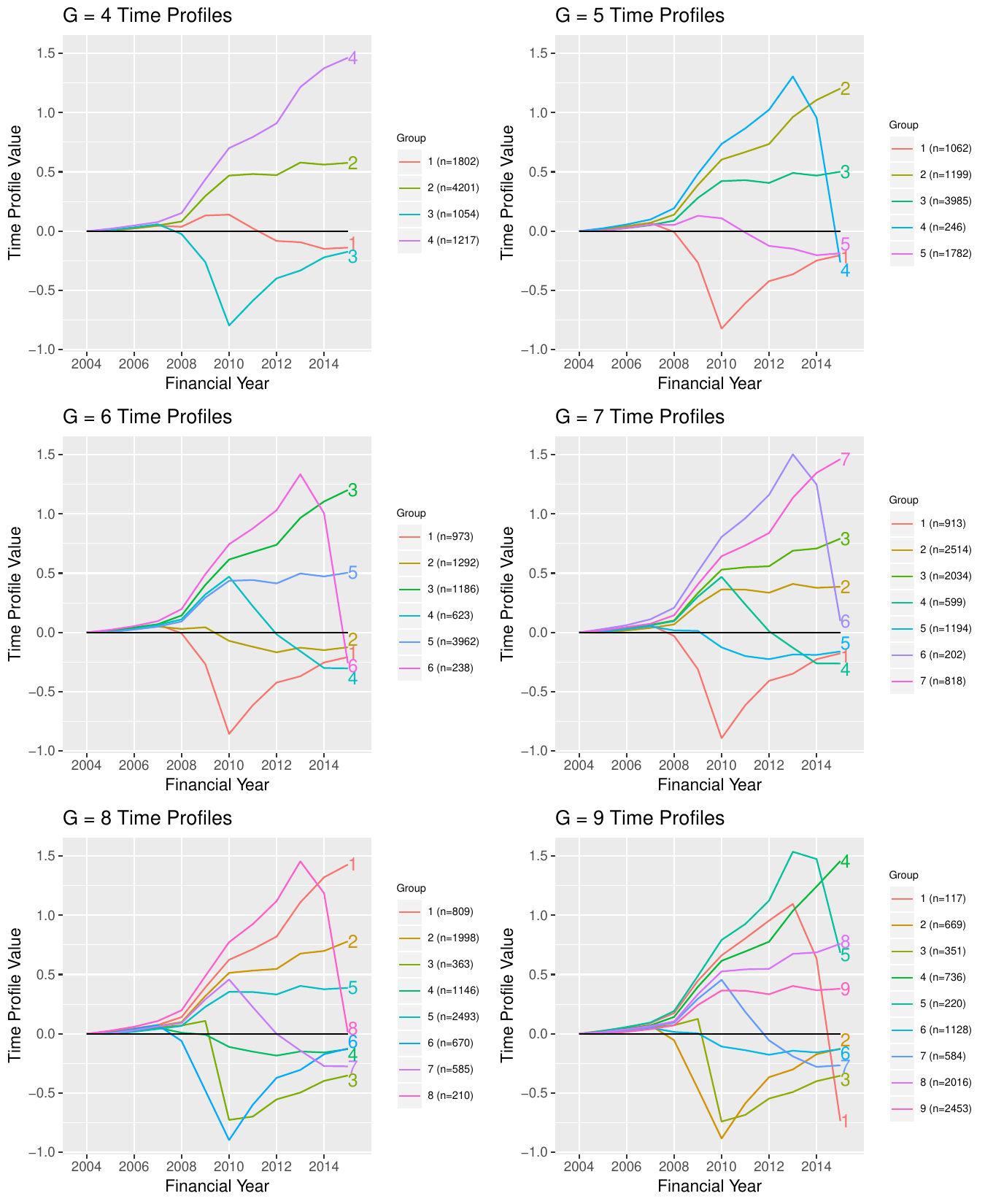}
	
\end{figure}

\begin{figure}
	\caption{\label{fig:baln-subs-Geq7-Time-Profile-Estimates-with-CIs}Fully balanced
		subsample -- Point estimates and 95\% confidence intervals from analytical
		formula for effects of group-level time-varying unobservable heterogeneity
		on log regular drawdown rates assuming $G=7$. Shaded regions denote 95\% element-wise confidence intervals constructed
		using standard errors derived from fixed-$T$ variance estimate formula.
		Time-demeaned group time profiles shifted to begin at 0 on the vertical
		axis.}
	
	\includegraphics[width=1\textwidth]{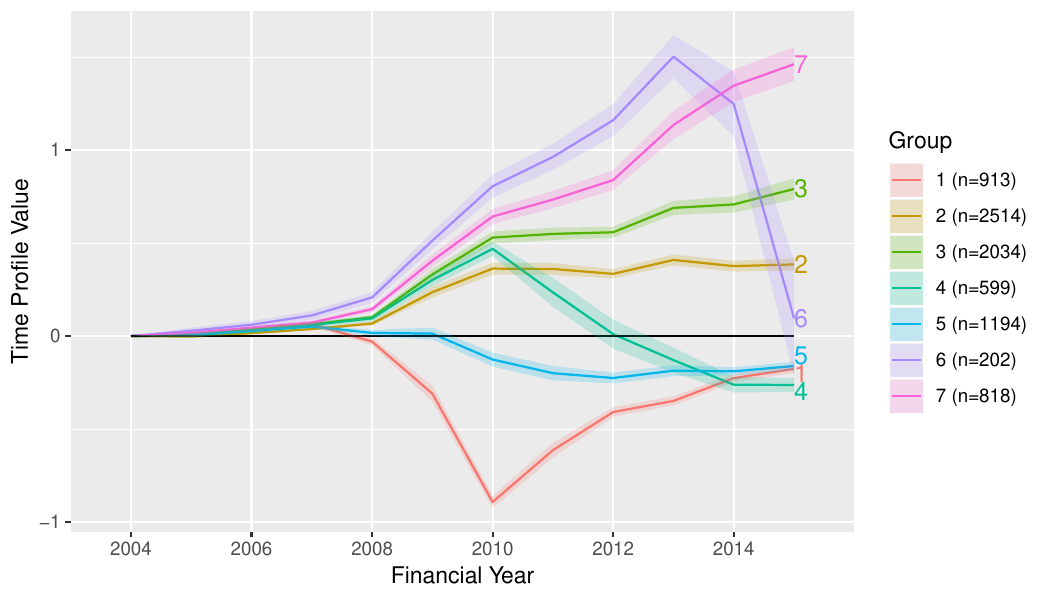}
	
\end{figure}

Figure \ref{fig:baln-subs-Geq4-to-9-time-profile-plots} plots the time
profiles for $G$ = 4--9 using the fully balanced subsample.
Figure \ref{fig:baln-subs-Geq7-Time-Profile-Estimates-with-CIs} shows the time
profiles for $G=7$, with element-wise confidence intervals computed using
standard error estimates derived from the fixed-$T$ variance estimate
formula.
While the numerical values differ, the economic interpretation of these results is similar to the interpretation of the main results in the paper.

\subsection{Estimation methodology}

\subsubsection{\label{subsec:Number-of-Starting}Number of starting values}

The number of possible allocations of $N$ units into $G$ groups is $G^{N}$, and solutions found by the GFE procedure depend on starting values for the algorithm \citep{bonhomme2015grouped}; hence,  as $N$ and $G$ increase, the GFE procedure is more likely to find a local, rather than global, optimum.
This may require an increasing number of randomly selected starting values for the algorithm
to adequately explore the solution space, if individual runs become
trapped in regions around local optima.

\begin{figure}
	\caption{\label{fig:Time-Profiles-1-million-starting-values}One million starting
		values -- Point estimates and 95\% confidence intervals from analytical
		formula for effects of group-level time-varying unobservable heterogeneity
		on log regular drawdown rates assuming $G=7$. Shaded regions denote 95\% element-wise confidence intervals derived
		from fixed-$T$ variance estimate formula. Time-demeaned group time
		profiles shifted to begin at 0 on the vertical axis.}
	
	\includegraphics[width=1\textwidth]{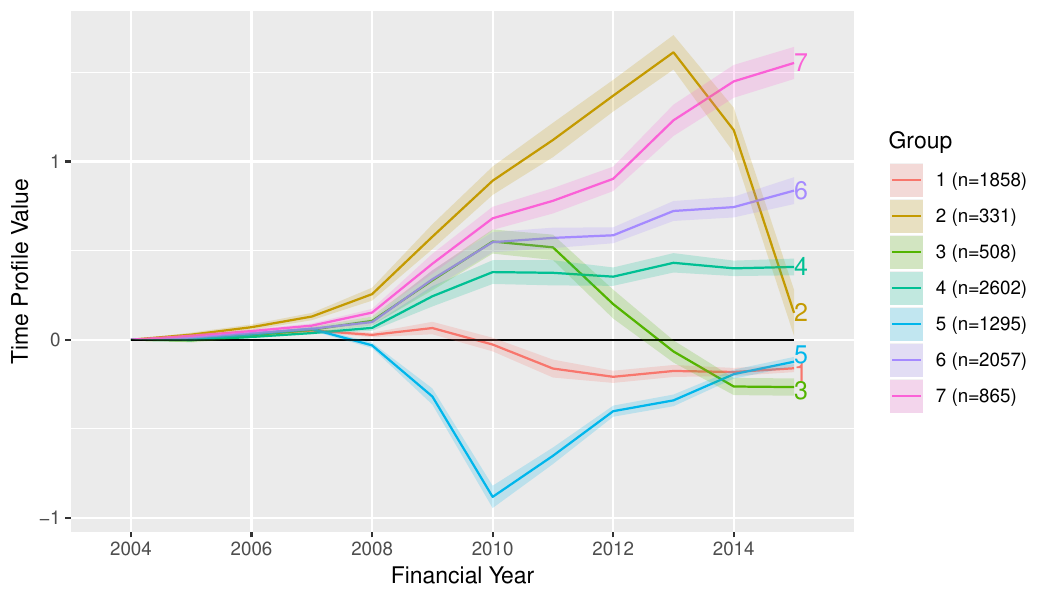}
	
\end{figure}

To test the sensitivity of the main results to the choice of 1000 starting
values, Figure \ref{fig:Time-Profiles-1-million-starting-values} provides
the estimated group time profiles from the equivalent estimation using
1 million starting values. With 1000 starting values, the objective function value is 2673.732,
while with 1 million starting values the value is 2673.716. This change in the sum of squared errors across 107,935 data points is small, suggesting there is little gained from running the estimation procedure for more than 1000 starting values; the following paragraphs investigate whether there are any economically meaningful differences in the results with 1 million starting values compared to the main results. 

The time profiles are nearly identical
to those from the main results in the paper. The largest absolute difference in any pair of corresponding
time profile value estimates is approximately $1.49\times10^{-3}$.
With 1 million starting values, using the group labels as in the main results in the paper, group 4 contains four more people, group
7 has one person less and group 5 has three fewer people, compared to the run with 1000 starting values.

Table \ref{tab:Covariate-Effects-=0020131-million-starting-values} also provides the covariate effect estimates for the estimation
with 1 million starting values.
The estimates and standard errors are identical to those in the main results
to three decimal places. 
The largest difference in estimated effect magnitudes is approximately
$2.87\times10^{-4}$, for the coefficient on the log minimum drawdown rate variable.
These results suggest that the economic interpretation of the main results in the paper are robust to the number of starting values used; however, it is still possible that there exists a more optimal solution with materially different results for the covariate effects
or time profiles that was not found using 1 million starting values.

\begin{table}
	
	\caption{\label{tab:Covariate-Effects-=0020131-million-starting-values}One million
		starting values -- Point estimates and 95\% confidence
		intervals for partial effects of log minimum drawdown rate and log account balance covariates
		on log regular drawdown rate, controlling for group-level time-varying
		unobservable heterogeneity assuming $G=6$. Standard errors are derived from fixed-$T$ variance estimate
		formula.}
	\centering
	\begin{tabular}{|c|c|}
		\hline 
		Covariate & Estimate \\&(Standard Error)\tabularnewline
		\hline 
		Log Minimum Drawdown Rate & 0.144 \\&(0.0248)\tabularnewline
		\hline 
		Log Account Balance & $-$0.147 \\&(0.0139)\tabularnewline
		\hline 
	\end{tabular}
	
\end{table}

\subsubsection{Unmodified estimation procedure}

The paper uses a modified estimation
method for unbalanced data, having the property that for $G=1$,
the results align precisely with those obtained by running a standard
two-way fixed-effects regression model. Comparison output is presented here, using the unmodified algorithm to test the sensitivity of the
main results in the paper to this alternative procedure.

\begin{figure}
	\caption{\label{fig:UNF-covrt-ests-vs-G-1-to-16}Unmodified estimation procedure
		-- Point estimates and 95\% confidence
		intervals for partial effects of log minimum drawdown rate and log account balance covariates on log
		regular drawdown rate, controlling for group-level time-varying unobservable
		heterogeneity assuming $G=1,2,...,16$. Shaded regions denote confidence intervals constructed
		using standard errors derived from fixed-$T$ variance estimate formula.}
	
	\includegraphics[width=1\textwidth]{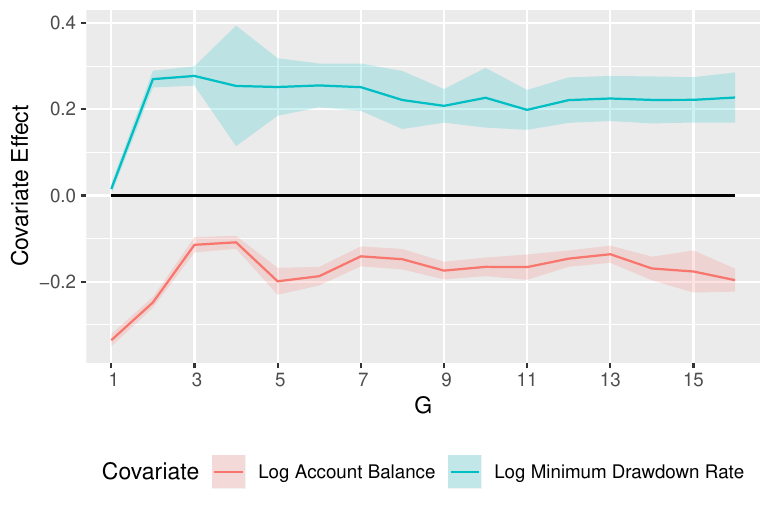}
	
\end{figure}

Figure \ref{fig:UNF-covrt-ests-vs-G-1-to-16} shows how the covariate
effect estimates evolve with $G$. Overall, the results for the log
account balance variable appear almost identical to those in the main
results. Although the results for the log minimum drawdown rate differ
more significantly in magnitude to those in the main results, their economic
implications are similar. 

\begin{figure}
	\caption{\label{fig:UNF-Geq4-to-9-time-profile-plots}Unmodified estimation
		procedure -- Point estimates for effects of group-level time-varying
		unobservable heterogeneity on log regular drawdown rates assuming
		$G=4,5,...,9$. Estimated time-demeaned group time profiles shifted
		to begin at 0 on the vertical axis.}
	
	\includegraphics[width=1\textwidth]{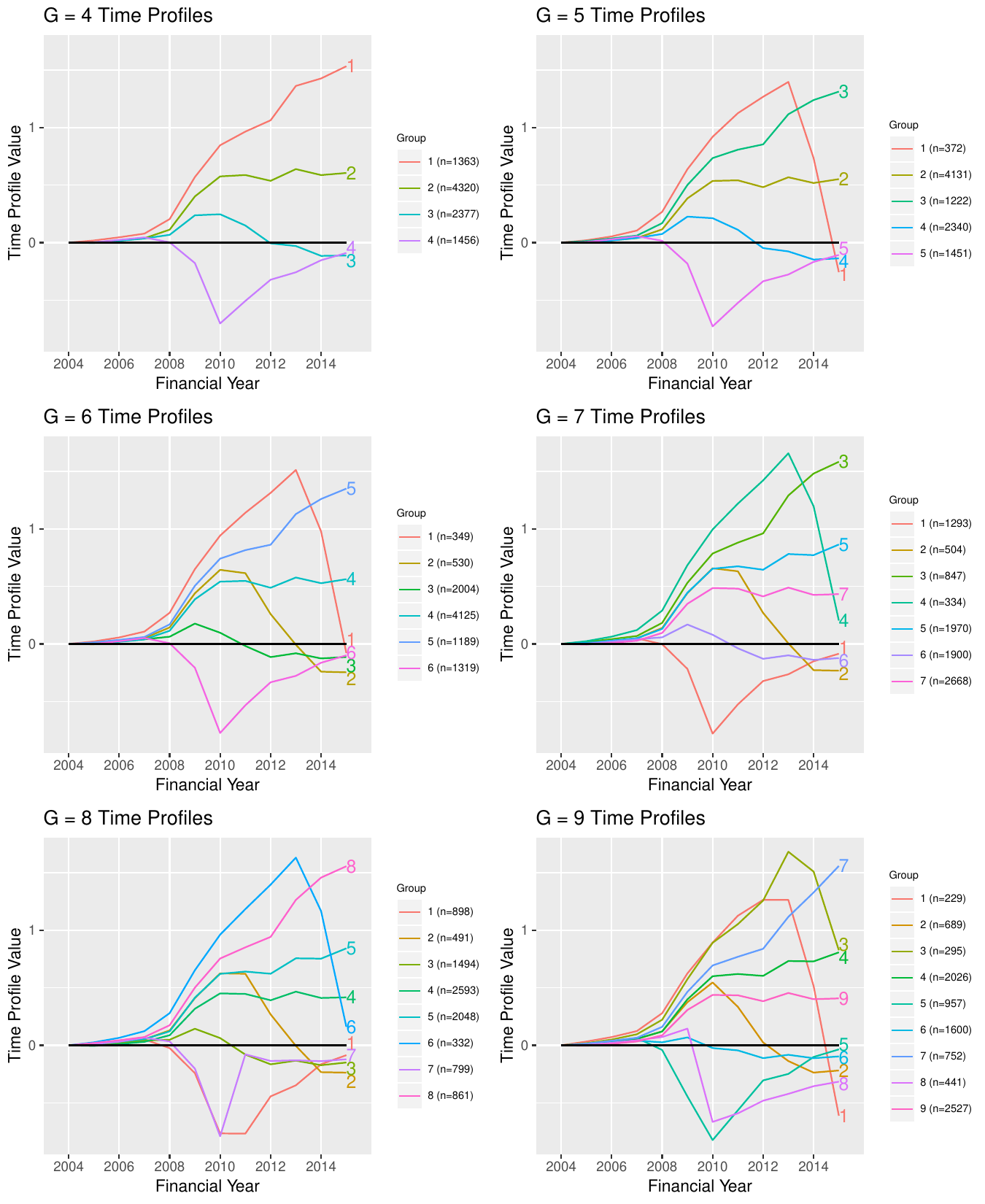}
	
\end{figure}

\begin{figure}
	\caption{\label{fig:UNF-Geq7-Time-Profile-Estimates-with-CIs}Unmodified estimation
		procedure -- Point estimates and 95\% confidence intervals from analytical
		formula for effects of group-level time-varying unobservable heterogeneity
		on log regular drawdown rates assuming $G=7$. Shaded regions denote 95\% element-wise confidence intervals constructed
		using standard errors derived from fixed-$T$ variance estimate formula.
		Time-demeaned group time profiles shifted to begin at 0 on the vertical
		axis.}
	
	\includegraphics[width=1\textwidth]{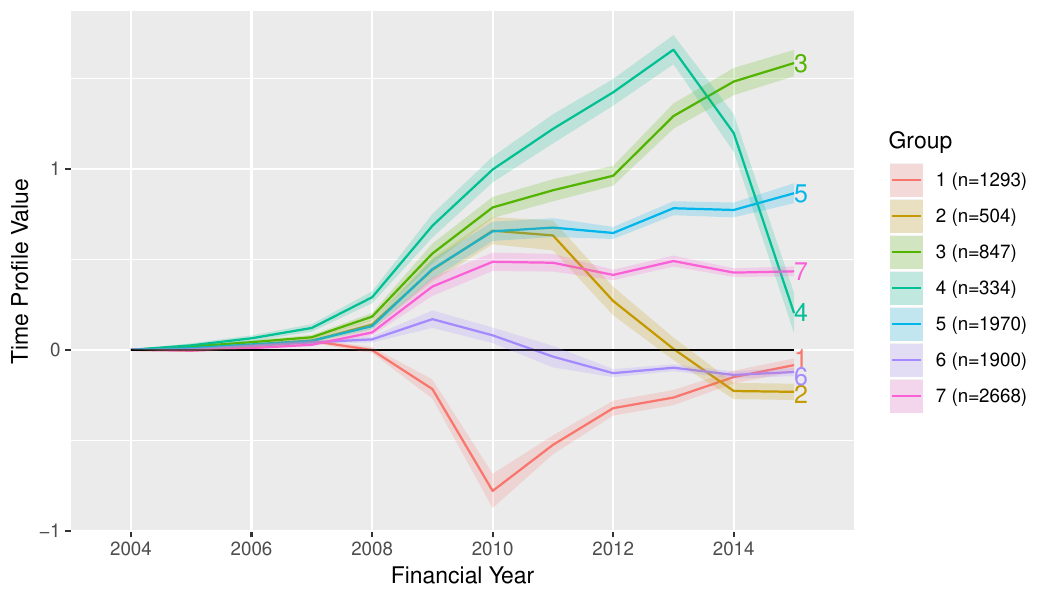}
	
\end{figure}

Figure \ref{fig:UNF-Geq4-to-9-time-profile-plots} presents time profile
point estimates for $G=4,5,...,9$. Figure \ref{fig:UNF-Geq7-Time-Profile-Estimates-with-CIs}
shows the time profile plot for $G=7$, including 95\% element-wise confidence intervals
constructed using standard errors derived from the fixed-$T$ variance
estimate formula. The plots show that the time profiles obtained
from both estimation strategies are similar. 

\begin{figure}
	\caption{\label{fig:UNF-Geq1-Time-Profile-Estimates-with-CIs}Unmodified estimation
		procedure -- Point estimates and 95\% confidence intervals from analytical
		formula for effects of group-level time-varying unobservable heterogeneity
		on log regular drawdown rates assuming $G=1$. Shaded regions (indistinguishable from point estimates in plot) denote 95\% element-wise confidence intervals constructed
		using standard errors derived from fixed-$T$ variance estimate formula.
		Time-demeaned group time profiles shifted to begin at 0 on the vertical
		axis.}
	
	\includegraphics[width=1\textwidth]{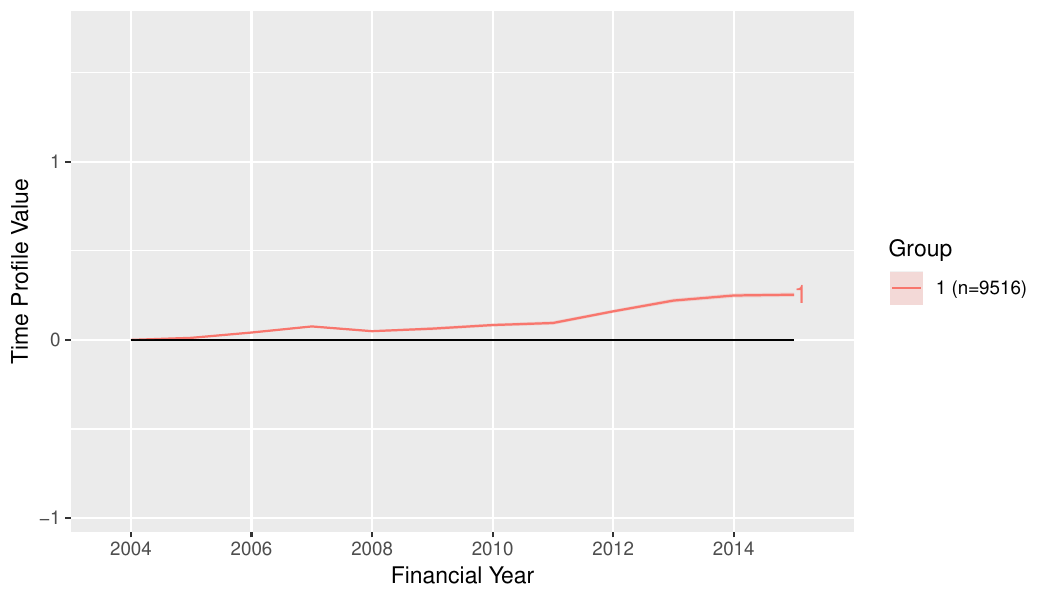}
	
\end{figure}

Figure \ref{fig:UNF-Geq1-Time-Profile-Estimates-with-CIs}
shows the time profile plot for $G=1$, with axes identical to the
corresponding time profile plot in the main results section of the
paper. Comparing the plots reveals that the point estimates
differ depending on the algorithm used, and the economic interpretations
of time effects around financial year 2008 are different. Using
the unmodified procedure suggests a small downward effect in 2008 followed
by a gradual rise, while this initial drop is absent in the corresponding
plot created using the modified algorithm in the paper.

\section{\label{sec:Simulation-Exercise}Simulation exercise}

This section uses simulation evidence to argue for the validity of
the GFE procedure in applications with data resembling the superannuation dataset. It also uses the proposed method for matching labels between estimations to
investigate how standard errors derived from the fixed-$T$ variance
estimate formula and the bootstrap compare to the simulated standard
errors estimated across simulations.

\subsection{\label{subsec:Methodology}Simulation methodology}

Creating the simulated datasets and 
a framework for analysing the results are now described.

\subsubsection{\label{subsec:Generating-Simulated-Data}Generating simulated data}

Consider the data generating process (DGP)
\begin{equation}
	\dot{y}_{it}^{\star}:=\dot{x}_{it}^{\star'}\widehat{\theta}+\widehat{\dot{\alpha}}_{g_{i}t}+\dot{v}_{it}^{\star},\label{eq:data-generating-process-model} 
\end{equation}
where:
\begin{itemize}
	\item $\dot{x}_{it}^{\star}=(\dot{{x}}_{1,it}^{\star},\dot{{x}}_{2,it}^{\star})'$
	is a column vector of simulated covariate values for unit $i$ at
	time $t$;
	\item $\widehat{\theta}$ and $\widehat{\dot{\alpha}}_{g_{i}t}$ are the GFE estimates
	for the covariate effects and time-demeaned group time profiles from
	the main results in the paper, respectively;
	\item the simulated covariates have mean zero and there is no time-constant
	individual-specific fixed effect -- that is, the generated data resembles the true data after time-demeaning;
	\item $\dot{x}_{k,it}^{\star}\sim N(0,\widehat{\sigma}_{\dot{x}_{k}}^{2})$,
	where $\widehat{\sigma}_{\dot{x}_{k}}^{2}$ is the sample variance
	of all values of the $\dot{{x}}_{k,it}$, for $k=1,2$;
	\item $\dot{v}_{it}^{\star}\sim N(0,\widehat{\sigma}_{\widehat{\dot{v}}}^{2})$,
	where $\widehat{\sigma}_{\widehat{\dot{v}}}^{2}$ is the sample
	variance of the empirical residuals $\widehat{\dot{v}}_{it}:=\dot{y}_{it}-\dot{x}_{it}'\widehat{\theta}-\widehat{\dot{\alpha}}_{g_{i}t}$; 
	\item $\dot{x}_{k,it}^{\star}$ are generated using a method that induces
	correlation between the time profile values $\widehat{\dot{\alpha}}_{g_{i}t}$
	and the covariates $\dot{x}_{k,it}^{\star}$, for $k=1,2$; this approximates
	the correlation observed in the data. The details of this are given below.
\end{itemize}
A simulation exercise using data simulated from (\ref{eq:data-generating-process-model})
with the $\dot{x}_{k,it}^{\star}$ uncorrelated with $\widehat{\dot{\alpha}}_{g_{i}t}$
would be unfaithful to the challenges involved
in estimating the model on the original data. Recall that the GFE
method allows for arbitrary correlation between the covariates and
the unobserved grouped fixed effects. Moreover, in the absence of
correlation between $\dot{x}_{k,it}^{\star}$ and $\widehat{\dot{\alpha}}_{g_{i}t}$,
a standard two-way fixed-effects regression of $\dot{y}$ on the $\dot{x}$-es directly obtains unbiased estimates $\widehat{\theta}$. Thus, recovering
accurate estimates using the simulated data is unrealistically easy if the
covariates are uncorrelated with the time profiles.

A realistic exercise simulates correlation between
the covariates and the time profiles to match that observed in the
data. Using the data to estimate the correlation
statistics $\widehat{\rho}_{k,g}$, for all $\left(k,g\right)$, allows a flexible correlation structure. The $\widehat{\rho}_{k,g}$ 
values are the correlations between values of $\dot{x}_{k,it}^{\star}$
and values of $\widehat{\dot{\alpha}}_{g_{i}t}$; i.e., $\widehat{\rho}_{k,g_{i}}$
is the correlation between the value of covariate $k$ and the value
contributed to the dependent variable by individual $i$'s group time
profile. The $\widehat{\rho}_{k,g}$ are estimated for all $\left(k,g\right)$ by:
\begin{enumerate}
	\item filtering the observed data to keep only records where $g_{i}=g$;
	\item computing the sample correlation statistic between the observed values
	of $\dot{x}_{k,it}^{\star}$ and corresponding estimated values of
	$\widehat{\dot{\alpha}}_{g_{i}t}$; call this value $\widehat{\rho}_{k,g}$.
\end{enumerate}
Having estimated the correlation statistics, the aim is to generate
Gaussian random variables $\dot{x}_{k,it}^{\star}$ which have correlation $\widehat{\rho}_{k,g_{i}}$
with the $\widehat{\dot{\alpha}}_{g_{i}t}$. 
The following procedure induces this correlation structure, treating the model estimates of $\widehat{\dot{\alpha}}_{gt}$ for $\left(g,t\right) \in \{1,2,...,G \} \times \{1,2,...,T \}$ as if they had been drawn from a Gaussian distribution. For all $\left(k, i,t \right) \in  \{1,2 \}  \times \{1,2,...,N \} \times \{1,2,...,T \} $:
\begin{enumerate}
	\item set $W_{1,kit}=\widehat{\dot{\alpha}}_{g_{i}t}/\widehat{\sigma}_{\alpha}$, where $\widehat{\sigma}_{\alpha}$ is estimated using the sample standard deviation of the set of $G\times T$ estimated values $\widehat{\dot{\alpha}}_{gt}$;
	\item draw $W_{2,kit}\sim N(0,1)$;
	\item set $W_{3,kit}=\widehat{\rho}_{k,g_{i}}W_{1,kit}+\sqrt{1-\widehat{\rho}_{k,g_{i}}^{2}}W_{2,kit}$;
	\item set $\dot{x}_{k,it}^{\star}=\widehat{\sigma}_{\dot{x}_{k}}W_{3,kit}$,
	where $\widehat{\sigma}_{\dot{x}_{k}}$ is the sample standard deviation of all observed values of covariate
	$\dot{x}_{k}$.
\end{enumerate}   

The error term $\dot{v}_{it}^{\star}\sim N(0,\widehat{\sigma}_{\widehat{\dot{v}}}^{2})$,
and (\ref{eq:data-generating-process-model}) gives the simulated values of the dependent variable $\dot{y}^{\star}$. The GFE procedure with $G=7$ is then run on a large number of simulated datasets. Comparing the GFE estimation results to the DGP values checks the validity of the GFE procedure applied to this setting and the code implementing the method.

\subsubsection{\label{subsec:Interpreting-Results}Framework for interpreting results}

The simulation exercise investigates the following properties of the GFE estimator:
\begin{enumerate}
	\item Whether the GFE procedure applied to a known DGP, constructed from
	the results of applying the GFE procedure to the superannuation dataset, estimates
	the DGP accurately in simulated datasets whose dimensions are the same as
	in the superannuation dataset.
	\item The closeness of the confidence intervals derived from the fixed-$T$ variance estimate formula to the `simulated' confidence intervals---the intervals obtained by matching time profile estimates across a large
	number of simulated datasets and observing the empirical spread of
	estimates.
	\item The closeness of the simulated confidence intervals to the bootstrap confidence intervals---the intervals obtained by matching time profile estimates across a large number
	of bootstrap samples drawn from the first simulated dataset and observing the empirical spread of the estimates.
\end{enumerate}
The following steps are used to generate the simulation results:
\begin{enumerate}
	\item $M=1000$ datasets are generated independently from the DGP, each with
	$N=9516$ units and covering $T=12$ time periods. The GFE procedure is then run on each of these using 1000 random starting values
	for each estimation, and the standard errors are obtained from the
	fixed-$T$ variance estimate formula.
	\item Using the resulting set of $M$ estimates of $\widehat{\theta}$ and
	$\widehat{\dot{\alpha}}:=\{ \widehat{\dot{\alpha}}_{gt} \}_{\left(g,t\right)\in \{1,2,...,G\} \times \{1,2,...,T\} }$ as samples $\{\widehat{\theta}^{(m)}\}_{m=1}^{M}$
	and $\{\widehat{\dot{\alpha}}^{(m)}\}_{m=1}^{M}$ from the sampling
	distributions of the estimators, simulated
	standard errors and 95\% element-wise confidence intervals are estimated.
	\item All bootstrap results are obtained by creating $B=1000$ bootstrap
	replicate datasets using the method outlined in the methodology section
	of the paper---except that here the first simulated dataset is treated
	as the source dataset for the bootstrap sampling. The GFE procedure with $G=7$ is then run on each of the resulting bootstrap replicate datasets. The bootstrap results that follow are different to the results obtained
	in the main results section of the paper, which use the observed
	data as the source dataset for bootstrap sampling.
	\item Estimated time profiles are compared to the
	DGP time profiles after shifting all time
	profiles to begin at 0.
\end{enumerate}

\subsection{\label{subsec:Results}Results}

\begin{figure}
	\caption{\label{fig:DGP-time-profiles-vs-first-simulated-dataset-estimates}DGP
		time profiles vs. first simulated dataset estimates. Red series represent DGP values, blue series are time
		profile estimates derived from the first simulated dataset. Time-demeaned
		group time profiles shifted to begin at 0 on the vertical axis.}
	
	\includegraphics[width=1\textwidth]{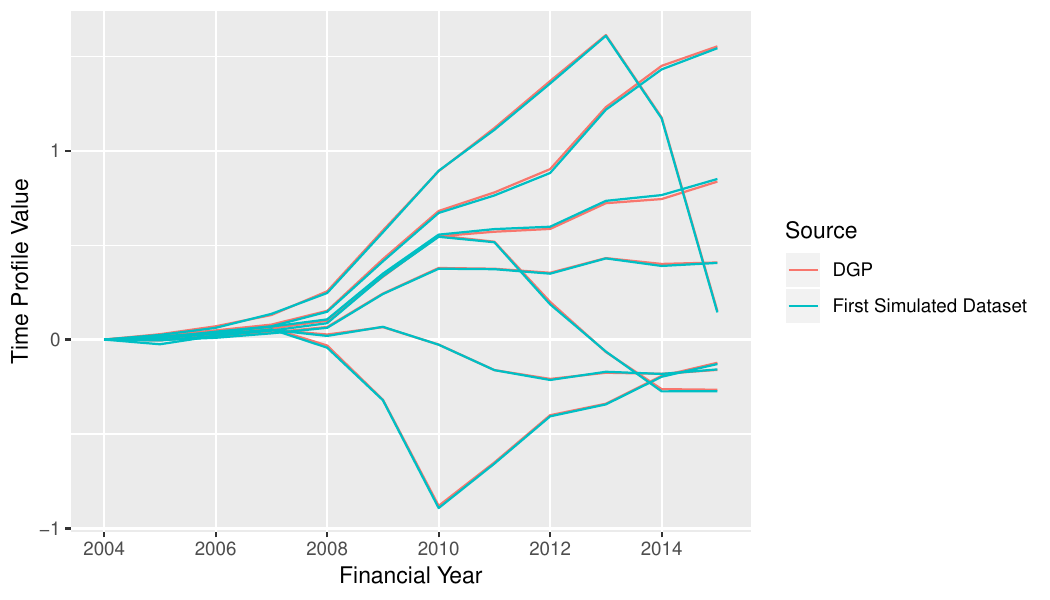}
	
\end{figure}

For the first property listed in Section \ref{subsec:Interpreting-Results},
the input values used for the DGP are compared to those estimated
using the GFE procedure on the first simulated dataset. Figure \ref{fig:DGP-time-profiles-vs-first-simulated-dataset-estimates}
shows this comparison for the time profiles; Table \ref{tab:DGP-covariate-effects-vs-first-simulated-dataset-estimates}
compares the covariate effects numerically. The results are close and suggest that the GFE procedure recovers the parameters
of the DGP with a high degree of accuracy.

\begin{table}
	\caption{\label{tab:DGP-covariate-effects-vs-first-simulated-dataset-estimates}DGP
		covariate effects vs. first simulated dataset estimates.}
	
	\begin{tabular}{|c|c|c|}
		\hline 
		& \textbf{Log Minimum Drawdown Rate} & \textbf{Log Account Balance}\tabularnewline
		\hline 
		DGP& 0.1436 & $-$0.1472\tabularnewline
		\hline 
		First simulated dataset & 0.1445 & $-$0.1490\tabularnewline
		\hline 
	\end{tabular}
\end{table}

\begin{figure}
	\caption{\label{fig:Time-profiles-with-true-95p-CIs}DGP time profiles and
		simulated 95\% CIs. Time-demeaned group time profile values are the inputs
		to the DGP and here shifted to begin at 0 on the vertical axis. Shaded
		regions denote 95\% element-wise confidence intervals computed from empirical percentiles
		of the shifted, time-demeaned group time profile estimates across
		1000 simulated datasets.}
	
	\includegraphics[width=1\textwidth]{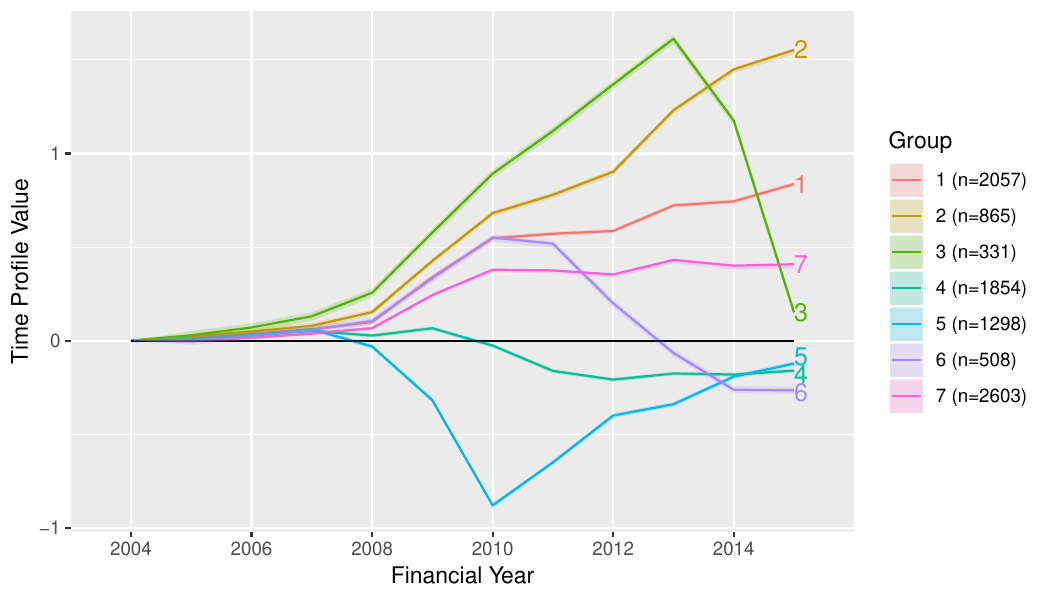}
	
\end{figure}

Figure \ref{fig:Time-profiles-with-true-95p-CIs} summarises the distribution of time profile
estimates across all 1000 simulated datasets.
The 95\% element-wise confidence interval bounds represent the empirical $2.5$
and $97.5$ percentiles of each estimated value. The tightness
of these confidence intervals around the DGP values suggests that in most simulated datasets, the time profile
estimates are numerically close to the true values, and economically
indistinguishable---meaning the interpretations of the time profiles are the same. The DGP time profile values for group 7
tend to be close to the upper bounds of the simulated empirical 95\%
intervals, and for the terminal time period, are slightly above the
upper bound.

Table \ref{tab:DGP-covariate-effects-and-true-95p-CIs} makes the
corresponding comparison for the covariate effects.
The simulated confidence intervals surround the DGP values; however,
for both covariates, the DGP estimates are relatively close to the
upper bounds of the intervals.

\begin{table}
	\caption{\label{tab:DGP-covariate-effects-and-true-95p-CIs}DGP covariate effects
		and simulated 95\% CIs.}

	\begin{tabular}{|c|c|c|}
		\hline 
		& \textbf{Log Minimum Drawdown Rate} & \textbf{Log Account Balance}\tabularnewline
		\hline 
		DGP value & 0.1436 & $-$0.1472\tabularnewline
		\hline 
		Simulated 95\% CIs & {[}0.1340, 0.1459{]} & {[}$-$0.1522, $-$0.1471{]}\tabularnewline
		\hline 
	\end{tabular}

	{\vspace{4pt} \small{} 
		Simulated 95\% CI bounds represent empirical $2.5$
		and $97.5$ percentiles of the estimated covariate effects across
		1000 simulated datasets.}{\small\par}
\end{table}

\begin{figure}
	\caption{\label{fig:Time-profile-estimates-from-first-simulated-dataset-CIs-from-formula}Time
		profile estimates from first simulated dataset, CIs from formula. Time-demeaned group time profile values from GFE estimation
		on first simulated dataset and shifted to begin at 0 on the vertical
		axis. Shaded regions denote 95\% element-wise confidence intervals derived from
		fixed-$T$ variance estimate formula.}
	
	\includegraphics[width=1\textwidth]{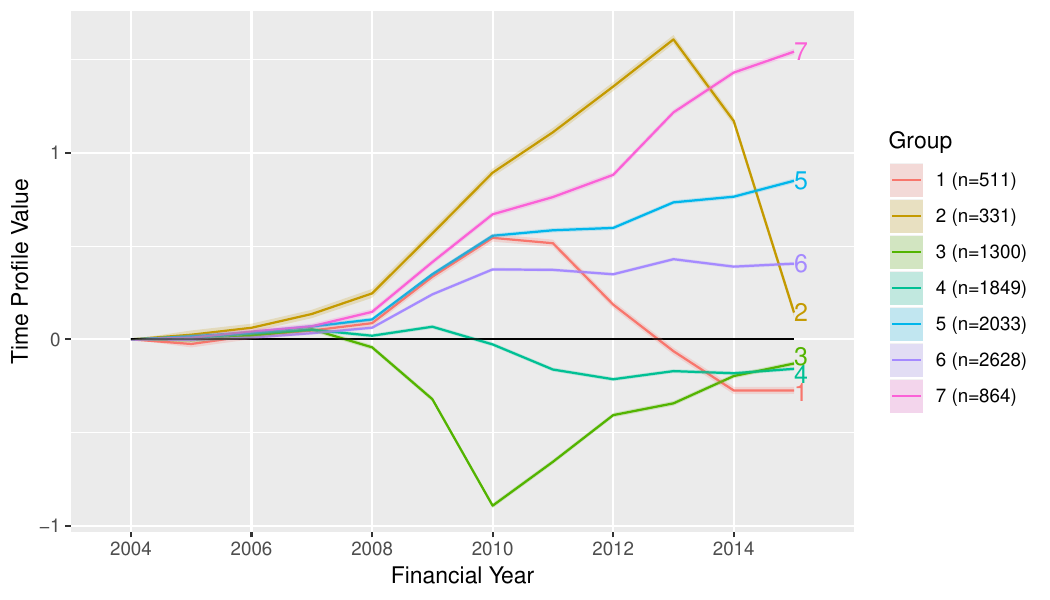}
	
\end{figure}

For the second property listed in Section \ref{subsec:Interpreting-Results}, Figure \ref{fig:Time-profiles-with-true-95p-CIs} is compared to Figure \ref{fig:Time-profile-estimates-from-first-simulated-dataset-CIs-from-formula}.
Figure \ref{fig:Time-profile-estimates-from-first-simulated-dataset-CIs-from-formula} shows time profile estimates and 95\% element-wise confidence intervals
constructed from standard errors derived from the fixed-$T$ variance
estimate formula, where the input data is the first simulated dataset.
The plots are nearly indistinguishable up to group relabelling, suggesting that the fixed-$T$ variance estimate formula applied
to the first simulated dataset estimates the true standard
errors with reasonable precision. Table \ref{tab:Covariate-effect-estimates-from-first-simulated-dataset-CIs-from-formula}
gives a similar comparison for the covariate effects, and has the same interpretation.

\begin{table}
	
	\caption{\label{tab:Covariate-effect-estimates-from-first-simulated-dataset-CIs-from-formula}Covariate
		effect estimates from first simulated dataset, CIs from formula.}
	\begin{tabular}{|c|c|c|}
		\hline 
		& \textbf{Log Minimum Drawdown Rate} & \textbf{Log Account Balance}\tabularnewline
		\hline 
		Covariate Effect Estimate & 0.1445 & $-$0.1490\tabularnewline
		\hline 
		95\% CI based on formula & {[}0.1415, 0.1475{]} & {[}$-$0.1515, $-$0.1465{]}\tabularnewline
		\hline 
	\end{tabular}
	
	{\vspace{4pt} \small{} 
		95\% CIs derived from fixed-$T$ variance estimate
		formula.}{\small\par}
\end{table}

As with the previous comparison, all 1000 simulated
datasets are considered next, by comparing the empirical distribution of estimated standard
errors---derived from the results of applying the fixed-$T$ variance formula to
1000 simulated datasets---to the simulated standard errors---computed
by calculating the sample standard deviation of parameter estimates across
the 1000 simulated datasets. 
Table \ref{tab:Lookup-Table-=002013Standard-Error-Distributions-for-Time-Profile-Estimates}
provides the figure references for the standard error distribution plots
by group of the
time-demeaned group time profile values, shifted to begin at zero
in the first time period, corresponding to the financial year ended
30 June 2004. For each group, the plots show the empirical distribution of
the standard errors for estimates corresponding to financial years
2005 to 2015, inclusive.
The group labels follow Figure \ref{fig:Time-profiles-with-true-95p-CIs},
which shows the group time profiles with element-wise confidence intervals derived from the simulated standard errors. These
are the same group labels presented in the figures for the main results in the paper.
In general, the simulated standard error value is in an area of nontrivial density in the corresponding empirical standard error distribution.

\begin{table}
	\caption{\label{tab:Lookup-Table-=002013Standard-Error-Distributions-for-Time-Profile-Estimates}Lookup
		table -- Standard error distributions for time profile estimates.
		Groups labelled as per Figure \ref{fig:Time-profiles-with-true-95p-CIs}.}
	\centering
	\begin{tabular}{|c|c|}
		\hline 
		Group Label & Figure\tabularnewline
		\hline 
		1 & \ref{fig:Group-1-Time-Profile-Analytical-SE-Distributions-across-Simulated-Datasets}\tabularnewline
		\hline 
		2 & \ref{fig:Group-2-Time-Profile-Analytical-SE-Distributions-across-Simulated-Datasets}\tabularnewline
		\hline 
		3 & \ref{fig:Group-3-Time-Profile-Analytical-SE-Distributions-across-Simulated-Datasets}\tabularnewline
		\hline 
		4 & \ref{fig:Group-4-Time-Profile-Analytical-SE-Distributions-across-Simulated-Datasets}\tabularnewline
		\hline 
		5 & \ref{fig:Group-5-Time-Profile-Analytical-SE-Distributions-across-Simulated-Datasets}\tabularnewline
		\hline 
		6 & \ref{fig:Group-6-Time-Profile-Analytical-SE-Distributions-across-Simulated-Datasets}\tabularnewline
		\hline 
		7 & \ref{fig:Group-7-Time-Profile-Analytical-SE-Distributions-across-Simulated-Datasets}\tabularnewline
		\hline 
	\end{tabular}
\end{table}

\begin{figure}
	\caption{\label{fig:Covariate-Effect-Analytical-SE-Distributions-across-Simulated-Datasets}Covariate
		effect analytical SE distributions across simulated datasets. Black lines plot the kernel density estimate for standard
		errors derived from the fixed-$T$ variance estimate formula after
		estimating the GFE model on 1000 simulated datasets. Red vertical
		lines represent the value of the simulated standard error.}
	
	\includegraphics[width=1\textwidth]{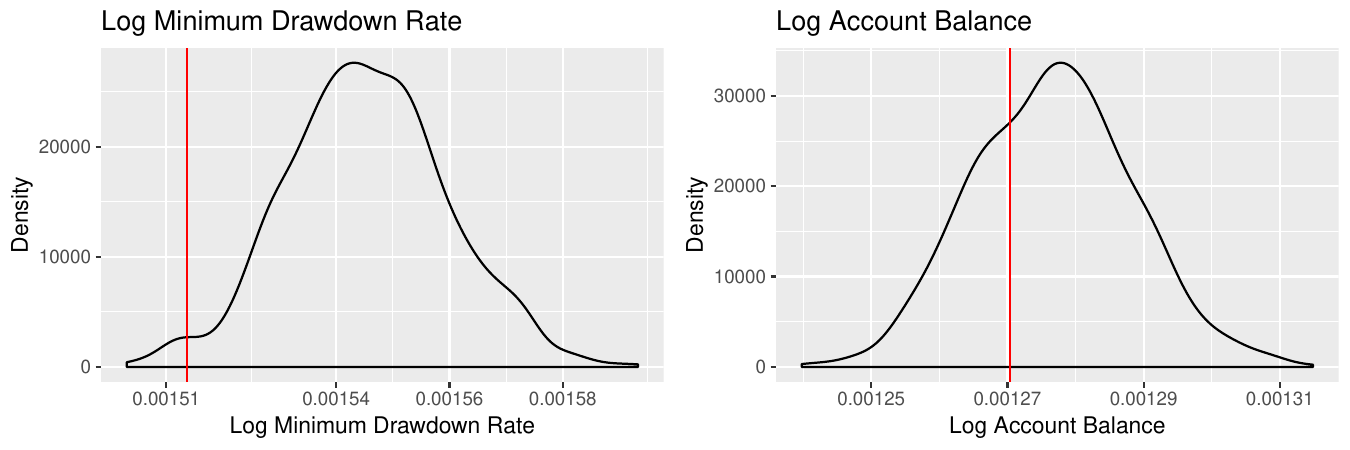}
	
\end{figure}

Figure \ref{fig:Covariate-Effect-Analytical-SE-Distributions-across-Simulated-Datasets} provides the corresponding plots for standard errors of the covariate
estimates.
The fixed-$T$ variance estimate formula tends to overestimate the true standard error for the first covariate; for the second covariate,
the true standard error is more centrally located in the distribution of empirical standard errors.

\begin{figure}
	\caption{\label{fig:Time-profile-estimates-from-first-simulated-dataset-CIs-from-bootstrap}Time
		profile estimates from first simulated dataset, CIs from bootstrap. Results from point estimates aggregated over 1000 bootstrap replications
		using the first simulated dataset to generate bootstrap replicate
		datasets. Time-demeaned group time profile values from GFE estimation
		on first simulated dataset and shifted to begin at 0 on the vertical
		axis. Shaded regions denote 95\% element-wise confidence intervals computed from
		empirical percentiles across 1000 bootstrap replications.}
	
	\includegraphics[width=1\textwidth]{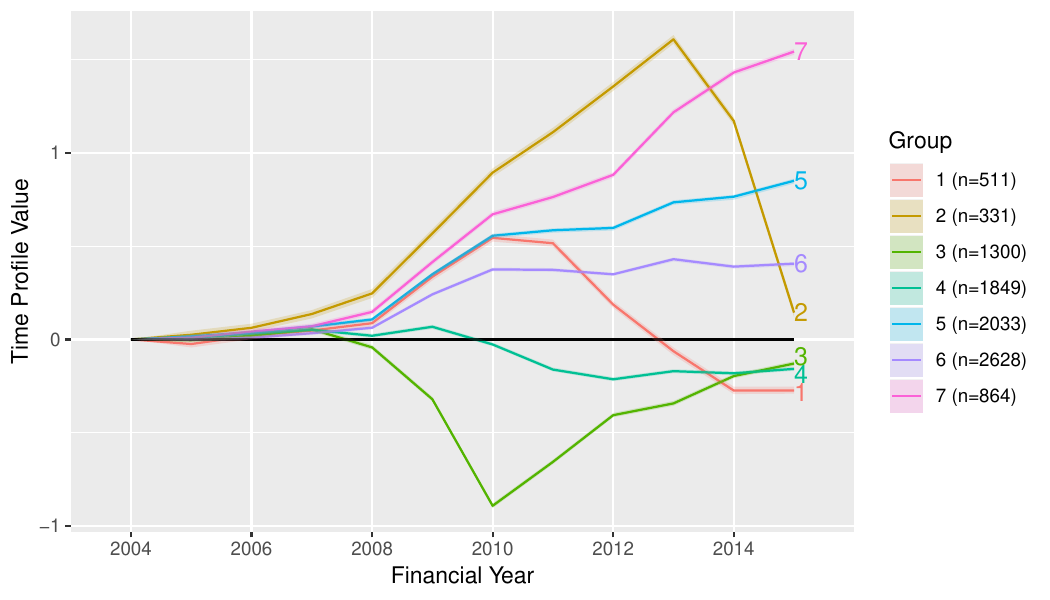}

\end{figure}

Figures \ref{fig:Time-profiles-with-true-95p-CIs} and \ref{fig:Time-profile-estimates-from-first-simulated-dataset-CIs-from-bootstrap} are compared for the third property listed in Section \ref{subsec:Interpreting-Results}.
Figure \ref{fig:Time-profile-estimates-from-first-simulated-dataset-CIs-from-bootstrap} gives time profile estimates and 95\% element-wise confidence intervals
constructed from the empirical distribution of estimates across 1000
bootstrap replications, where the input data is the first simulated
dataset.
The results are again nearly indistinguishable up to group relabelling. This suggests that inference conducted using the bootstrap
procedure gives almost identical results to that using the fixed-$T$
variance estimate formula, which approximates the true standard
errors well. Ideally, the bootstrap procedure would be performed on
all 1000 simulated datasets, and the resulting distributions of bootstrap
standard errors compared to the simulated standard errors, as done for the fixed-$T$ variance estimate formula. However, computational constraints prevent this.

Considering the simulation evidence, the fixed-$T$ variance estimate formula performs well in datasets
simulated from the generating process implied by the GFE estimation
results for the superannuation dataset. This suggests that if the GFE
assumptions are satisfied, the analytical formula may be adequate in datasets of similar size to the superannuation dataset, not requiring
the bootstrap procedure for inference on the parameter
estimates. This is important as using our code to perform
the bootstrap on the superannuation dataset is currently too computationally intensive to
run on a standard machine; it requires access to a high-performance
computing cluster. As for the bootstrap results on the simulated data,
these suggest that the bootstrap may also perform comparably well;
however, due to computational constraints, our implementation is unable to test this as rigorously as for the analytical
formula.

\section{Superannuation drawdowns dataset}

\subsection{More on covariate selection}

Possible covariates for the analysis included in the dataset are limited
to age, account balance and gender. From the available information,
two derived covariates are also constructed: the minimum drawdown
rate, and a crude estimate of an individual's risk appetite over the
observation period. For each person--year
observation, the age of the member maps to the relevant minimum drawdown rate the retiree
is constrained by, with concessional reductions in
the rates applying to the 2009--2013 financial years.

As the risk appetite variable is only used descriptively,
and does not enter into the model estimation, its preciseness
does not affect the main results. Its construction involves observing movements in account balances and comparing
these with the amounts drawn down and contributed to the funds by retirees.
From this, ignoring administrative fees on the accounts, one can roughly
estimate the return on assets. As the source data is at a monthly
frequency, this return is computed monthly and then annualised; comparing
it to the S\&P/ASX 200 market index over matching time periods gives
an approximate measure of sensitivity to market returns. Taking the
magnitude of the average of these sensitivities then serves as a proxy
for risk appetite. 

Applying the within transformation---centering all variables around their time averages for each individual---prevents estimating the effect of any time-invariant covariates,
which do not show within-individual variation; this includes gender
as well as the constructed risk appetite variable. Thus, gender and risk appetite do not enter in the GFE estimation, although they are used to qualitatively
characterise the groups that the GFE method finds in the data.

Furthermore, age is not included as a covariate because the focus is
on estimating group effects for each time period; these time effects cannot easily
be separated from the effect of ageing after within-transforming the data.

\subsection{\label{subsec:Exploratory-Data-Analysis}Exploratory data analysis}

The remainder of this section presents a preliminary descriptive analysis
of the dataset used to generate the main results.

\subsubsection{\label{subsec:Summary-Statistics}Summary statistics}

Table \ref{tab:agg-time-invariant-variables}
summarises characteristics that vary across individuals but not time. 
The age at 31 December 2015 represents the individual's cohort, equivalent
to measuring a year-of-birth variable. The median retiree in the sample
was born in 1936, with more than 50\% of the sample born in an interval
capturing four years on either side.

\begin{table}
	\caption{\label{tab:agg-time-invariant-variables}Summary statistics -- Time-invariant
		variables.}
	
	\resizebox{\textwidth}{!}{
		
		\begin{tabular}{|c|c|c|c|c|}
			\hline 
			& Age at 31 December 2015 & Age at Account Open & Sex: Male & Risk Appetite\tabularnewline
			\hline 
			Mean & 79.4 & 63.57 & 0.56 & 0.41\tabularnewline
			\hline 
			SD & 5.22 & 4.17 & 0.5 & 0.21\tabularnewline
			\hline 
			Median & 79.78 & 64.23 & 1 & 0.46\tabularnewline
			\hline 
			Q1 & 76.37 & 60.9 & 0 & 0.25\tabularnewline
			\hline 
			Q3 & 83.03 & 65.39 & 1 & 0.54\tabularnewline
			\hline 
			Min & 60.66 & 48.48 & 0 & 0\tabularnewline
			\hline 
			Max & 101.46 & 85.44 & 1 & 1.88\tabularnewline
			\hline 
			Count & 9516 & 9516 & 9516 & 9507\tabularnewline
			\hline 
		\end{tabular}
		
	}
\end{table}

The age at account opening is the age when the retiree
initiates a phased withdrawal product and begins drawing down from the
account. In the superannuation dataset, the median retiree was aged 64 when opening
their account. In general, opening an account before age 65 requires
an individual to cease employment. 

The sex indicator variable equals 1 if the the retiree
is male, and 0 otherwise. The mean value of 0.56 represents the
proportion of retirees in the sample that are male.

The risk appetite variable is a proxy for the returns in the account relative
to the reference S\&P/ASX 200 index. The median retiree earned approximately
46\% of the index returns in the sample period, with 75\% of the
sample earning less than 55\% of the index returns. This variable
suggests that most retirees have asset mixes that are conservative
or balanced, with few retirees seeking aggressive returns in these
accounts.

Table \ref{tab:agg-time-varying-variables} summarises the time-varying variables in the dataset. 
The regular drawdown rate is the dependent variable of interest. The
median drawdown rate in the sample is 9\% of the account balances
annually, while the mean is 12\%. In absolute terms, the median regular
drawdown amount is \$4800 while the average is \$6436.

\begin{table}
	\caption{\label{tab:agg-time-varying-variables}Summary statistics -- Time-varying
		variables.}
	
	\resizebox{\textwidth}{!}{
		
		\begin{tabular}{|c|c|c|c|c|c|c|}
			\hline 
			& Regular Drawdown & Regular Drawdown & Ad-hoc Drawdown & Ad-hoc Drawdown & Ad-hoc Drawdown & Account Balance\tabularnewline & Rate & Amount & Indicator & Rate & Amount &  \tabularnewline
			\hline 
			Mean & 0.12 & 6435.91 & 0.07 & 0.15 & 10,216.56 & 72,686.55\tabularnewline
			\hline 
			SD & 0.12 & 6121.76 & 0.25 & 0.21 & 24,672.68 & 78,721.39\tabularnewline
			\hline 
			Median & 0.09 & 4800 & 0 & 0.07 & 4655.9 & 52,063\tabularnewline
			\hline 
			Q1 & 0.07 & 2992 & 0 & 0.02 & 1132.33 & 30,532.5\tabularnewline
			\hline 
			Q3 & 0.12 & 7728 & 0 & 0.18 & 10,000 & 87,427\tabularnewline
			\hline 
			Min & 0 & 1 & 0 & 0 & 1 & 1\tabularnewline
			\hline 
			Max & 2 & 166,695 & 1 & 0.9 & 600,000 & 2,427,083\tabularnewline
			\hline 
			Count & 107,935 & 107,975 & 108,717 & 7450 & 7454 & 108,635\tabularnewline
			\hline 
		\end{tabular}
		
	}
\end{table}

The ad-hoc drawdown indicator variable equals 1 if the retiree
made an ad-hoc withdrawal from their account balance during a given
financial year; its mean value of 0.07 indicates that 7\% of the observations
recorded contain an ad-hoc drawdown. An interpretation is
that the average retiree in the sample makes an ad-hoc drawdown roughly
once every 14 years. Conditional on making an ad-hoc drawdown, the
median ad-hoc drawdown rate is 7\% of the account balance at the start of
the year, while the mean is 15\%. In dollars, the median ad-hoc
drawdown is \$4656 and the average is \$10,217.

Median account balances, as measured at the beginning of each financial
year, are \$52,063, with roughly 50\% of balances lying in the interval
(\$30,000, \$87,000).

\subsubsection{\label{subsec:Visualisations}Histograms}

\begin{figure}
	\caption{\label{fig:agg-histograms-of}Histograms -- Time-invariant variables. These are graphical representations of the data summarised in Table \ref{tab:agg-time-invariant-variables}.}
	
	\includegraphics[width=1\textwidth]{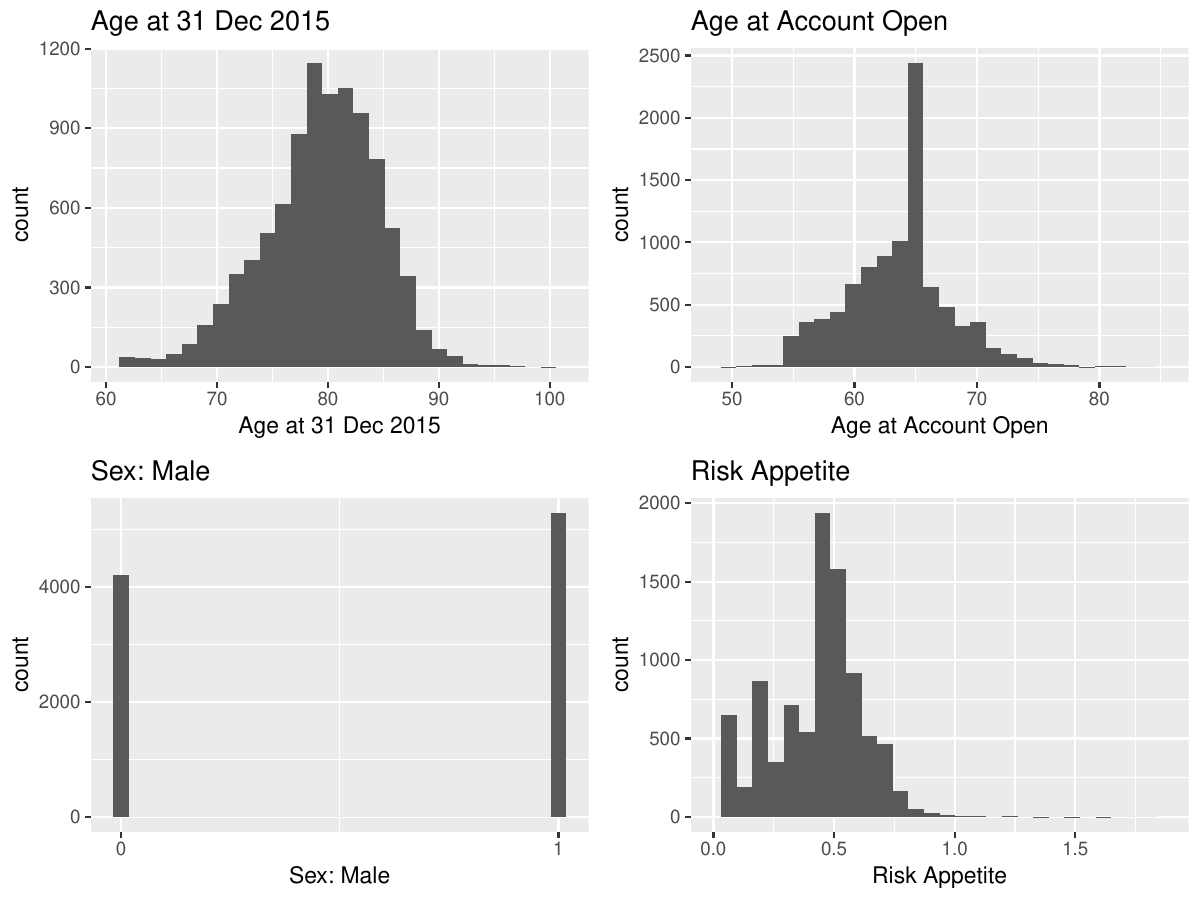}
\end{figure}

Figure \ref{fig:agg-histograms-of} plots the histograms
of the time-invariant covariates.
Most notable is the spike around age 65 for the account open age distribution, corresponding to the age at which individuals can
open a phased withdrawal account unconditionally.
Also instructive are the multiple peaks in the risk appetite distribution,
suggesting a bunching of retirees into distinct asset mix options.

\begin{figure}
	\caption{\label{fig:agg-histograms-of-time-varying-non-dummy-variables}Histograms
		-- Time-varying variables. These are graphical representations of the data summarised in Table \ref{tab:agg-time-varying-variables}.}
	
	\includegraphics[width=1\textwidth]{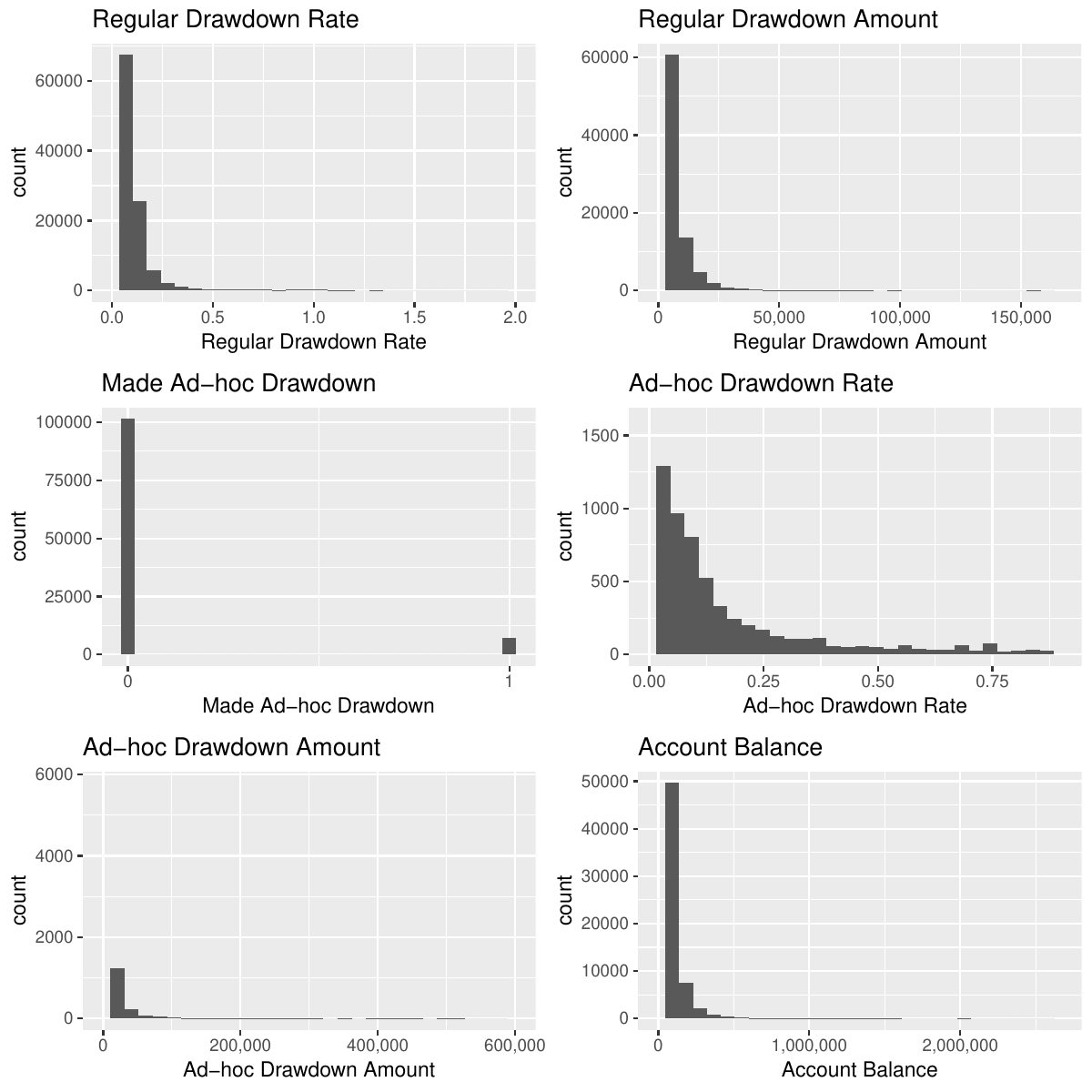}
\end{figure}

Figure \ref{fig:agg-histograms-of-time-varying-non-dummy-variables}
plots the histograms for the time-varying variables, which show some evidence of ad-hoc
drawdown rates bunching around several modes for the larger values.

\section{Characterising groups in the seven-group model}

The section provides summary statistics and histograms created for
each of the estimated groups from the main results. 
This allows comparison of each of the groups' characteristics
against other groups, using the following set of tables and plots.
Alternatively,
it is possible to compare group-level characteristics against the aggregate
sample, by comparing with the results in Section \ref{subsec:Exploratory-Data-Analysis} of this Supplement.
We also present panel plots to supplement those given in
the paper. Throughout the section, group labels follow the
main results section in the paper.

\subsection{Summary statistics by group}

\subsubsection{Time-invariant variables}

Table \ref{tab:Lookup-Table-=002013Summary-Statistics-for-Time-Invariant-Variables-by-Group}
lists table references for the summary statistics of the time-invariant
variables by group.

\begin{table}
	\caption{\label{tab:Lookup-Table-=002013Summary-Statistics-for-Time-Invariant-Variables-by-Group}Lookup
		table -- Summary statistics for time-invariant variables by group. Group labels follow the main results section in the paper.}
	\centering
	\begin{tabular}{|c|c|}
		\hline 
		Group & Table\tabularnewline
		\hline 
		1 & \ref{tab:g1-summ-stats-TI-vars}\tabularnewline
		\hline 
		2 & \ref{tab:g2-summ-stats-TI-vars}\tabularnewline
		\hline 
		3 & \ref{tab:g3-summ-stats-TI-vars}\tabularnewline
		\hline 
		4 & \ref{tab:g4-summ-stats-TI-vars}\tabularnewline
		\hline 
		5 & \ref{tab:g5-summ-stats-TI-vars}\tabularnewline
		\hline 
		6 & \ref{tab:g6-summ-stats-TI-vars}\tabularnewline
		\hline 
		7 & \ref{tab:g7-summ-stats-TI-vars}\tabularnewline
		\hline 
	\end{tabular}
\end{table}

\subsubsection{Time-varying variables}

Table \ref{tab:Lookup-Table-=002013Summary-Statistics-for-Time-Varying-Variables-by-Group}
lists table references for the summary statistics of the time-varying
variables by group.

\begin{table}
	\caption{\label{tab:Lookup-Table-=002013Summary-Statistics-for-Time-Varying-Variables-by-Group}Lookup
		table -- Summary statistics for time-varying variables by group. Group labels follow the main results section in the paper.}
	\centering
	\begin{tabular}{|c|c|}
		\hline 
		Group & Table\tabularnewline
		\hline 
		1 & \ref{tab:g1-summ-stats-TV-vars}\tabularnewline
		\hline 
		2 & \ref{tab:g2-summ-stats-TV-vars}\tabularnewline
		\hline 
		3 & \ref{tab:g3-summ-stats-TV-vars}\tabularnewline
		\hline 
		4 & \ref{tab:g4-summ-stats-TV-vars}\tabularnewline
		\hline 
		5 & \ref{tab:g5-summ-stats-TV-vars}\tabularnewline
		\hline 
		6 & \ref{tab:g6-summ-stats-TV-vars}\tabularnewline
		\hline 
		7 & \ref{tab:g7-summ-stats-TV-vars}\tabularnewline
		\hline 
	\end{tabular}
\end{table}

\subsection{Histograms by group}

\subsubsection{Time-invariant variables}

Table \ref{tab:Lookup-Table-=002013Histograms-of-Time-Invariant-Variables-by-Group}
lists figure references for the histograms of the time-invariant variables
by group.

\begin{table}
	\caption{\label{tab:Lookup-Table-=002013Histograms-of-Time-Invariant-Variables-by-Group}Lookup
		table -- Histograms of time-invariant variables by group. Group labels follow the main results section in the paper.}
	\centering
	\begin{tabular}{|c|c|}
		\hline 
		Group & Figure\tabularnewline
		\hline 
		1 & \ref{fig:g1-hist-TI-vars}\tabularnewline
		\hline 
		2 & \ref{fig:g2-hist-TI-vars}\tabularnewline
		\hline 
		3 & \ref{fig:g3-hist-TI-vars}\tabularnewline
		\hline 
		4 & \ref{fig:g4-hist-TI-vars}\tabularnewline
		\hline 
		5 & \ref{fig:g5-hist-TI-vars}\tabularnewline
		\hline 
		6 & \ref{fig:g6-hist-TI-vars}\tabularnewline
		\hline 
		7 & \ref{fig:g7-hist-TI-vars}\tabularnewline
		\hline 
	\end{tabular}
\end{table}

\subsubsection{Time-varying variables}

Table \ref{tab:Lookup-Table-=002013Histograms-of-Time-Varying-Variables-by-Group}
lists figure references for the histograms of the time-varying variables
by group.

\begin{table}
	\caption{\label{tab:Lookup-Table-=002013Histograms-of-Time-Varying-Variables-by-Group}Lookup
		table -- Histograms of time-varying variables by group. Group labels follow the main results section in the paper.}
\centering	
	\begin{tabular}{|c|c|}
		\hline 
		Group & Figure\tabularnewline
		\hline 
		1 & \ref{fig:g1-hist-TV-vars}\tabularnewline
		\hline 
		2 & \ref{fig:g2-hist-TV-vars}\tabularnewline
		\hline 
		3 & \ref{fig:g3-hist-TV-vars}\tabularnewline
		\hline 
		4 & \ref{fig:g4-hist-TV-vars}\tabularnewline
		\hline 
		5 & \ref{fig:g5-hist-TV-vars}\tabularnewline
		\hline 
		6 & \ref{fig:g6-hist-TV-vars}\tabularnewline
		\hline 
		7 & \ref{fig:g7-hist-TV-vars}\tabularnewline
		\hline 
	\end{tabular}
\end{table}

\subsection{Time-demeaned panel plots by group}

Table \ref{tab:Lookup-Table-=002013Panel-Plots-by-Group} lists figure
references for time-demeaned panel plots by group. Each figure shows
four variables after time-demeaning by unit: the log regular drawdown
rate; the log regular drawdown dollar amount; the log account balance
at the financial year start; composite residuals from the estimation.
We define the composite residuals as $\widehat{\dot{\alpha}}_{g_{i}t}+\widehat{\dot{v}}_{it}:=\dot{y}_{it}-\dot{x}_{it}'\widehat{\theta}$; a composite residual is the estimated group time profile value
plus the model residual, obtained by subtracting the estimated
effect of covariates from the dependent variable.
The black line represents the estimated time-demeaned
group time profile values $\widehat{\dot{\alpha}}_{gt}$. The composite residual plots provide information similar to that of a residual plot of $\widehat{\dot{v}}_{it}$, but also convey a sense of how large the estimated idiosyncratic unobserved component $\widehat{\dot{v}}_{it}$ is compared to the estimated systematic unobserved component $\widehat{\dot{\alpha}}_{g_{i}t}$.

\begin{table}
	\caption{\label{tab:Lookup-Table-=002013Panel-Plots-by-Group}Lookup table
		-- Time-demeaned (TD) panel plots by group. Group labels follow the main results section in the paper.}
	\centering
	\begin{tabular}{|c|c|}
		\hline 
		Group & Figure\tabularnewline
		\hline 
		1 & \ref{fig:g1-sixplots}\tabularnewline
		\hline 
		2 & \ref{fig:g2-sixplots}\tabularnewline
		\hline 
		3 & \ref{fig:g3-sixplots}\tabularnewline
		\hline 
		4 & \ref{fig:g4-sixplots}\tabularnewline
		\hline 
		5 & \ref{fig:g5-sixplots}\tabularnewline
		\hline 
		6 & \ref{fig:g6-sixplots}\tabularnewline
		\hline 
		7 & \ref{fig:g7-sixplots}\tabularnewline
		\hline 
	\end{tabular}
\end{table}

\section{Panel plots for the two-group model}

\begin{figure}
	\caption{\label{fig:Geq2-model-=002013group-1-panel-plots}$G=2$ model --
		Group 1 time-demeaned (TD) panel plots. Account balances as at financial year start. The black
		series in the bottom-right panel represents estimated time-demeaned
		group time profile values.}
	
	\includegraphics[width=1\textwidth]{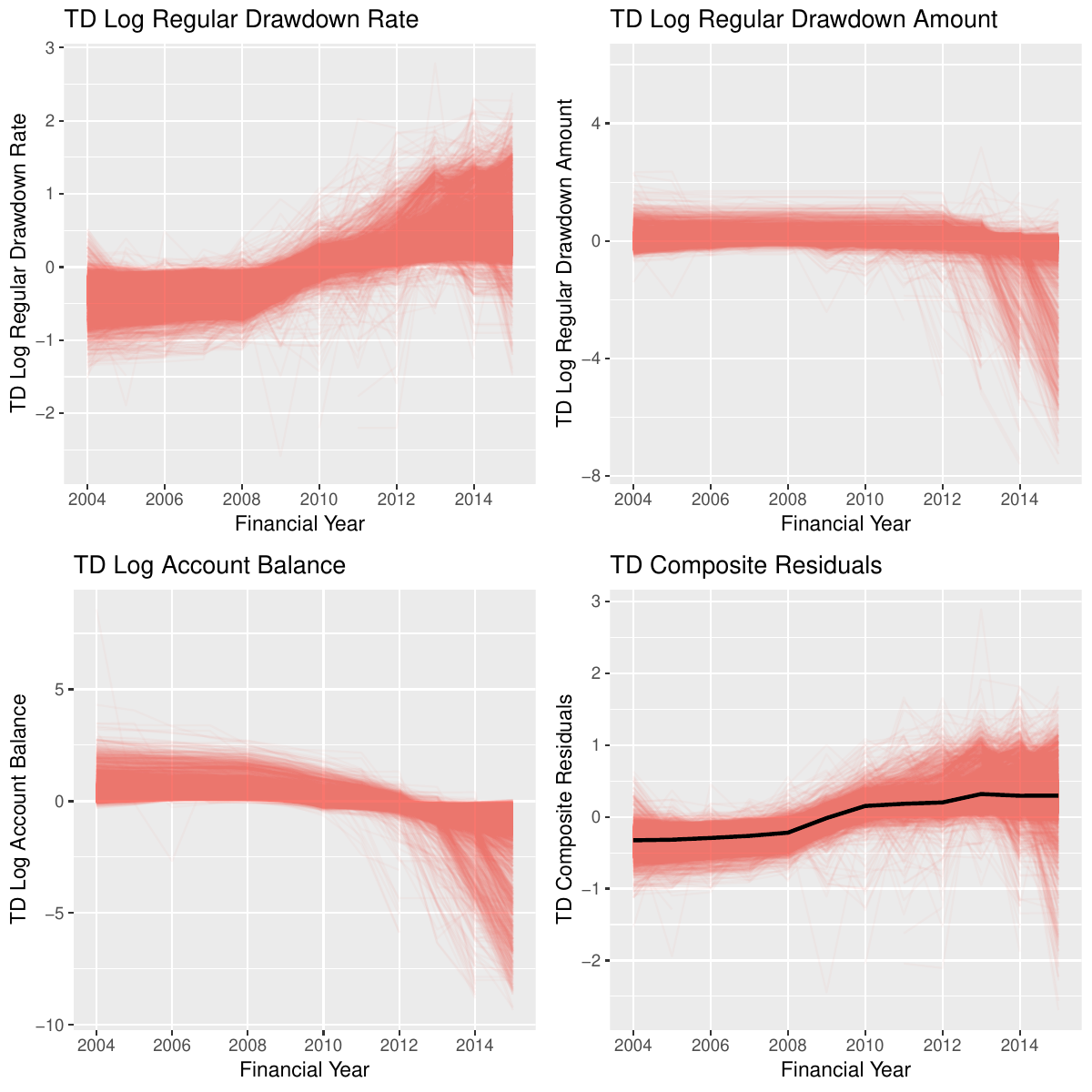}
	
\end{figure}

\begin{figure}
	\caption{\label{fig:Geq2-model-=002013group-2-panel-plots}$G=2$ model --
		Group 2 time-demeaned (TD) panel plots. Account balances as at financial year start. The black
		series in the bottom-right panel represents estimated time-demeaned
		group time profile values.}
	
	\includegraphics[width=1\textwidth]{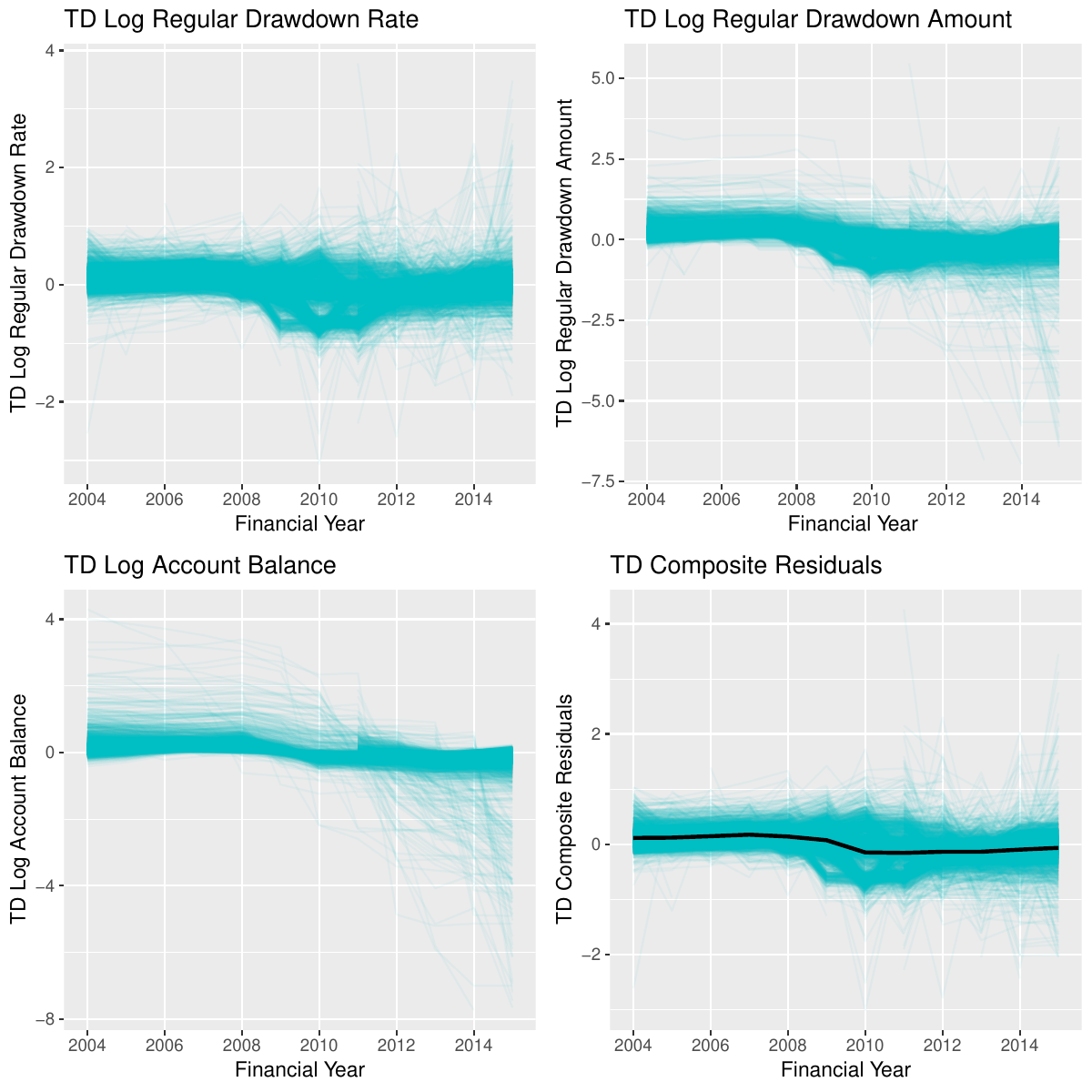}
	
\end{figure}

Figures \ref{fig:Geq2-model-=002013group-1-panel-plots} and \ref{fig:Geq2-model-=002013group-2-panel-plots}
show panel plots of time-demeaned variables for the two groups in the two-group model.
Group labels in these plots follow the two-group model results in the
paper.

\begin{table}
	\caption{\label{tab:g1-summ-stats-TI-vars}Group 1 summary statistics -- Time-invariant
		variables.}
	
	\resizebox{\textwidth}{!}{
		
		\begin{tabular}{|c|c|c|c|c|}
			\hline 
			& Age at 31 Dec 2015 & Age at Account Open & Sex: Male & Risk Appetite\tabularnewline
			\hline 
			Mean & 81.53 & 64.67 & 0.62 & 0.43\tabularnewline
			\hline 
			SD & 4.22 & 3.84 & 0.49 & 0.18\tabularnewline
			\hline 
			Median & 81.76 & 65 & 1 & 0.46\tabularnewline
			\hline 
			Q1 & 78.87 & 62.69 & 0 & 0.34\tabularnewline
			\hline 
			Q3 & 84.57 & 66.67 & 1 & 0.54\tabularnewline
			\hline 
			Min & 67.67 & 50.45 & 0 & 0\tabularnewline
			\hline 
			Max & 94.7 & 80.22 & 1 & 0.92\tabularnewline
			\hline 
			Count & 2057 & 2057 & 2057 & 2055\tabularnewline
			\hline 
		\end{tabular}
		
	}
\end{table}

\begin{table}
	\caption{\label{tab:g2-summ-stats-TI-vars}Group 2 summary statistics -- Time-invariant
		variables.}
	
	\resizebox{\textwidth}{!}{
		
		\begin{tabular}{|c|c|c|c|c|}
			\hline 
			& Age at 31 December 2015 & Age at Account Open & Sex: Male & Risk Appetite\tabularnewline
			\hline 
			Mean & 80.03 & 63.36 & 0.62 & 0.41\tabularnewline
			\hline 
			SD & 5.23 & 5 & 0.48 & 0.2\tabularnewline
			\hline 
			Median & 79.41 & 63.67 & 1 & 0.46\tabularnewline
			\hline 
			Q1 & 76.42 & 59.82 & 0 & 0.24\tabularnewline
			\hline 
			Q3 & 83.54 & 65.57 & 1 & 0.54\tabularnewline
			\hline 
			Min & 67.53 & 50.26 & 0 & 0.01\tabularnewline
			\hline 
			Max & 96.23 & 81.78 & 1 & 1.02\tabularnewline
			\hline 
			Count & 865 & 865 & 865 & 863\tabularnewline
			\hline 
		\end{tabular}
		
	}
\end{table}

\begin{table}
	\caption{\label{tab:g3-summ-stats-TI-vars}Group 3 summary statistics -- Time-invariant
		variables.}
	
	\resizebox{\textwidth}{!}{
		
		\begin{tabular}{|c|c|c|c|c|}
			\hline 
			& age\_at\_31DEC15 & age\_at\_account\_open & sex\_male & risk\_appetite\tabularnewline
			\hline 
			Mean & 81.75 & 64.66 & 0.69 & 0.35\tabularnewline
			\hline 
			SD & 4.43 & 4.29 & 0.46 & 0.22\tabularnewline
			\hline 
			Median & 81.39 & 64.94 & 1 & 0.4\tabularnewline
			\hline 
			Q1 & 79.24 & 61.93 & 0 & 0.17\tabularnewline
			\hline 
			Q3 & 83.88 & 65.74 & 1 & 0.5\tabularnewline
			\hline 
			Min & 65.82 & 55.92 & 0 & 0\tabularnewline
			\hline 
			Max & 99.87 & 85.44 & 1 & 1.52\tabularnewline
			\hline 
			Count & 331 & 331 & 331 & 331\tabularnewline
			\hline 
		\end{tabular}
		
	}
\end{table}

\begin{table}
	\caption{\label{tab:g4-summ-stats-TI-vars}Group 4  summary statistics -- Time-invariant
		variables.}
	
	\resizebox{\textwidth}{!}{
		
		\begin{tabular}{|c|c|c|c|c|}
			\hline 
			& Age at 31 December 2015 & Age at Account Open & Sex: Male & Risk Appetite\tabularnewline
			\hline 
			Mean & 78.17 & 63.54 & 0.51 & 0.44\tabularnewline
			\hline 
			SD & 6.09 & 4.19 & 0.5 & 0.21\tabularnewline
			\hline 
			Median & 78.93 & 64.1 & 1 & 0.47\tabularnewline
			\hline 
			Q1 & 74.25 & 60.79 & 0 & 0.31\tabularnewline
			\hline 
			Q3 & 82.57 & 65.46 & 1 & 0.57\tabularnewline
			\hline 
			Min & 60.66 & 51.3 & 0 & 0\tabularnewline
			\hline 
			Max & 97.06 & 79.79 & 1 & 1.59\tabularnewline
			\hline 
			Count & 1854 & 1854 & 1854 & 1852\tabularnewline
			\hline 
		\end{tabular}
		
	}
\end{table}

\begin{table}
	\caption{\label{tab:g5-summ-stats-TI-vars}Group 5  summary statistics -- Time-invariant
		variables.}
	
	\resizebox{\textwidth}{!}{
		
		\begin{tabular}{|c|c|c|c|c|}
			\hline 
			& Age at 31 December 2015 & Age at Account Open & Sex: Male & Risk Appetite\tabularnewline
			\hline 
			Mean & 78.12 & 63.92 & 0.49 & 0.47\tabularnewline
			\hline 
			SD & 6.17 & 4.25 & 0.5 & 0.21\tabularnewline
			\hline 
			Median & 78.8 & 64.35 & 0 & 0.49\tabularnewline
			\hline 
			Q1 & 73.35 & 61.48 & 0 & 0.35\tabularnewline
			\hline 
			Q3 & 82.77 & 65.93 & 1 & 0.6\tabularnewline
			\hline 
			Min & 60.81 & 51.58 & 0 & 0\tabularnewline
			\hline 
			Max & 101.46 & 81.91 & 1 & 1.88\tabularnewline
			\hline 
			Count & 1298 & 1298 & 1298 & 1298\tabularnewline
			\hline 
		\end{tabular}
		
	}
\end{table}

\begin{table}
	\caption{\label{tab:g6-summ-stats-TI-vars}Group 6  summary statistics -- Time-invariant
		variables.}
	
	\resizebox{\textwidth}{!}{
		
		\begin{tabular}{|c|c|c|c|c|}
			\hline 
			& Age at 31 December 2015 & Age at Account Open & Sex: Male & Risk Appetite\tabularnewline
			\hline 
			Mean & 78.19 & 63.16 & 0.53 & 0.45\tabularnewline
			\hline 
			SD & 5.7 & 4.16 & 0.5 & 0.21\tabularnewline
			\hline 
			Median & 78.99 & 63.78 & 1 & 0.48\tabularnewline
			\hline 
			Q1 & 74.56 & 60.5 & 0 & 0.34\tabularnewline
			\hline 
			Q3 & 82.34 & 65.29 & 1 & 0.58\tabularnewline
			\hline 
			Min & 61.74 & 52.68 & 0 & 0\tabularnewline
			\hline 
			Max & 94.21 & 76.01 & 1 & 1.09\tabularnewline
			\hline 
			Count & 508 & 508 & 508 & 508\tabularnewline
			\hline 
		\end{tabular}
		
	}
\end{table}

\begin{table}
	\caption{\label{tab:g7-summ-stats-TI-vars}Group 7  summary statistics -- Time-invariant
		variables.}
	
	\resizebox{\textwidth}{!}{
		
		\begin{tabular}{|c|c|c|c|c|}
			\hline 
			& Age at 31 December 2015 & Age at Account Open & Sex: Male & Risk Appetite\tabularnewline
			\hline 
			Mean & 78.97 & 62.56 & 0.54 & 0.35\tabularnewline
			\hline 
			SD & 3.93 & 3.77 & 0.5 & 0.2\tabularnewline
			\hline 
			Median & 79.08 & 63.06 & 1 & 0.39\tabularnewline
			\hline 
			Q1 & 76.49 & 60.15 & 0 & 0.22\tabularnewline
			\hline 
			Q3 & 81.83 & 65.02 & 1 & 0.5\tabularnewline
			\hline 
			Min & 65.27 & 48.48 & 0 & 0\tabularnewline
			\hline 
			Max & 92.5 & 78.32 & 1 & 0.88\tabularnewline
			\hline 
			Count & 2603 & 2603 & 2603 & 2600\tabularnewline
			\hline 
		\end{tabular}
		
	}
\end{table}

\begin{table}
	\caption{\label{tab:g1-summ-stats-TV-vars}Group 1 summary statistics -- Time-varying
		variables.}
	
	\resizebox{\textwidth}{!}{
		
		\begin{tabular}{|c|c|c|c|c|c|c|}
			\hline 
			& Regular Drawdown & Regular Drawdown & Ad-hoc Drawdown & Ad-hoc Drawdown & Ad-hoc Drawdown & Account Balance\tabularnewline & Rate & Amount & Indicator & Rate & Amount &  \tabularnewline
			\hline 
			Mean & 0.13 & 6687.54 & 0.05 & 0.14 & 10434.4 & 60246.23\tabularnewline
			\hline 
			SD & 0.07 & 5615.68 & 0.21 & 0.16 & 20751.13 & 56052.49\tabularnewline
			\hline 
			Median & 0.11 & 5172 & 0 & 0.09 & 6000 & 44855\tabularnewline
			\hline 
			Q1 & 0.09 & 3360 & 0 & 0.05 & 3432 & 27314\tabularnewline
			\hline 
			Q3 & 0.15 & 7944 & 0 & 0.15 & 10013 & 73574.5\tabularnewline
			\hline 
			Min & 0.01 & 1 & 0 & 0 & 1 & 1\tabularnewline
			\hline 
			Max & 1.22 & 70800 & 1 & 0.9 & 500000 & 812479\tabularnewline
			\hline 
			Count & 24638 & 24644 & 24684 & 1144 & 1145 & 24675\tabularnewline
			\hline 
		\end{tabular}
		
	}
\end{table}

\begin{table}
	\caption{\label{tab:g2-summ-stats-TV-vars}Group 2 summary statistics -- Time-varying
		variables.}
	
	\resizebox{\textwidth}{!}{
		
		\begin{tabular}{|c|c|c|c|c|c|c|}
			\hline 
			& Regular Drawdown & Regular Drawdown & Ad-hoc Drawdown & Ad-hoc Drawdown & Ad-hoc Drawdown & Account Balance\tabularnewline & Rate & Amount & Indicator & Rate & Amount &  \tabularnewline
			\hline 
			Mean & 0.25 & 6933.94 & 0.1 & 0.18 & 10224.05 & 46450.67\tabularnewline
			\hline 
			SD & 0.23 & 5594.9 & 0.3 & 0.17 & 19495.22 & 47647.52\tabularnewline
			\hline 
			Median & 0.15 & 5460 & 0 & 0.12 & 5000 & 33775\tabularnewline
			\hline 
			Q1 & 0.11 & 3372 & 0 & 0.06 & 2000 & 15689.5\tabularnewline
			\hline 
			Q3 & 0.3 & 8928 & 0 & 0.25 & 10075 & 62153.75\tabularnewline
			\hline 
			Min & 0.01 & 2 & 0 & 0 & 1 & 1\tabularnewline
			\hline 
			Max & 2 & 42000 & 1 & 0.9 & 277831 & 543370\tabularnewline
			\hline 
			Count & 10344 & 10345 & 10372 & 1114 & 1115 & 10368\tabularnewline
			\hline 
		\end{tabular}
		
	}
\end{table}

\begin{table}
	\caption{\label{tab:g3-summ-stats-TV-vars}Group 3 summary statistics -- Time-varying
		variables.}
	
	\resizebox{\textwidth}{!}{
		
		\begin{tabular}{|c|c|c|c|c|c|c|}
			\hline 
			& Regular Drawdown & Regular Drawdown & Ad-hoc Drawdown & Ad-hoc Drawdown & Ad-hoc Drawdown & Account Balance\tabularnewline & Rate & Amount & Indicator & Rate & Amount &  \tabularnewline
			\hline 
			Mean & 0.34 & 6076.35 & 0.1 & 0.35 & 9789.94 & 33754.28\tabularnewline
			\hline 
			SD & 0.3 & 6305.35 & 0.3 & 0.3 & 19162.35 & 46781.99\tabularnewline
			\hline 
			Median & 0.2 & 4400 & 0 & 0.23 & 4013 & 19960\tabularnewline
			\hline 
			Q1 & 0.14 & 2291 & 0 & 0.09 & 42 & 5732.25\tabularnewline
			\hline 
			Q3 & 0.46 & 7481.25 & 0 & 0.6 & 10000 & 42861.5\tabularnewline
			\hline 
			Min & 0 & 1 & 0 & 0 & 1 & 1\tabularnewline
			\hline 
			Max & 1.38 & 48000 & 1 & 0.9 & 233305 & 632325\tabularnewline
			\hline 
			Count & 3716 & 3716 & 3944 & 461 & 461 & 3914\tabularnewline
			\hline 
		\end{tabular}
		
	}
\end{table}

\begin{table}
	\caption{\label{tab:g4-summ-stats-TV-vars}Group 4 summary statistics -- Time-varying
		variables.}
	
	\resizebox{\textwidth}{!}{
		
		\begin{tabular}{|c|c|c|c|c|c|c|}
			\hline 
			& Regular Drawdown & Regular Drawdown & Ad-hoc Drawdown & Ad-hoc Drawdown & Ad-hoc Drawdown & Account Balance\tabularnewline & Rate & Amount & Indicator & Rate & Amount &  \tabularnewline
			\hline 
			Mean & 0.07 & 6412.78 & 0.09 & 0.11 & 8635.62 & 90976.64\tabularnewline
			\hline 
			SD & 0.03 & 6524.65 & 0.29 & 0.2 & 25313.44 & 94553.22\tabularnewline
			\hline 
			Median & 0.07 & 4588.91 & 0 & 0.02 & 2253.33 & 65456\tabularnewline
			\hline 
			Q1 & 0.06 & 2712 & 0 & 0.01 & 732.5 & 39864\tabularnewline
			\hline 
			Q3 & 0.08 & 7560 & 0 & 0.09 & 7500 & 104543\tabularnewline
			\hline 
			Min & 0 & 30 & 0 & 0 & 1 & 15\tabularnewline
			\hline 
			Max & 0.94 & 99768 & 1 & 0.9 & 430000 & 1573153\tabularnewline
			\hline 
			Count & 19460 & 19472 & 19609 & 1831 & 1833 & 19597\tabularnewline
			\hline 
		\end{tabular}
		
	}
\end{table}

\begin{table}
	\caption{\label{tab:g5-summ-stats-TV-vars}Group 5 summary statistics -- Time-varying
		variables.}
	
	\resizebox{\textwidth}{!}{
		
		\begin{tabular}{|c|c|c|c|c|c|c|}
			\hline 
			& Regular Drawdown & Regular Drawdown & Ad-hoc Drawdown & Ad-hoc Drawdown & Ad-hoc Drawdown & Account Balance\tabularnewline & Rate & Amount & Indicator & Rate & Amount &  \tabularnewline
			\hline 
			Mean & 0.06 & 5898.44 & 0.11 & 0.11 & 9308.75 & 99415.18\tabularnewline
			\hline 
			SD & 0.04 & 6219.52 & 0.31 & 0.19 & 26711.86 & 103532.05\tabularnewline
			\hline 
			Median & 0.06 & 4201.49 & 0 & 0.03 & 2565.83 & 71760\tabularnewline
			\hline 
			Q1 & 0.05 & 2310 & 0 & 0 & 622.93 & 43445.75\tabularnewline
			\hline 
			Q3 & 0.07 & 7094.75 & 0 & 0.09 & 8705.42 & 115423.25\tabularnewline
			\hline 
			Min & 0 & 10 & 0 & 0 & 3 & 53\tabularnewline
			\hline 
			Max & 0.75 & 109320 & 1 & 0.9 & 600000 & 1514586.47\tabularnewline
			\hline 
			Count & 13184 & 13192 & 13294 & 1403 & 1403 & 13286\tabularnewline
			\hline 
		\end{tabular}
		
	}
\end{table}

\begin{table}
	\caption{\label{tab:g6-summ-stats-TV-vars}Group 6 summary statistics -- Time-varying
		variables.}
	
	\resizebox{\textwidth}{!}{
		
		\begin{tabular}{|c|c|c|c|c|c|c|}
			\hline 
			& Regular Drawdown & Regular Drawdown & Ad-hoc Drawdown & Ad-hoc Drawdown & Ad-hoc Drawdown & Account Balance\tabularnewline & Rate & Amount & Indicator & Rate & Amount &  \tabularnewline
			\hline 
			Mean & 0.1 & 6377.06 & 0.11 & 0.21 & 13530.6 & 73469.07\tabularnewline
			\hline 
			SD & 0.08 & 5853.41 & 0.31 & 0.27 & 32379.36 & 71795.16\tabularnewline
			\hline 
			Median & 0.08 & 4800 & 0 & 0.09 & 5000 & 53776\tabularnewline
			\hline 
			Q1 & 0.07 & 2670 & 0 & 0.02 & 1159.84 & 30503\tabularnewline
			\hline 
			Q3 & 0.11 & 8060 & 0 & 0.31 & 13318.59 & 89727\tabularnewline
			\hline 
			Min & 0 & 1 & 0 & 0 & 1 & 1\tabularnewline
			\hline 
			Max & 1.57 & 86688.13 & 1 & 0.9 & 455548 & 781753.8\tabularnewline
			\hline 
			Count & 5437 & 5438 & 5592 & 606 & 606 & 5585\tabularnewline
			\hline 
		\end{tabular}
		
	}
\end{table}

\begin{table}
	\caption{\label{tab:g7-summ-stats-TV-vars}Group 7 summary statistics -- Time-varying
		variables.}
	
	\resizebox{\textwidth}{!}{
		
		\begin{tabular}{|c|c|c|c|c|c|c|}
			\hline 
			& Regular Drawdown & Regular Drawdown & Ad-hoc Drawdown & Ad-hoc Drawdown & Ad-hoc Drawdown & Account Balance\tabularnewline & Rate & Amount & Indicator & Rate & Amount &  \tabularnewline
			\hline 
			Mean & 0.09 & 6366.74 & 0.03 & 0.16 & 12575.79 & 73117.21\tabularnewline
			\hline 
			SD & 0.03 & 6367.71 & 0.17 & 0.2 & 26520.99 & 76839.21\tabularnewline
			\hline 
			Median & 0.09 & 4764 & 0 & 0.09 & 6000 & 53267.5\tabularnewline
			\hline 
			Q1 & 0.07 & 3036 & 0 & 0.04 & 3000 & 33820\tabularnewline
			\hline 
			Q3 & 0.11 & 7416 & 0 & 0.18 & 12000 & 87103.25\tabularnewline
			\hline 
			Min & 0.02 & 12 & 0 & 0 & 31 & 13\tabularnewline
			\hline 
			Max & 0.92 & 166695 & 1 & 0.9 & 520802 & 2427083\tabularnewline
			\hline 
			Count & 31156 & 31168 & 31222 & 891 & 891 & 31210\tabularnewline
			\hline 
		\end{tabular}
		
	}
\end{table}

\begin{figure}
	\caption{\label{fig:Group-1-Time-Profile-Analytical-SE-Distributions-across-Simulated-Datasets}Group
		1 time profile analytical SE distributions across simulated datasets. Black lines plot the kernel density estimate for standard
		errors derived from the fixed-$T$ variance estimate formula after
		estimating the GFE model on 1000 simulated datasets. Red vertical
		lines represent the value of the simulated standard error.}
	
	\includegraphics[width=1\textwidth]{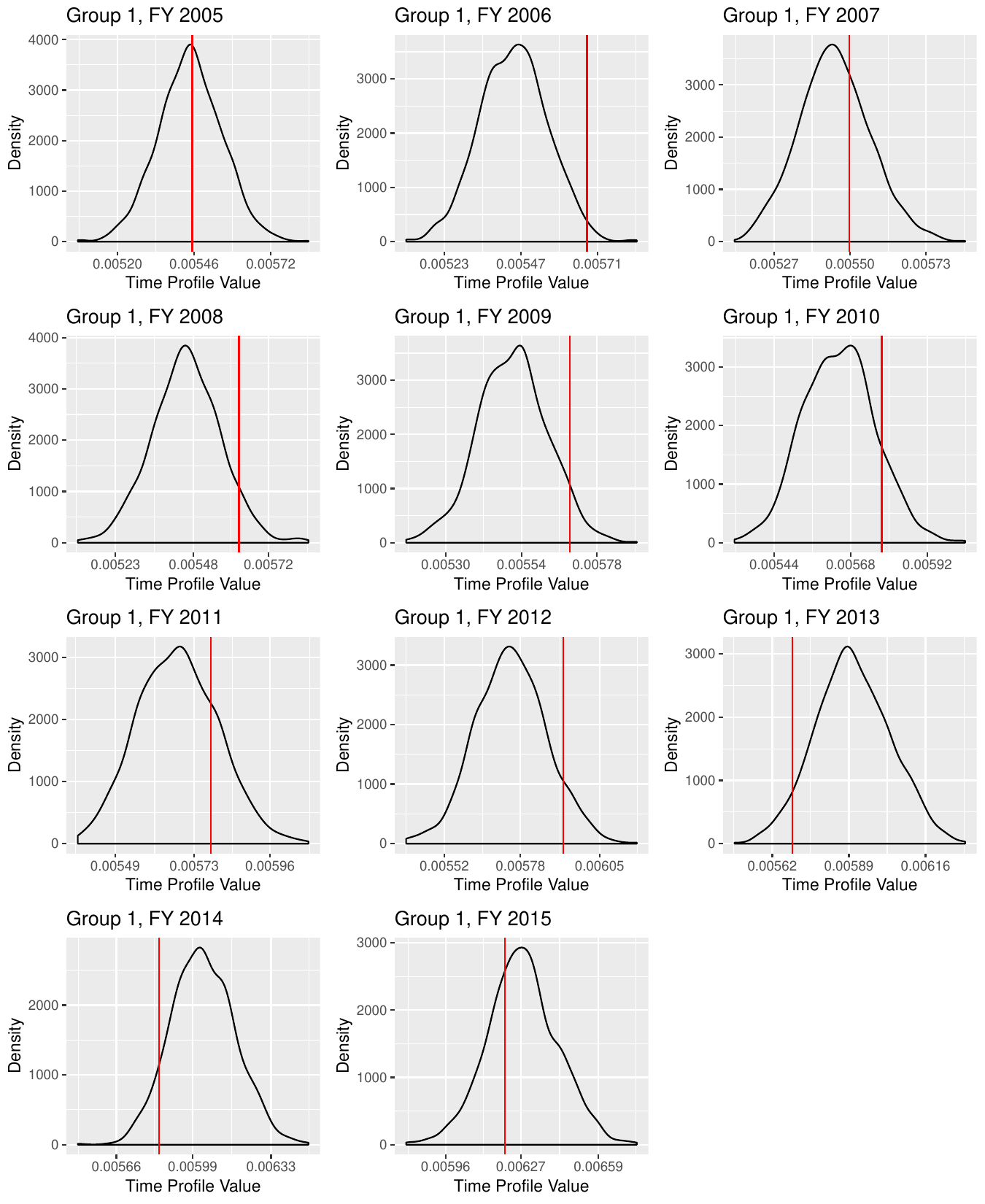}
	
\end{figure}
\begin{figure}
	\caption{\label{fig:Group-2-Time-Profile-Analytical-SE-Distributions-across-Simulated-Datasets}Group
		2 time profile analytical SE distributions across simulated datasets. Black lines plot the kernel density estimate for standard
		errors derived from the fixed-$T$ variance estimate formula after
		estimating the GFE model on 1000 simulated datasets. Red vertical
		lines represent the value of the simulated standard error.}
	
	\includegraphics[width=1\textwidth]{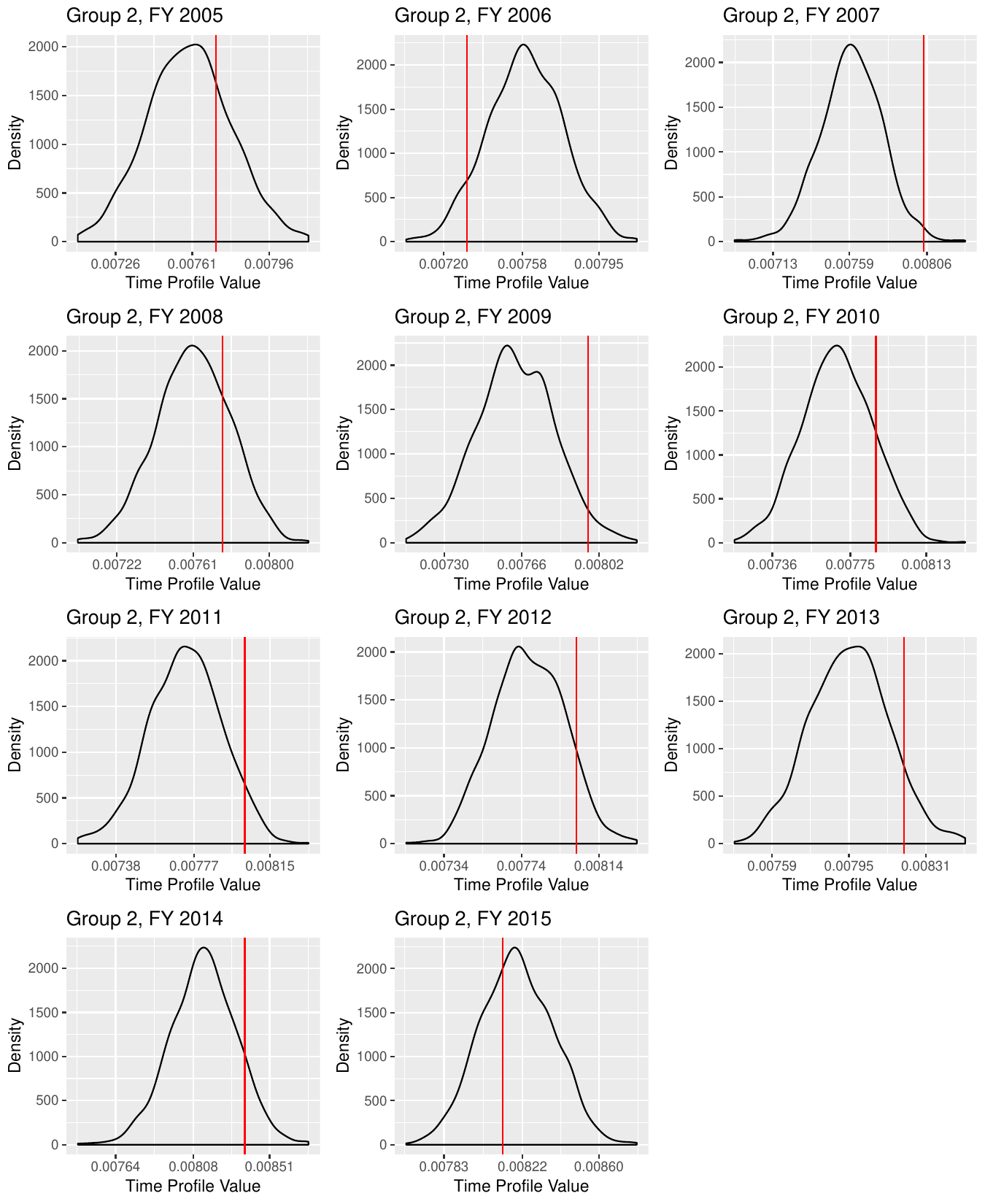}
	
\end{figure}
\begin{figure}
	\caption{\label{fig:Group-3-Time-Profile-Analytical-SE-Distributions-across-Simulated-Datasets}Group
		3 time profile analytical SE distributions across simulated datasets. Black lines plot the kernel density estimate for standard
		errors derived from the fixed-$T$ variance estimate formula after
		estimating the GFE model on 1000 simulated datasets. Red vertical
		lines represent the value of the simulated standard error.}
	
	\includegraphics[width=1\textwidth]{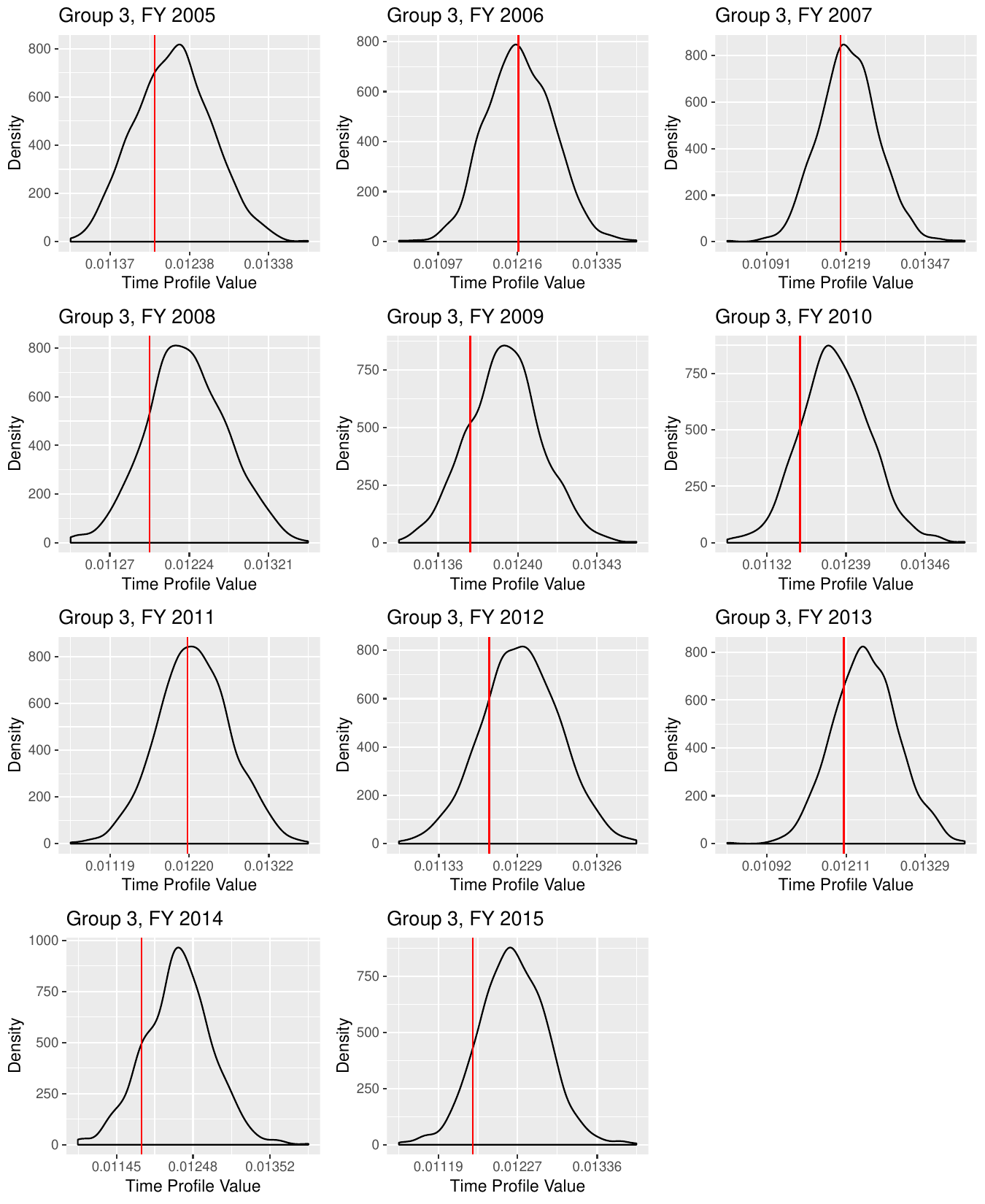}
	
\end{figure}
\begin{figure}
	\caption{\label{fig:Group-4-Time-Profile-Analytical-SE-Distributions-across-Simulated-Datasets}Group
		4 time profile analytical SE distributions across simulated datasets. Black lines plot the kernel density estimate for standard
		errors derived from the fixed-$T$ variance estimate formula after
		estimating the GFE model on 1000 simulated datasets. Red vertical
		lines represent the value of the simulated standard error.}
	
	\includegraphics[width=1\textwidth]{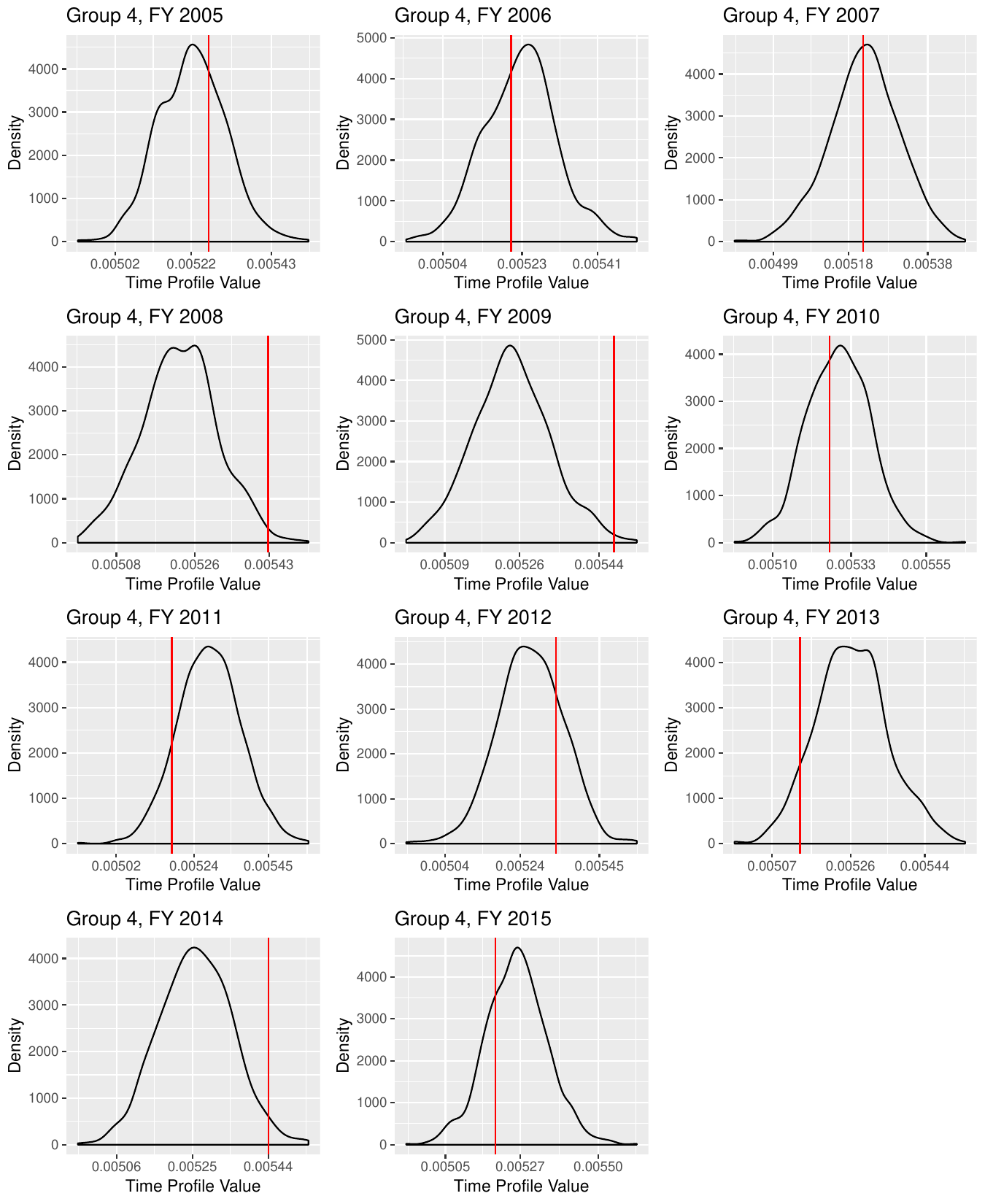}
	
\end{figure}
\begin{figure}
	\caption{\label{fig:Group-5-Time-Profile-Analytical-SE-Distributions-across-Simulated-Datasets}Group
		5 time profile analytical SE distributions across simulated datasets. Black lines plot the kernel density estimate for standard
		errors derived from the fixed-$T$ variance estimate formula after
		estimating the GFE model on 1000 simulated datasets. Red vertical
		lines represent the value of the simulated standard error.}
	
	\includegraphics[width=1\textwidth]{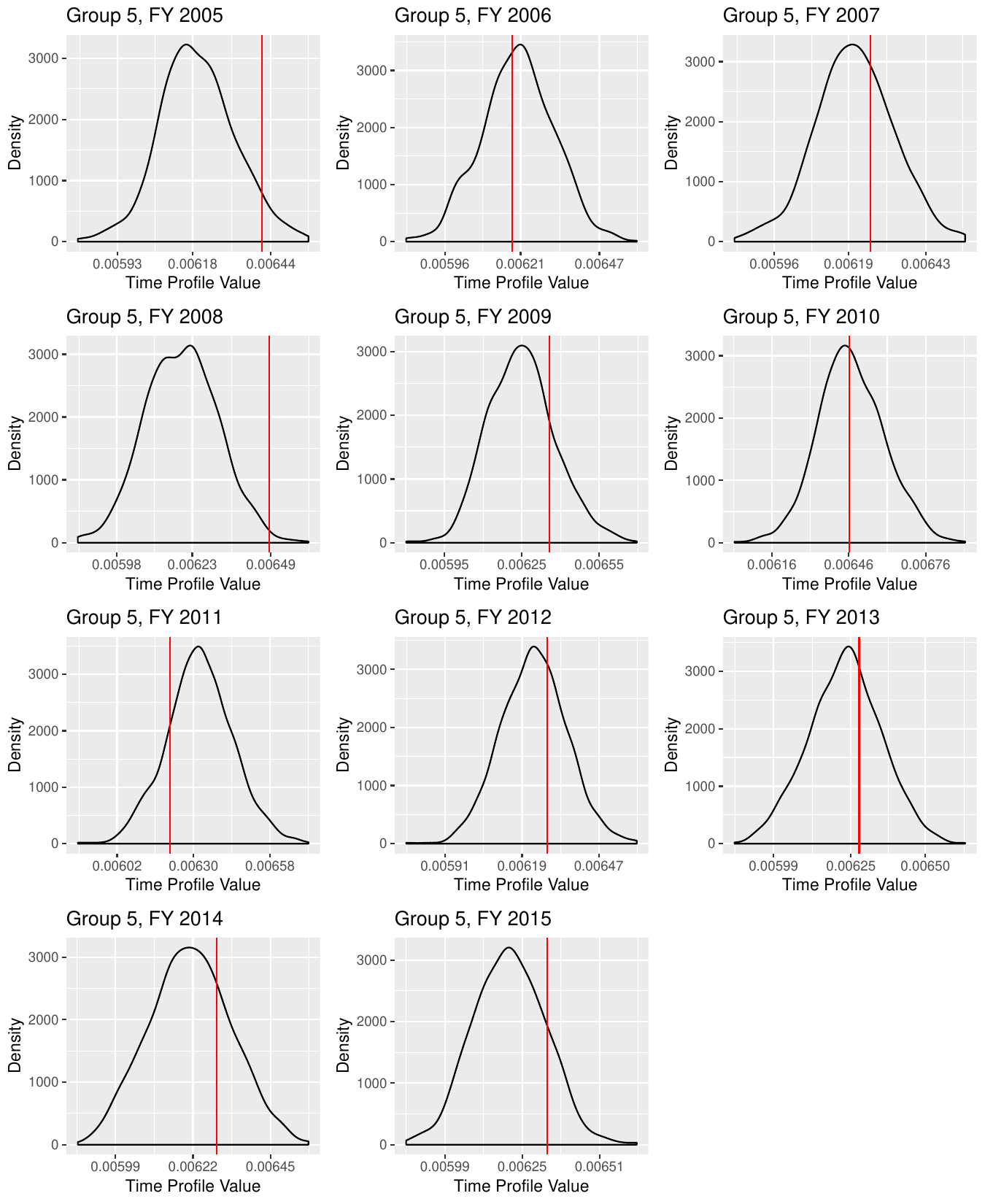}
	
\end{figure}
\begin{figure}
	\caption{\label{fig:Group-6-Time-Profile-Analytical-SE-Distributions-across-Simulated-Datasets}Group
		6 time profile analytical SE distributions across simulated datasets. Black lines plot the kernel density estimate for standard
		errors derived from the fixed-$T$ variance estimate formula after
		estimating the GFE model on 1000 simulated datasets. Red vertical
		lines represent the value of the simulated standard error.}
	
	\includegraphics[width=1\textwidth]{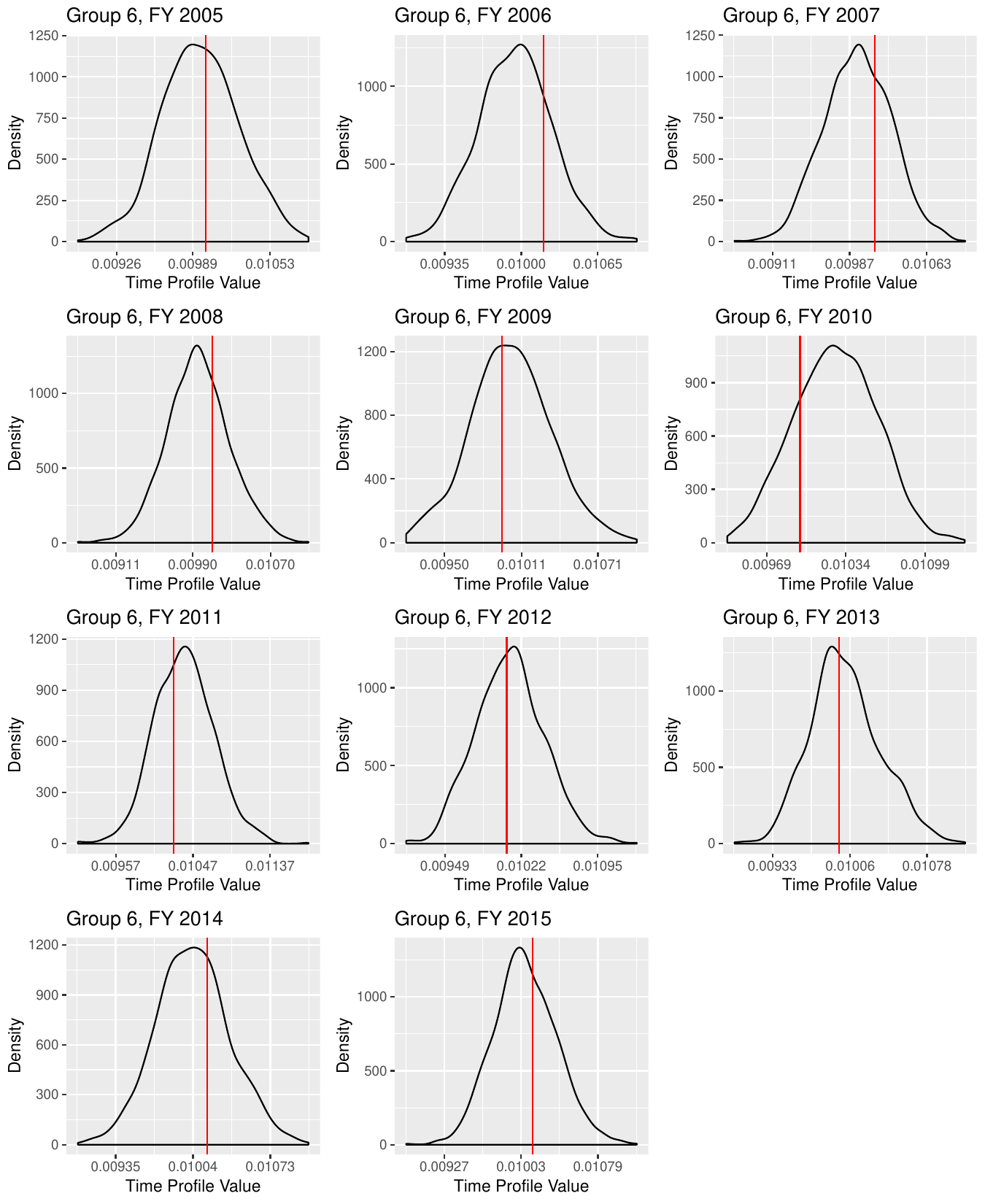}
	
\end{figure}

\begin{figure}
	\caption{\label{fig:Group-7-Time-Profile-Analytical-SE-Distributions-across-Simulated-Datasets}Group
		7 time profile analytical SE distributions across simulated datasets. Black lines plot the kernel density estimate for standard
		errors derived from the fixed-$T$ variance estimate formula after
		estimating the GFE model on 1000 simulated datasets. Red vertical
		lines represent the value of the simulated standard error.}
	
	\includegraphics[width=1\textwidth]{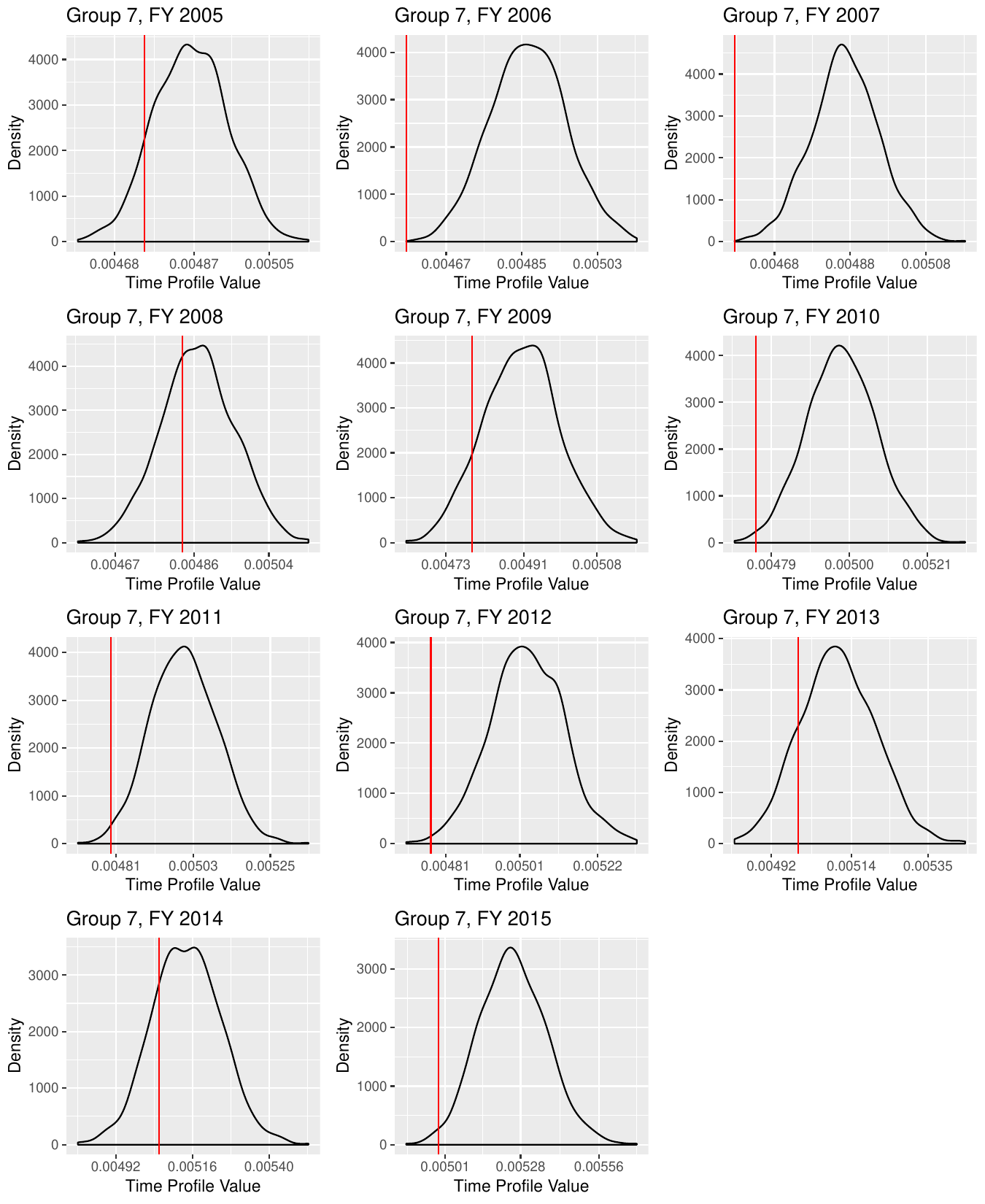}
	
\end{figure}

\begin{figure}
	\caption{\label{fig:g1-hist-TI-vars}Group 1 histograms -- Time-invariant
		variables.}
	
	\includegraphics[width=1\textwidth]{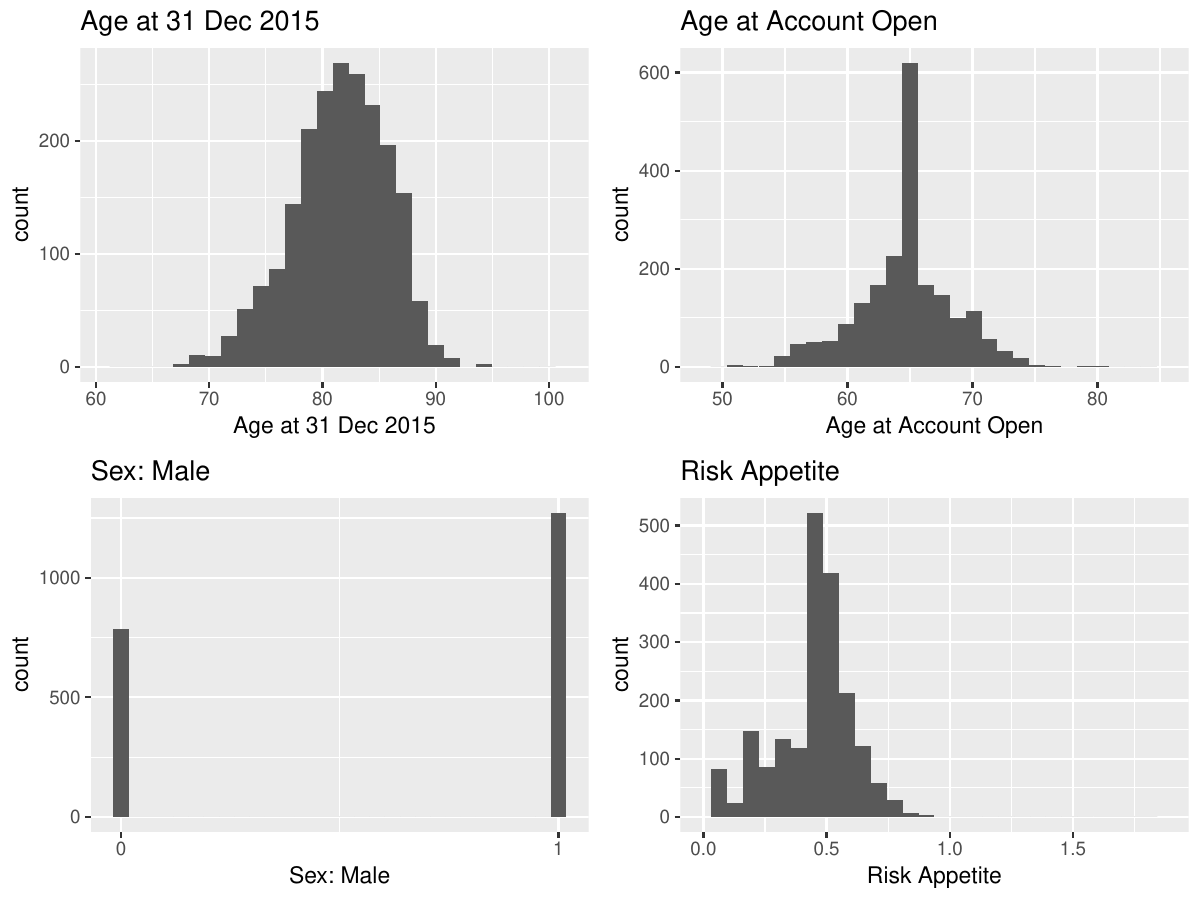}
\end{figure}

\begin{figure}
	\caption{\label{fig:g2-hist-TI-vars}Group 2 histograms -- Time-invariant
		variables.}
	
	\includegraphics[width=1\textwidth]{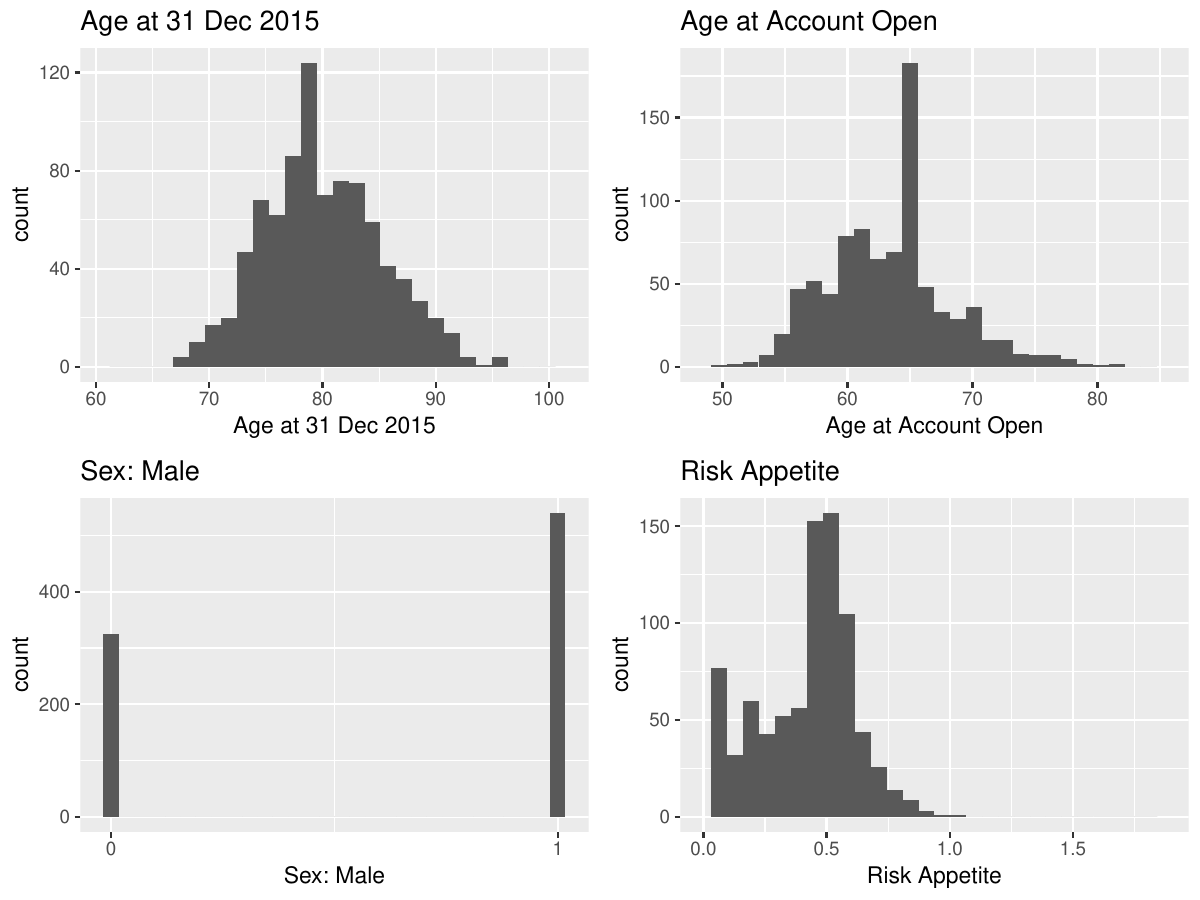}
\end{figure}

\begin{figure}
	\caption{\label{fig:g3-hist-TI-vars}Group 3 histograms -- Time-invariant
		variables.}
	
	\includegraphics[width=1\textwidth]{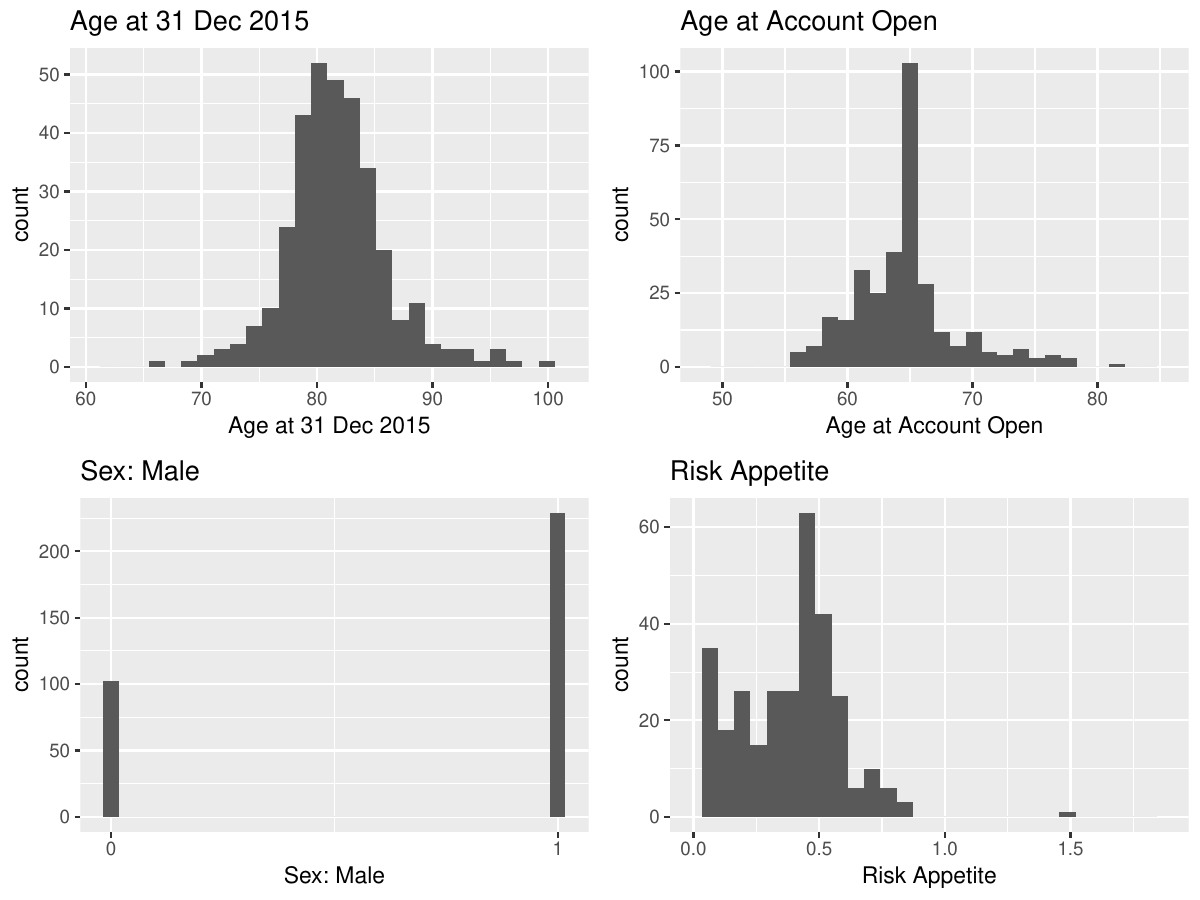}
\end{figure}

\begin{figure}
	\caption{\label{fig:g4-hist-TI-vars}Group 4 histograms -- Time-invariant
		variables.}
	
	\includegraphics[width=1\textwidth]{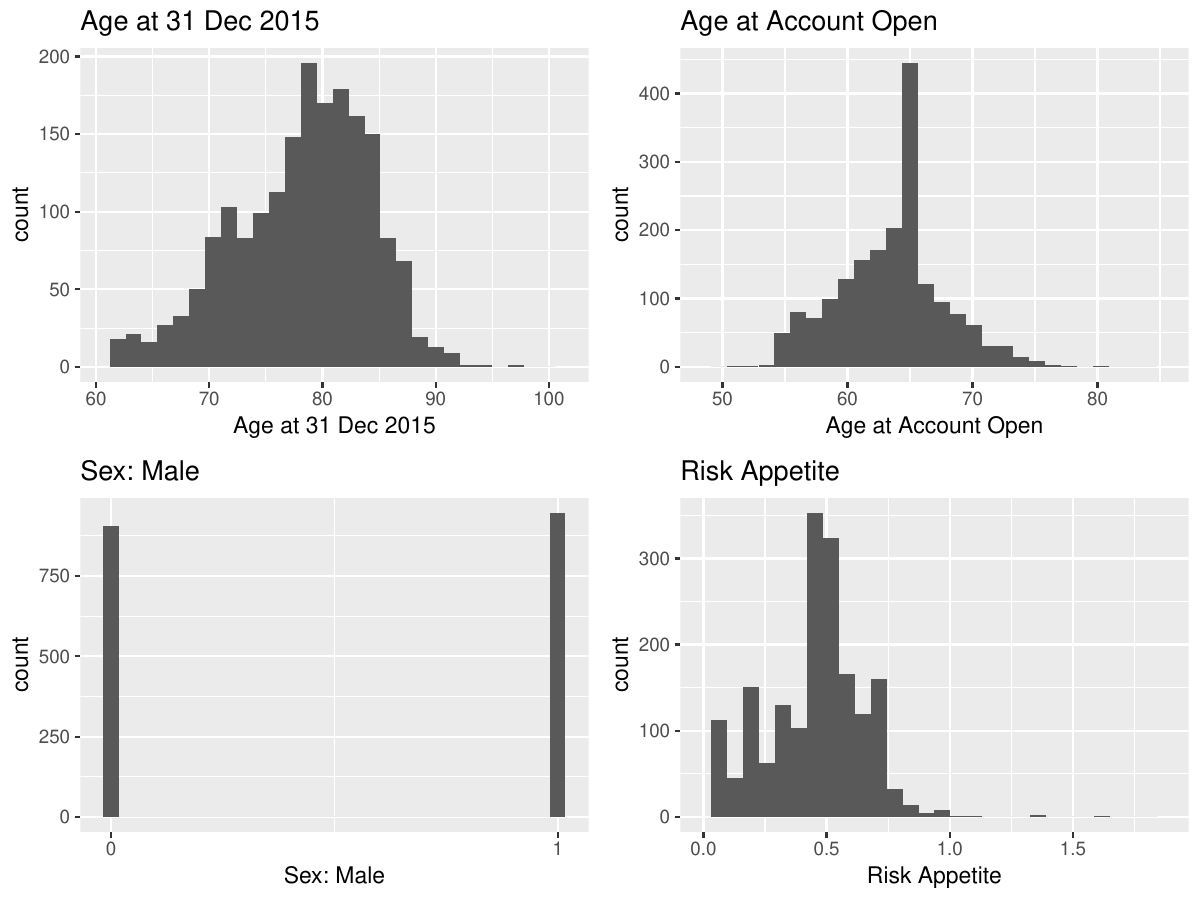}
\end{figure}

\clearpage

\begin{figure}
	\caption{\label{fig:g5-hist-TI-vars}Group 5 histograms -- Time-invariant
		variables.}
	
	\includegraphics[width=1\textwidth]{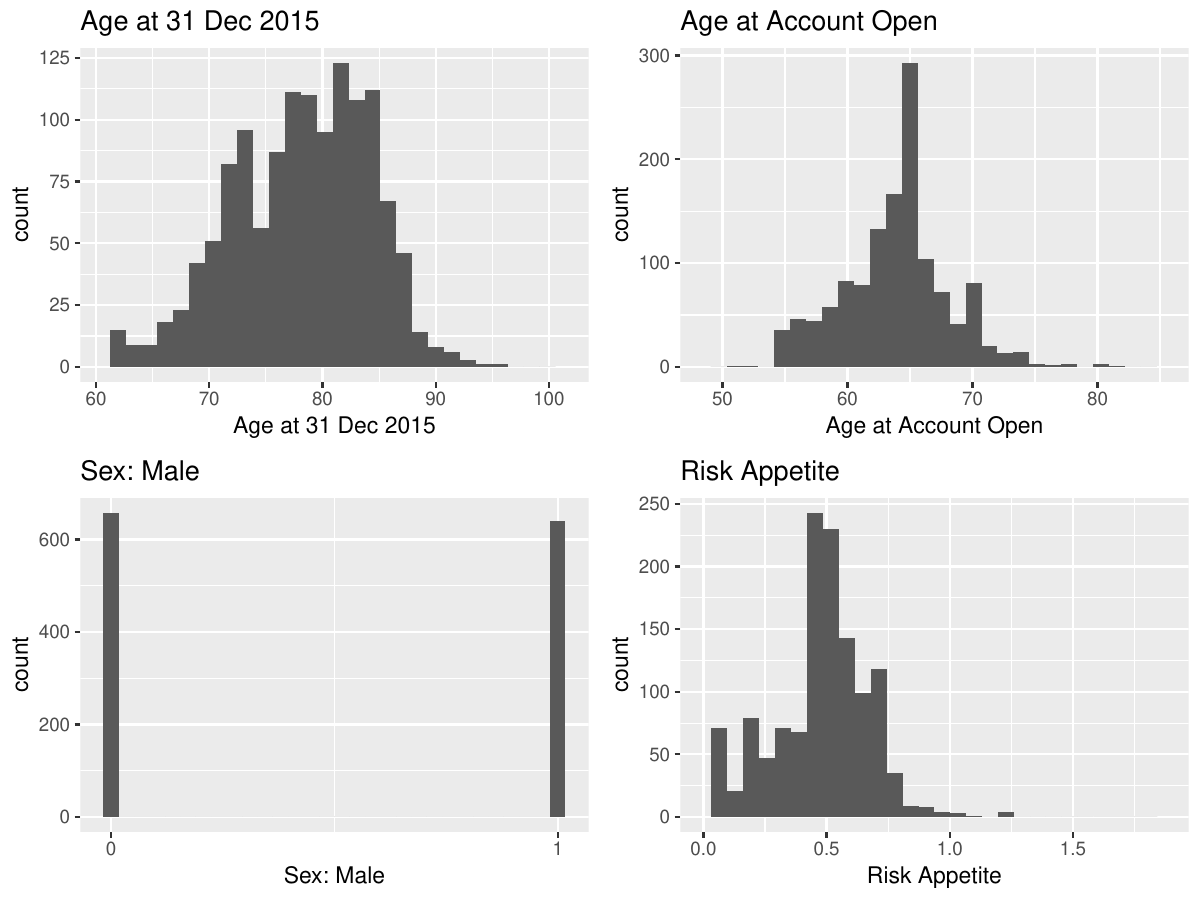}
\end{figure}

\begin{figure}
	\caption{\label{fig:g6-hist-TI-vars}Group 6 histograms -- Time-invariant
		variables.}
	
	\includegraphics[width=1\textwidth]{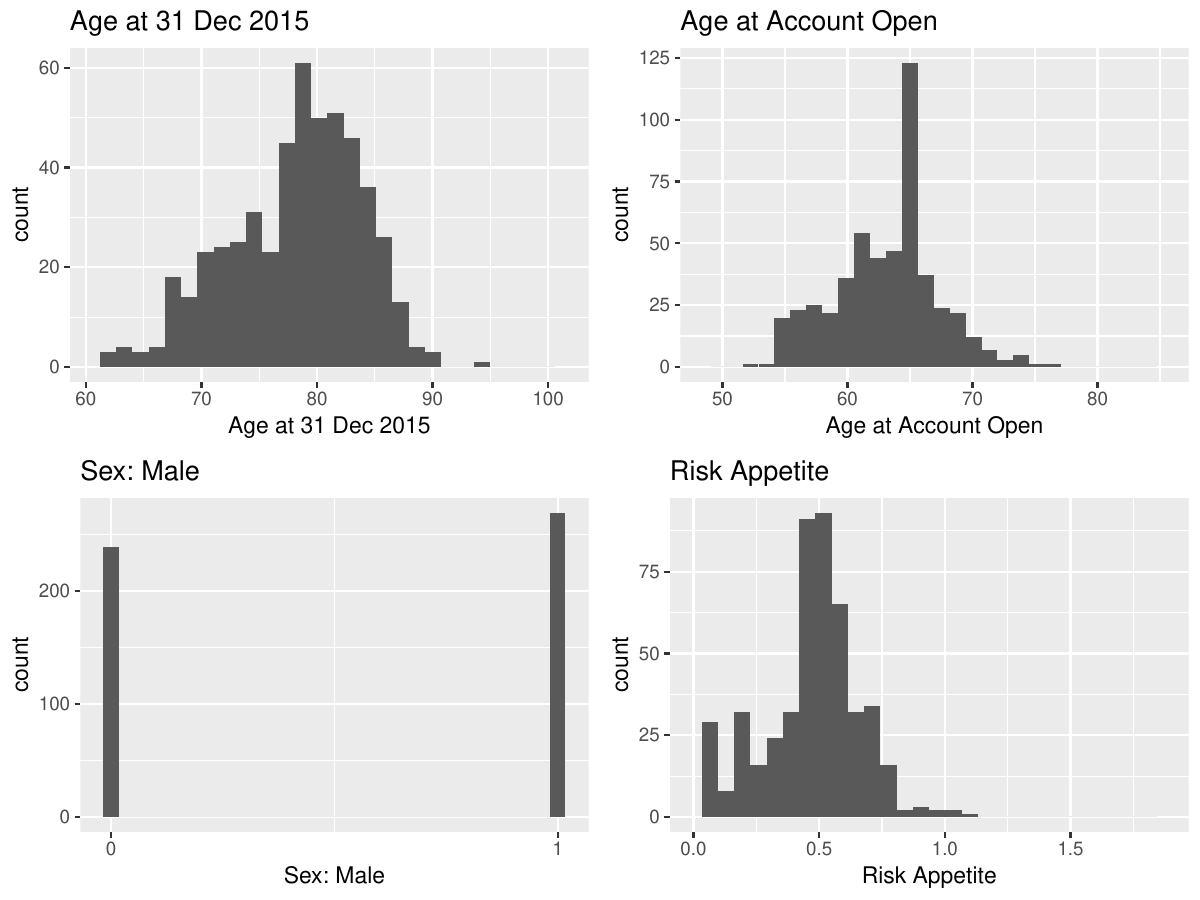}
\end{figure}

\begin{figure}
	\caption{\label{fig:g7-hist-TI-vars}Group 7 histograms -- Time-invariant
		variables.}
	
	\includegraphics[width=1\textwidth]{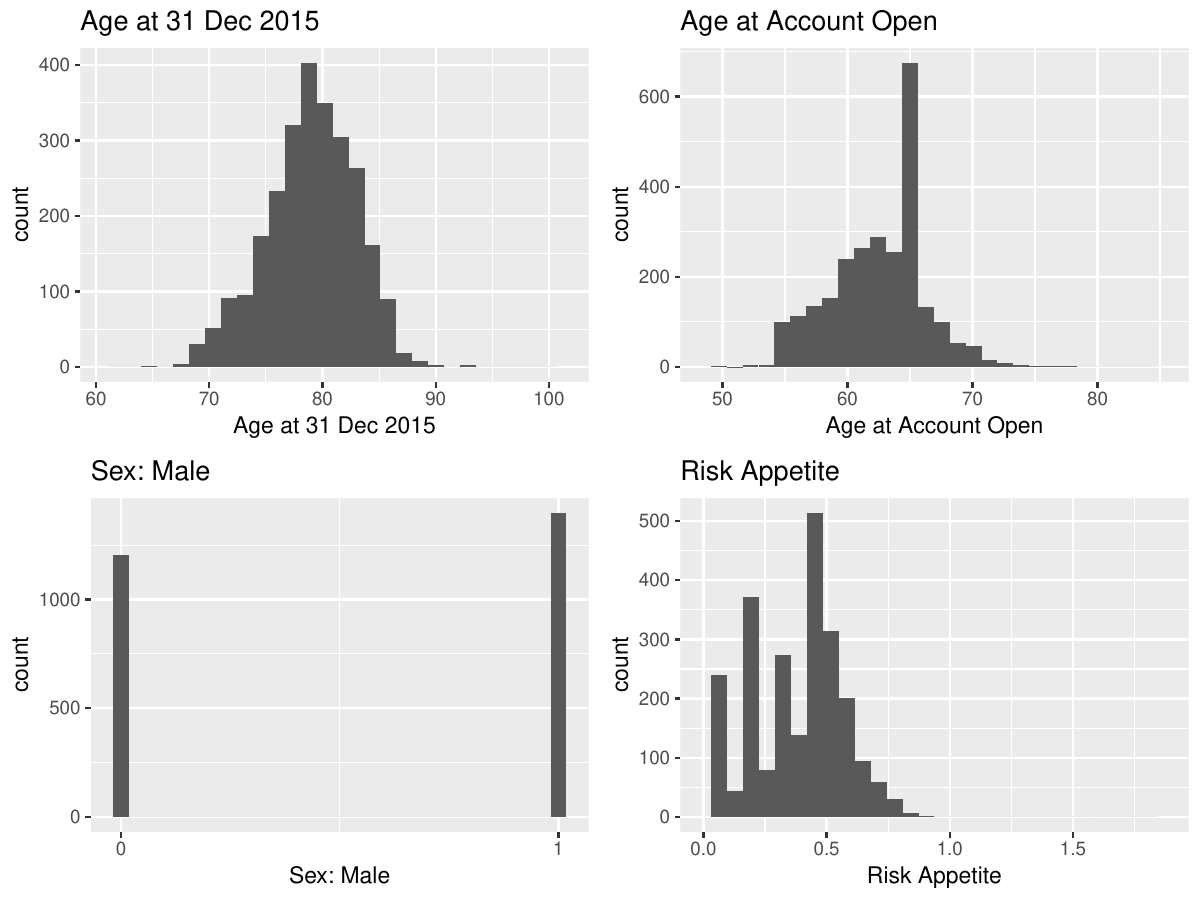}
\end{figure}

\begin{figure}
	\caption{\label{fig:g1-hist-TV-vars}Group 1 histograms -- Time-varying variables.}
	
	\includegraphics[width=1\textwidth]{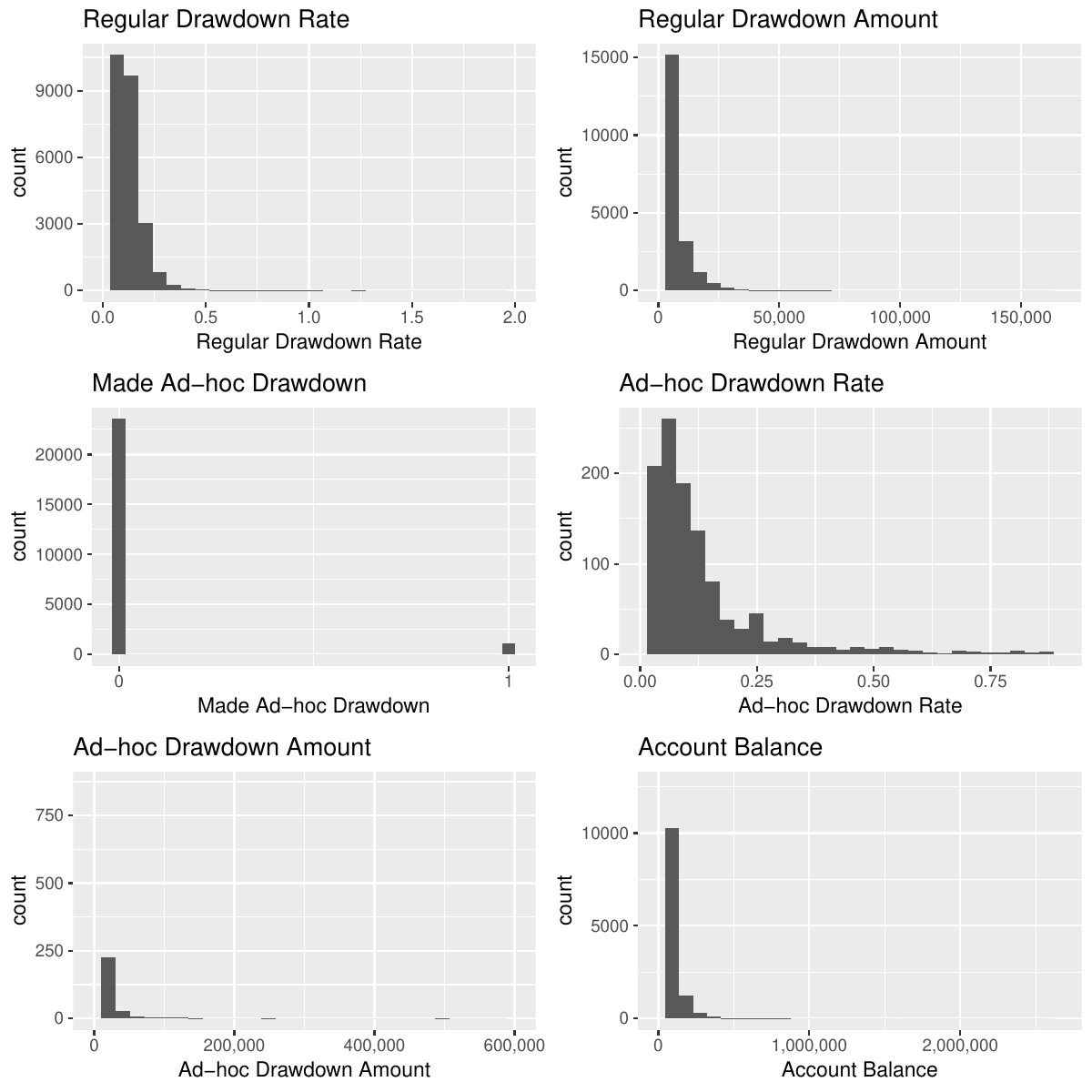}
\end{figure}

\begin{figure}
	\caption{\label{fig:g2-hist-TV-vars}Group 2 histograms -- Time-varying variables.}
	
	\includegraphics[width=1\textwidth]{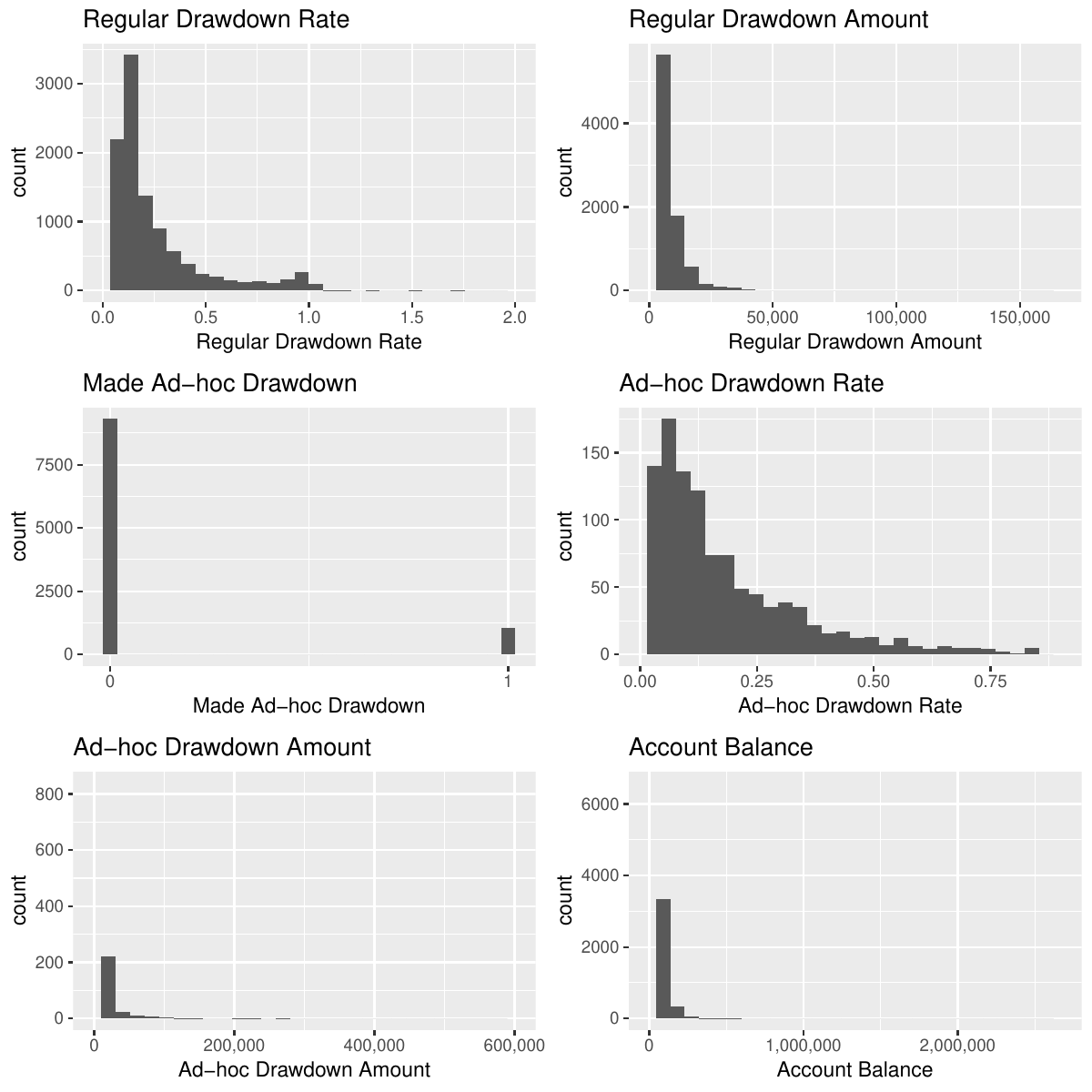}
\end{figure}

\begin{figure}
	\caption{\label{fig:g3-hist-TV-vars}Group 3 histograms -- Time-varying variables.}
	
	\includegraphics[width=1\textwidth]{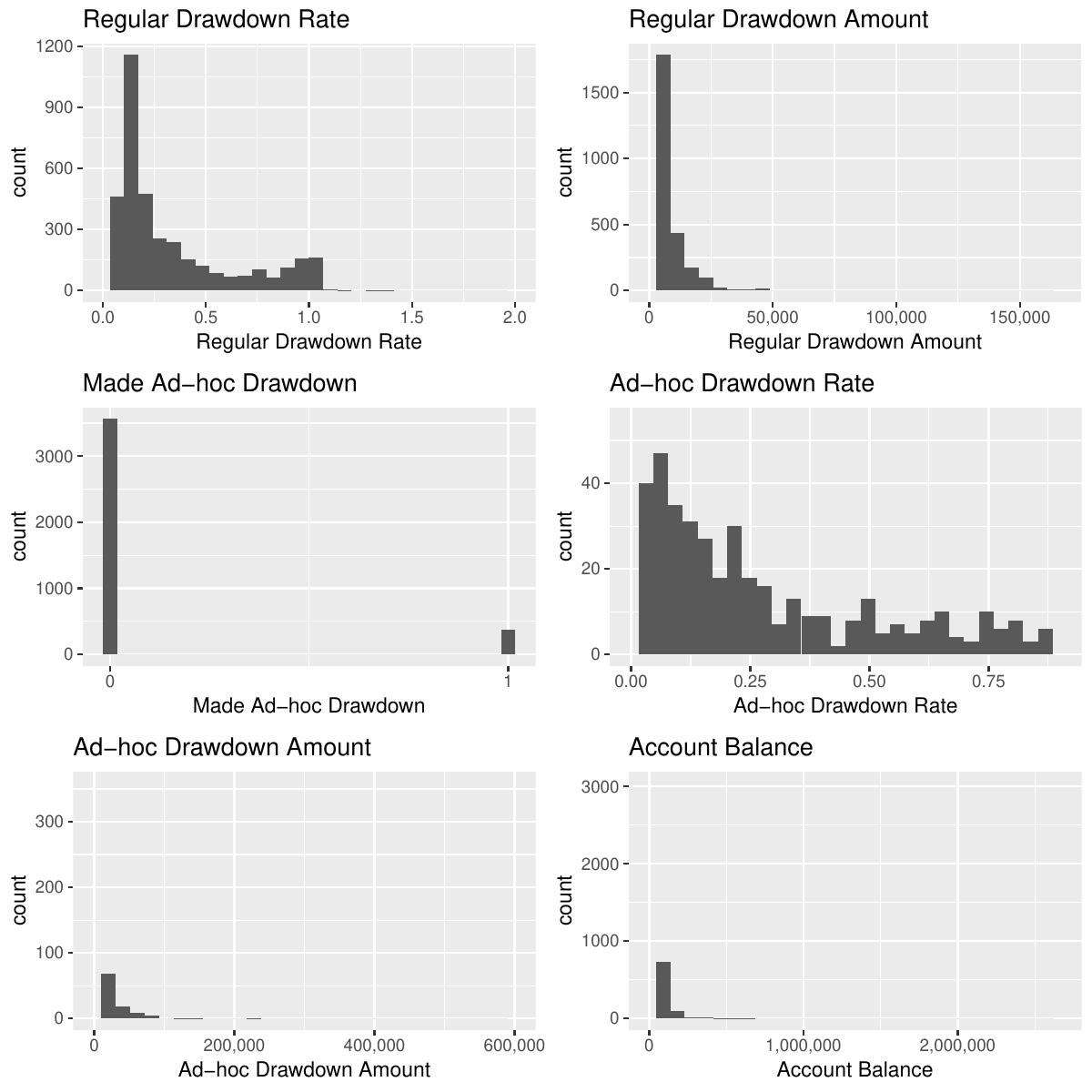}
\end{figure}

\begin{figure}
	\caption{\label{fig:g4-hist-TV-vars}Group 4 histograms -- Time-varying variables.}
	
	\includegraphics[width=1\textwidth]{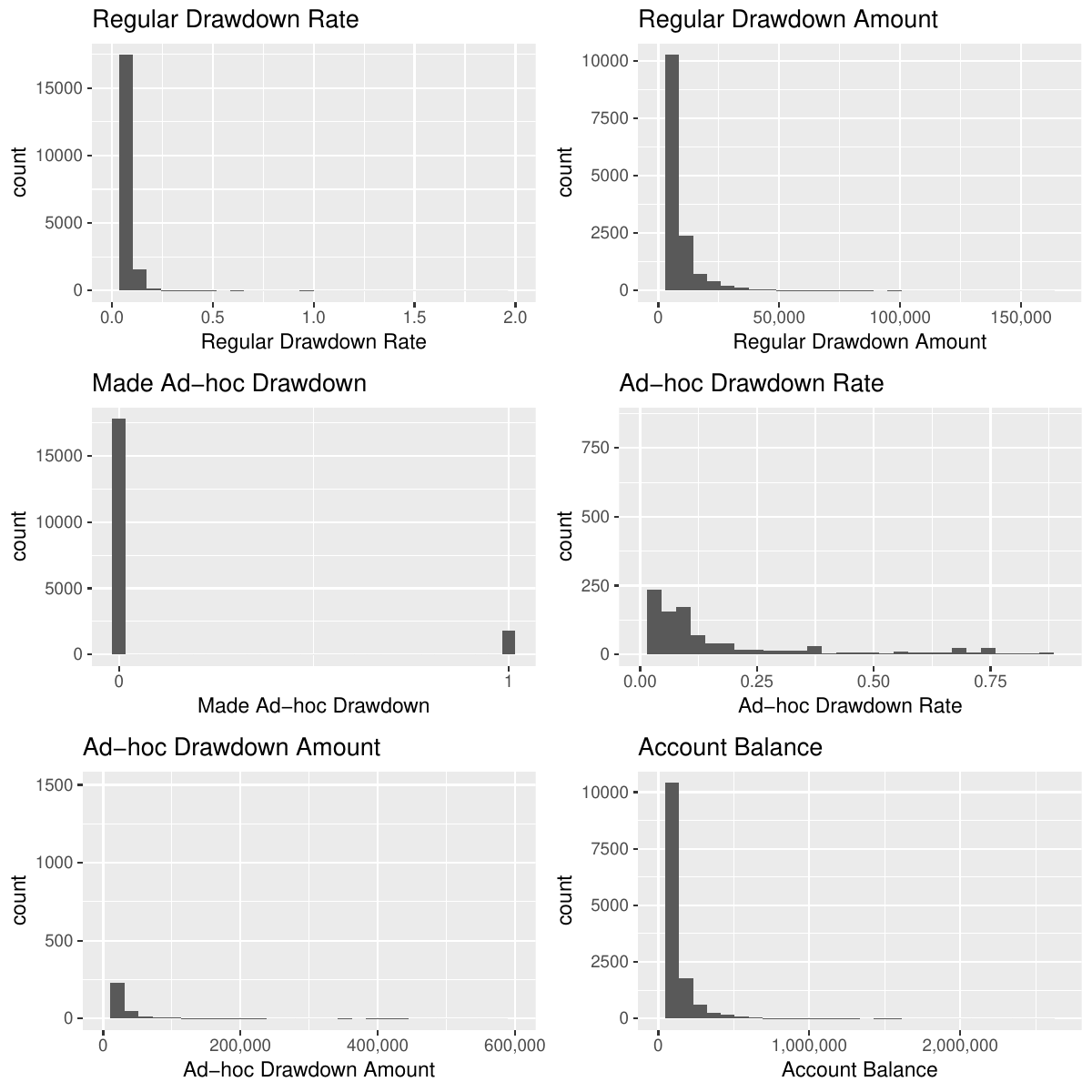}
\end{figure}

\begin{figure}
	\caption{\label{fig:g5-hist-TV-vars}Group 5 histograms -- Time-varying variables.}
	
	\includegraphics[width=1\textwidth]{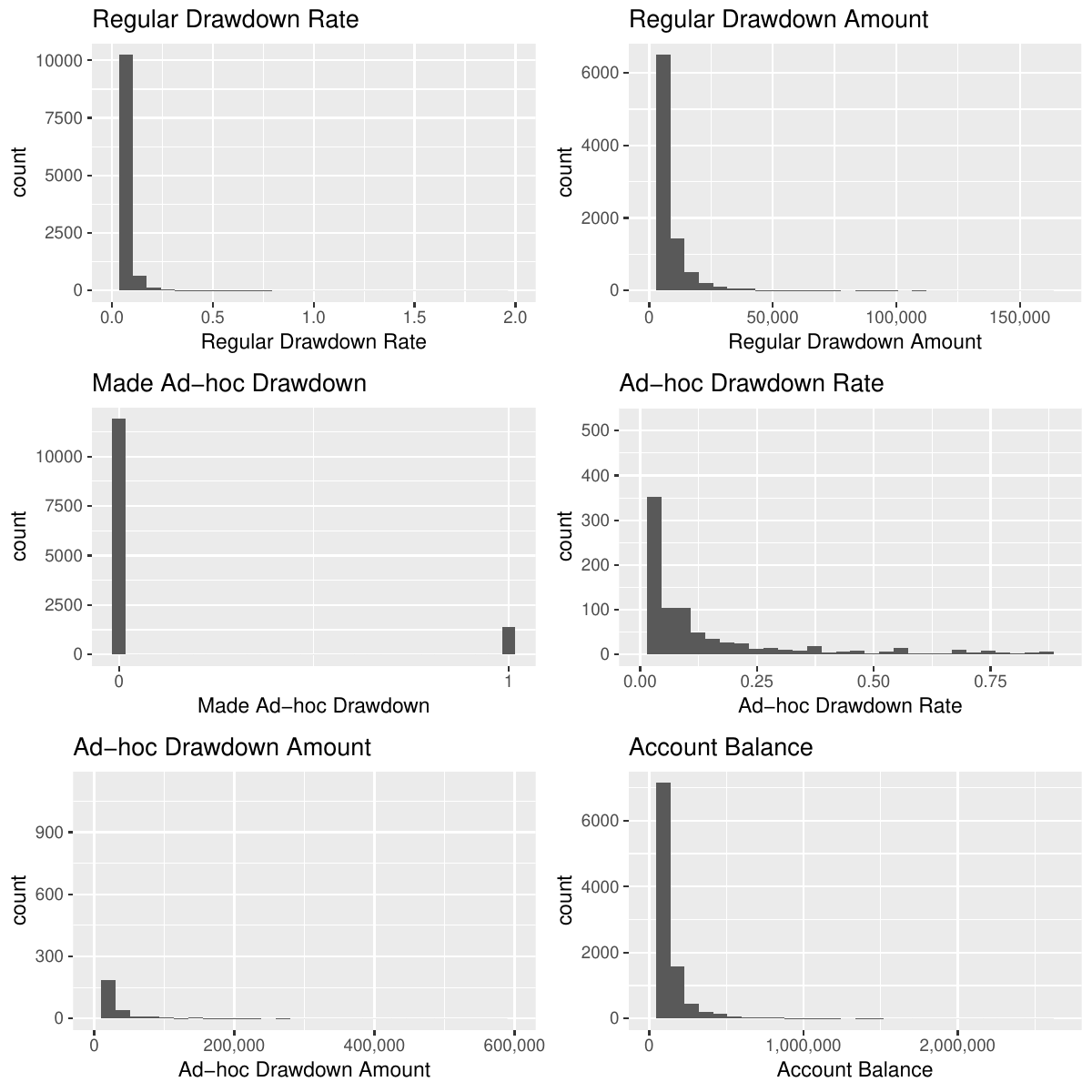}
\end{figure}

\begin{figure}
	\caption{\label{fig:g6-hist-TV-vars}Group 6 histograms -- Time-varying variables.}
	
	\includegraphics[width=1\textwidth]{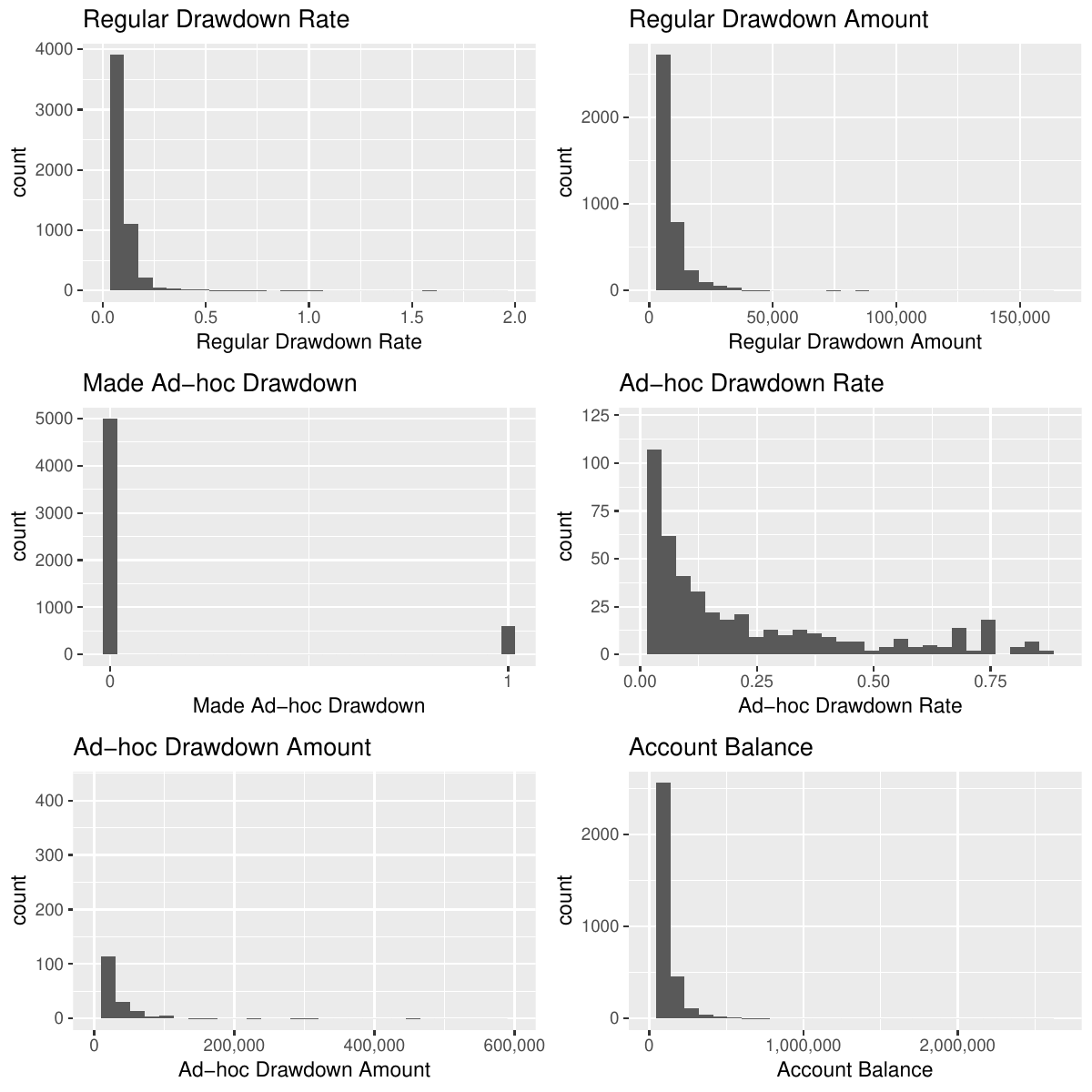}
\end{figure}

\begin{figure}
	\caption{\label{fig:g7-hist-TV-vars}Group 7 histograms -- Time-varying variables.}
	
	\includegraphics[width=1\textwidth]{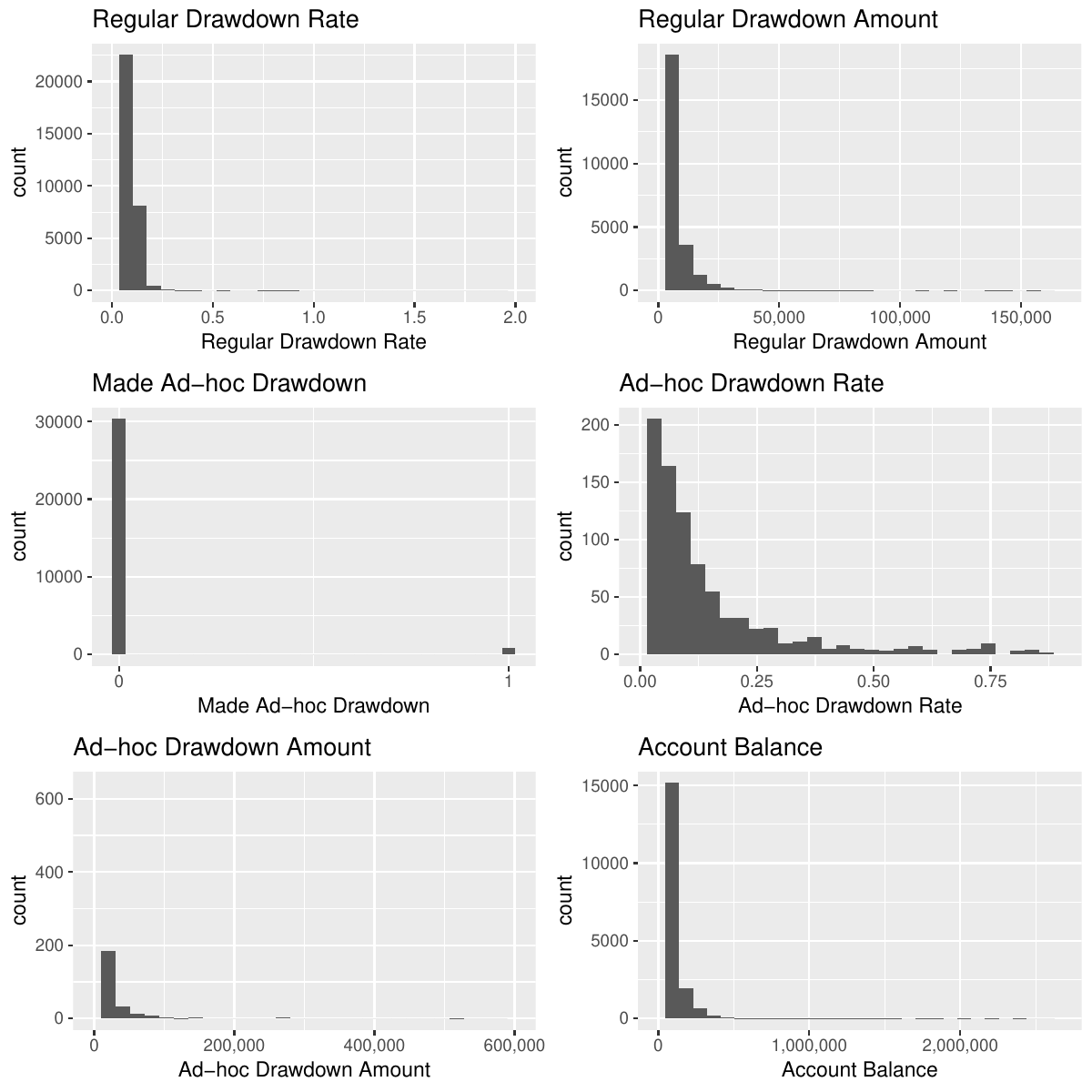}
\end{figure}

\begin{figure}
	\caption{\label{fig:g1-sixplots}$G=7$ model -- Group 1 time-demeaned (TD)
		panel plots. Account balances as at financial year start. The black
		series in the bottom-right panel represents estimated time-demeaned
		group time profile values.}
	
	\includegraphics[width=1\textwidth]{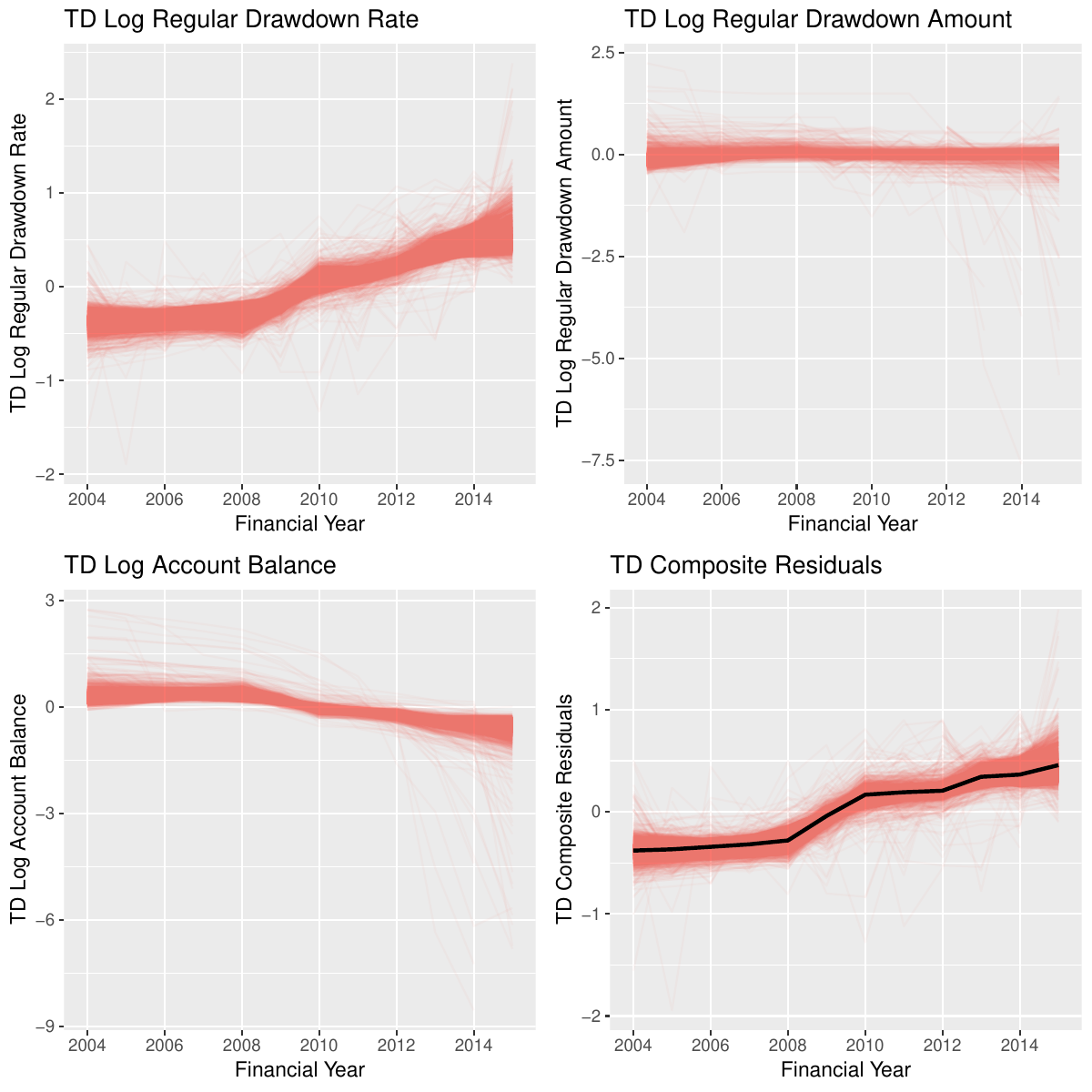}
	
\end{figure}

\begin{figure}
	\caption{\label{fig:g2-sixplots}$G=7$ model -- Group 2 time-demeaned (TD)
		panel plots. Account balances as at financial year start. The black
		series in the bottom-right panel represents estimated time-demeaned
		group time profile values.}
	
	\includegraphics[width=1\textwidth]{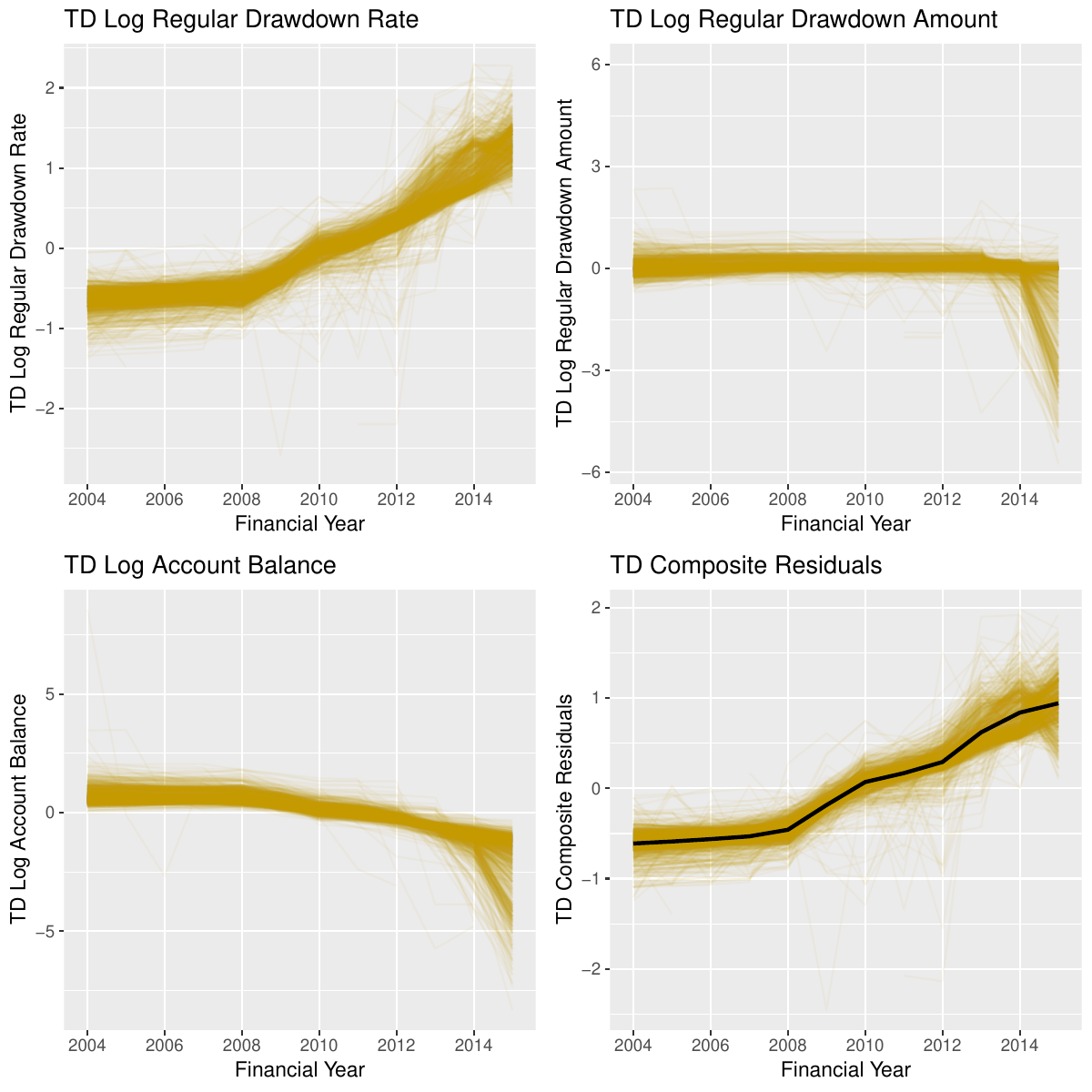}
	
\end{figure}

\begin{figure}
	\caption{\label{fig:g3-sixplots}$G=7$ model -- Group 3 time-demeaned (TD)
		panel plots. Account balances as at financial year start. The black
		series in the bottom-right panel represents estimated time-demeaned
		group time profile values.}
	
	\includegraphics[width=1\textwidth]{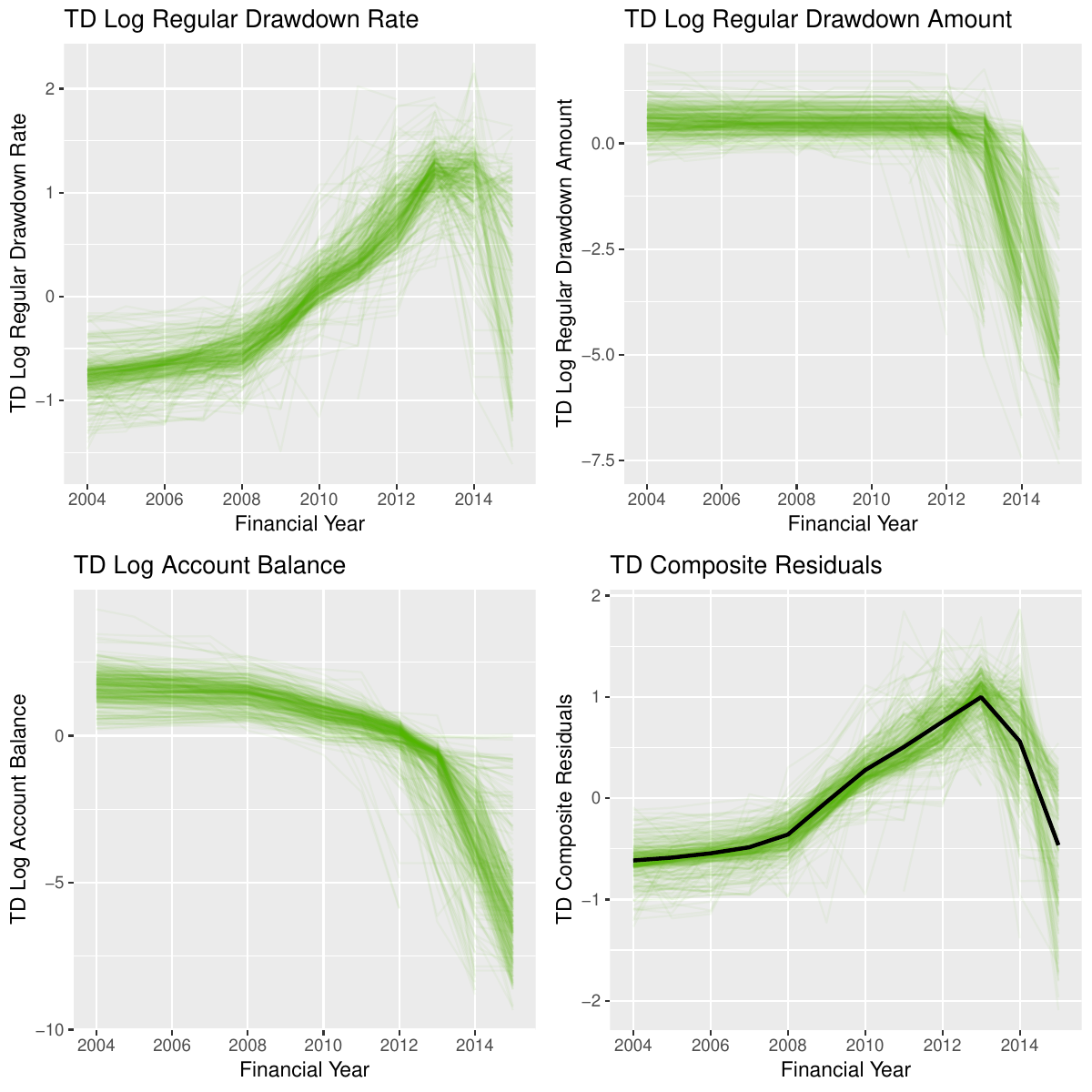}
	
\end{figure}

\begin{figure}
	\caption{\label{fig:g4-sixplots}$G=7$ model -- Group 4 time-demeaned (TD)
		panel plots. Account balances as at financial year start. The black
		series in the bottom-right panel represents estimated time-demeaned
		group time profile values.}
	
	\includegraphics[width=1\textwidth]{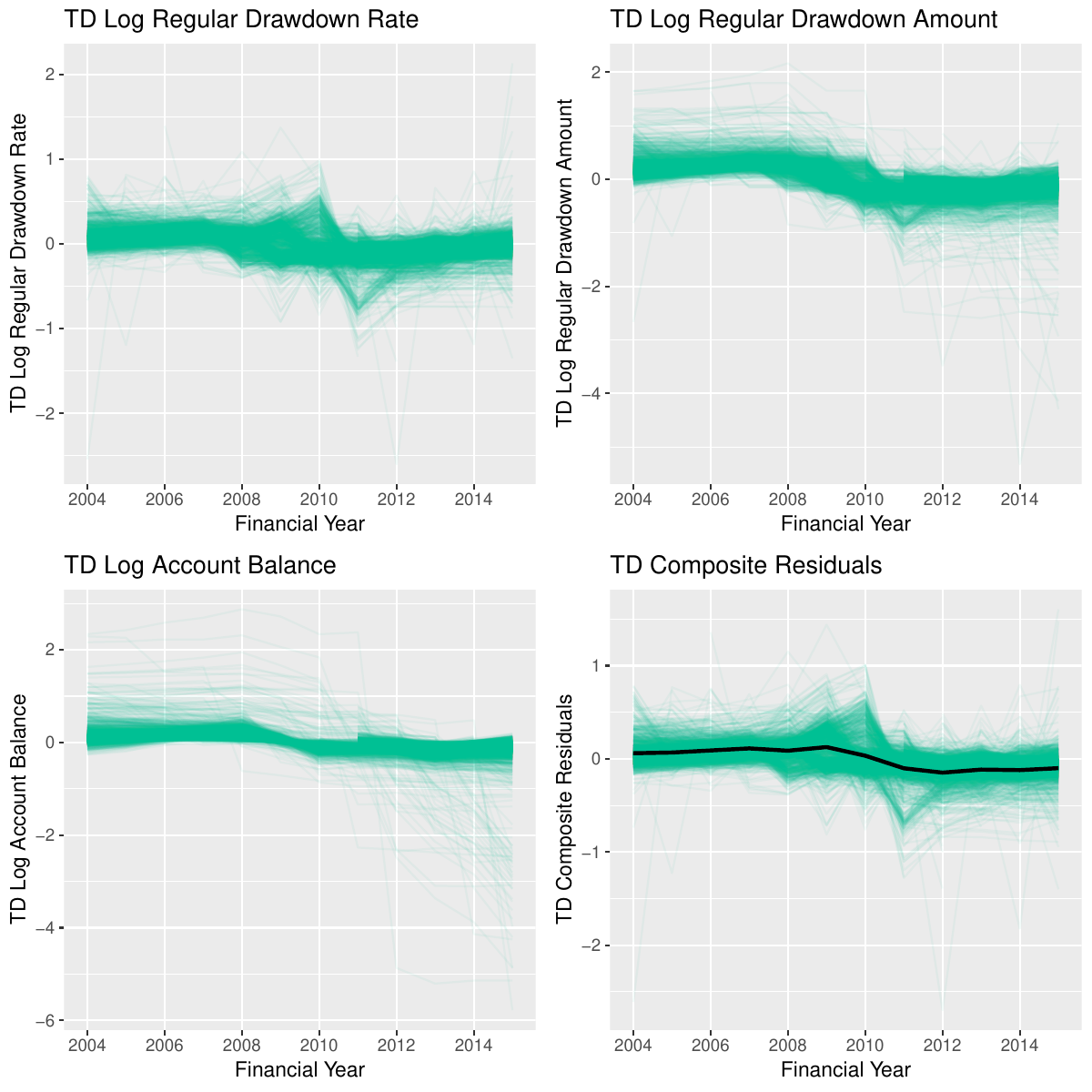}
	
\end{figure}

\begin{figure}
	\caption{\label{fig:g5-sixplots}$G=7$ model -- Group 5 time-demeaned (TD)
		panel plots. Account balances as at financial year start. The black
		series in the bottom-right panel represents estimated time-demeaned
		group time profile values.}
	
	\includegraphics[width=1\textwidth]{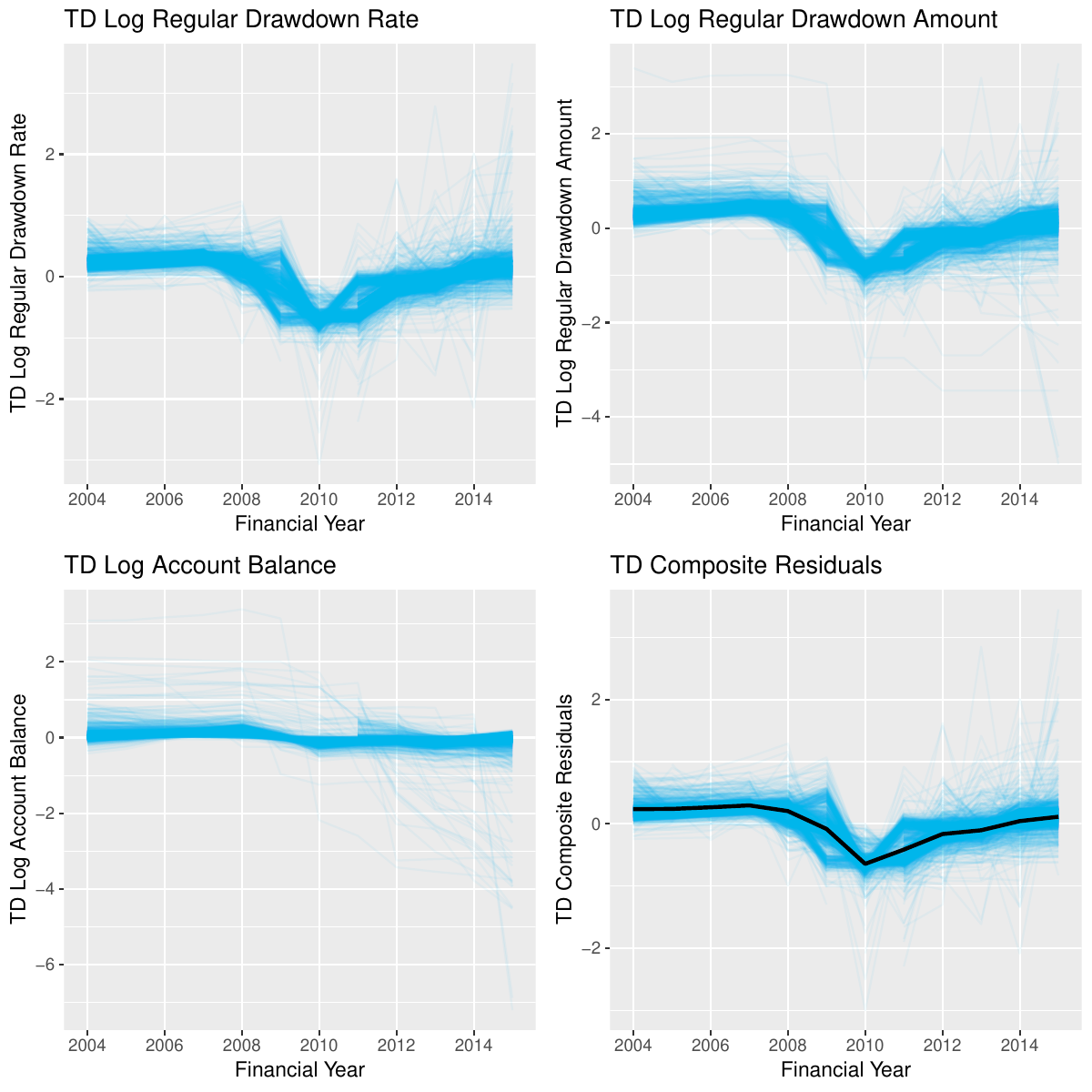}
	
\end{figure}

\begin{figure}
	\caption{\label{fig:g6-sixplots}$G=7$ model -- Group 6 time-demeaned (TD)
		panel plots. Account balances as at financial year start. The black
		series in the bottom-right panel represents estimated time-demeaned
		group time profile values.}
	
	\includegraphics[width=1\textwidth]{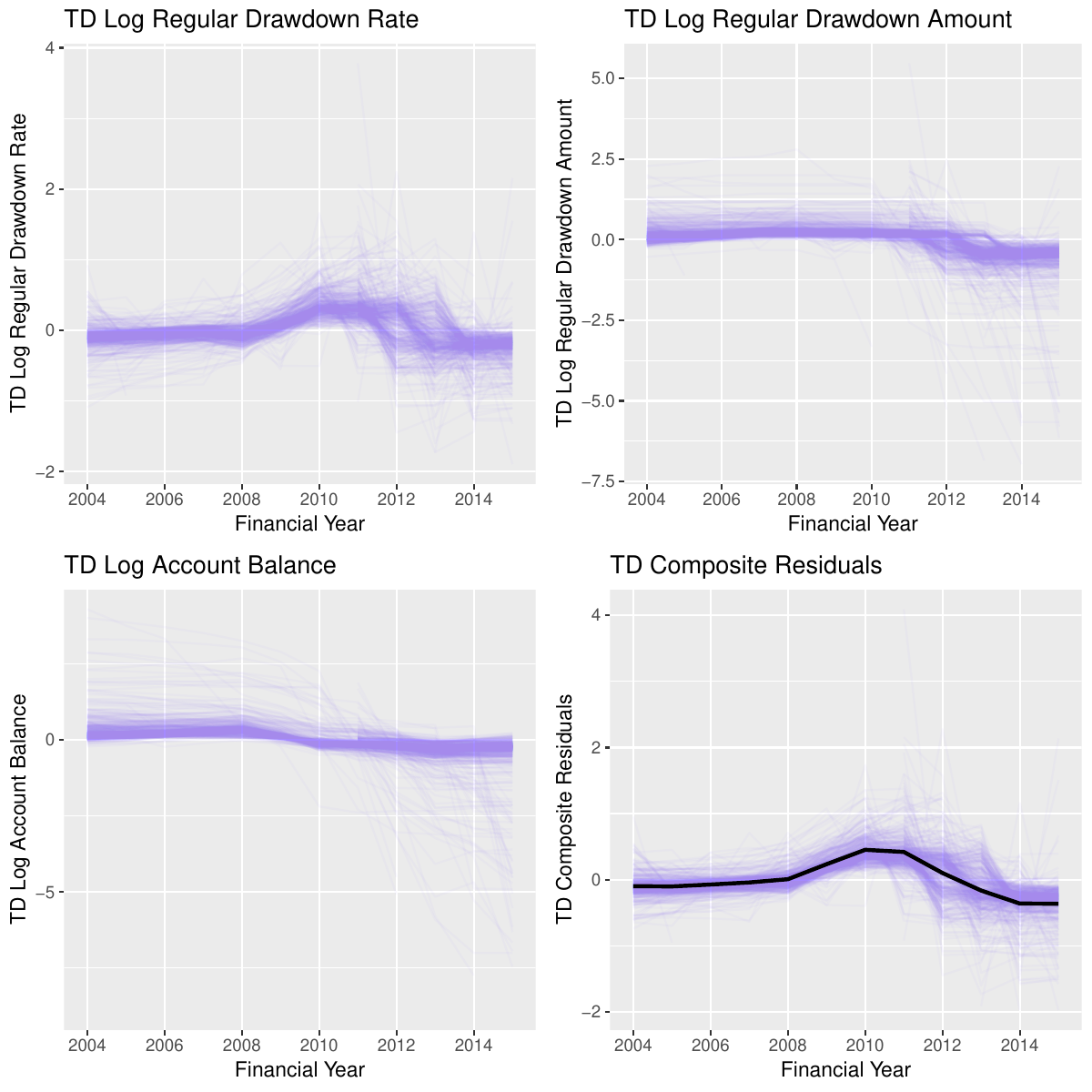}
	
\end{figure}

\begin{figure}
	\caption{\label{fig:g7-sixplots}$G=7$ model -- Group 7 time-demeaned (TD)
		panel plots. Account balances as at financial year start. The black
		series in the bottom-right panel represents estimated time-demeaned
		group time profile values.}
	
	\includegraphics[width=1\textwidth]{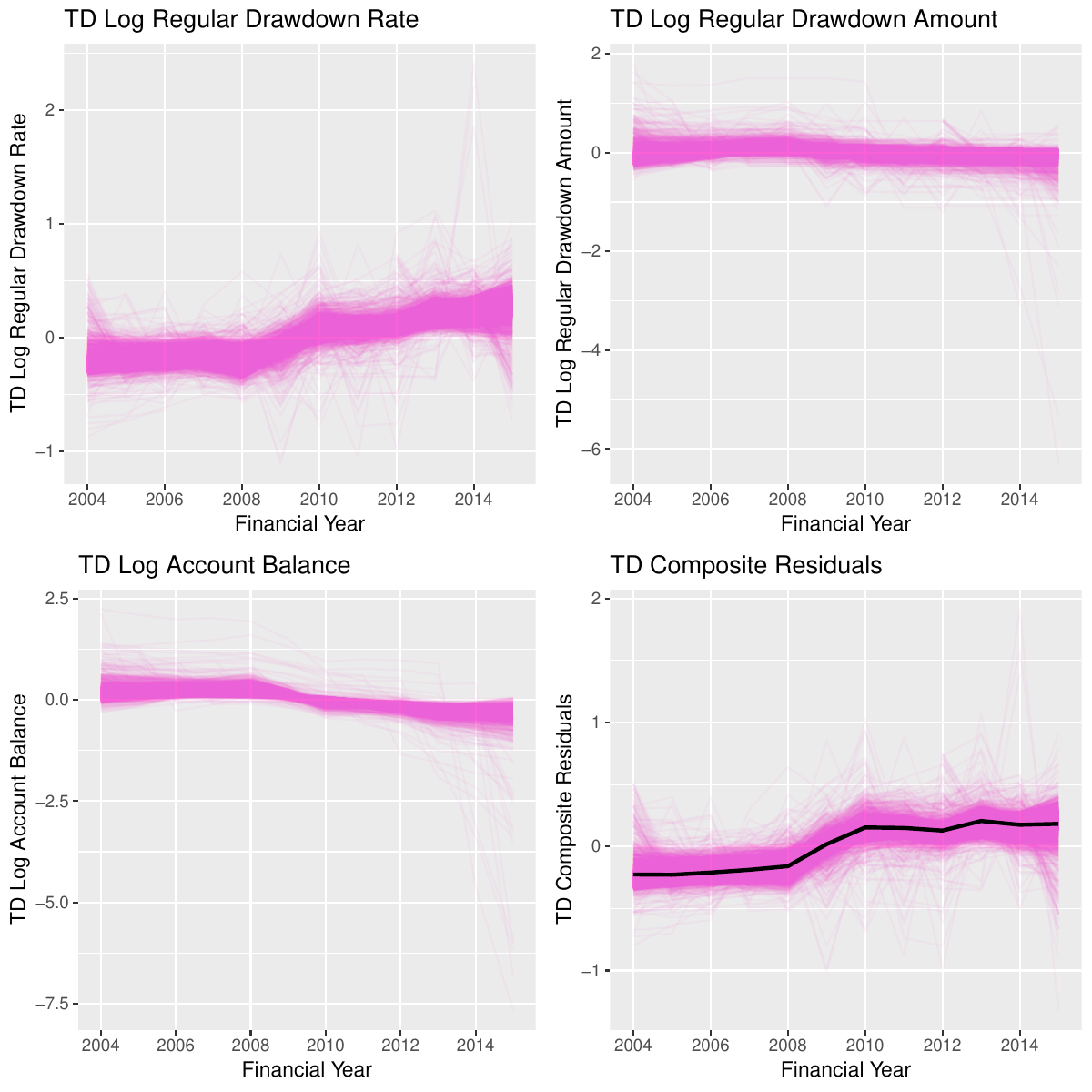}
	
\end{figure}

\clearpage

\phantomsection

\addcontentsline{toc}{section}{References}

\bibliographystyle{apacite} 

\bibliography{paper1subb-standalone}
